\documentclass[twocolumn,twocolappendix]{aastex631}

\usepackage[T1]{fontenc}
\usepackage{ae,aecompl}
\usepackage{multirow}
\usepackage{url}
\usepackage{enumitem}

\usepackage{graphicx}	
\usepackage{amsmath}	
\usepackage{amssymb}	
\usepackage{float}
\usepackage[caption=false]{subfig}

\defcitealias{mauerhofer2021}{M21}

\newcommand{\cII}{\ion{C}{2}}
\newcommand{\cIIl}{\ion{C}{2}~$\lambda1334$}
\newcommand{\cIIlstar}{\ion{C}{2}*~$\lambda1335$}

\newcommand{\siII}{\ion{Si}{2}}
\newcommand{\siIIl}{\ion{Si}{2}~$\lambda1260$}
\newcommand{\siIIlstar}{\ion{Si}{2}*~$\lambda1265$}

\newcommand{\classy}{\textsc{classy}}

\begin{document}

\title{Interpreting the Si II and C II line spectra from the COS Legacy Spectroscopic SurveY using a virtual galaxy from a high-resolution radiation-hydrodynamic simulation. }
\author[0000-0002-5659-4974]{Simon Gazagnes}
\affiliation{Department of Astronomy, The University of Texas at Austin, 2515 Speedway, Stop C1400, Austin, TX 78712-1205, USA}

\author[0000-0003-0595-9483]{Valentin Mauerhofer}
\affiliation{Kapteyn Astronomical Institute, University of Groningen, P.O Box 800, 9700 AV Groningen, The Netherlands}

\author[0000-0002-4153-053X]{Danielle A. Berg}
\affiliation{Department of Astronomy, The University of Texas at Austin, 2515 Speedway, Stop C1400, Austin, TX 78712-1205, USA}

\author[0000-0003-1609-7911]{Jeremy Blaizot}
\affiliation{Centre de Recherche Astrophysique de Lyon UMR5574, F-69230, Saint-Genis-Laval, France}

\author[0000-0002-2201-1865]{Anne Verhamme}
\affiliation{Department of Astronomy, University of Geneva, 51 Chemin Pegasi, 1290 Versoix, Switzerland}

\author[0000-0002-9613-9044]{Thibault Garel}
\affiliation{Department of Astronomy, University of Geneva, 51 Chemin Pegasi, 1290 Versoix, Switzerland}

\author[0000-0001-9714-2758]{Dawn K. Erb}
\affiliation{The Leonard E. Parker Center for Gravitation, Cosmology, and Astrophysics, Department of Physics,
University of Wisconsin-Milwaukee, 3135 N Maryland Avenue, Milwaukee, WI 53211, USA}

\author[0000-0002-2644-3518]{Karla Z. Arellano-C\'{o}rdova}
\affiliation{Department of Astronomy, The University of Texas at Austin, 2515 Speedway, Stop C1400, Austin, TX 78712-1205, USA}

\author[0000-0003-4359-8797]{Jarle Brinchmann}
\affiliation{Instituto de Astrofísica e Ciências do Espaço, Universidade do Porto, CAUP, Rua das Estrelas, PT4150-762 Porto, Portugal}

\author[0000-0002-0302-2577]{John Chisholm}
\affiliation{Department of Astronomy, The University of Texas at Austin, 2515 Speedway, Stop C1400, Austin, TX 78712-1205, USA}

\author[0000-0001-8587-218X]{Matthew Hayes}
\affiliation{Stockholm University, Department of Astronomy and Oskar Klein Centre for Cosmoparticle Physics, AlbaNova University Centre, SE-10691, Stockholm, Sweden}

\author[0000-0002-6586-4446]{Alaina Henry}
\affiliation{Space Telescope Science Institute, 3700 San Martin Drive, Baltimore, MD 21218, USA}

\author[0000-0003-4372-2006]{Bethan L. James}
\affiliation{AURA for ESA, Space Telescope Science Institute, 3700 San Martin Drive, Baltimore, MD 21218, USA}

\author[0000-0002-6790-5125]{Anne Jaskot}
\affiliation{Department of Astronomy, Williams College, USA}

\author[0000-0003-4270-5968]{Nika Jurlin}
\affiliation{Department of Astronomy, The University of Texas at Austin, 2515 Speedway, Stop C1400, Austin, TX 78712-1205, USA}

\author[0000-0001-9189-7818]{Crystal L. Martin}
\affiliation{Department of Physics, University of California, Santa Barbara, Santa Barbara, CA 93106, USA}

\author[0000-0003-0695-4414]{Michael Maseda}
\affiliation{Department of Astronomy, University of Wisconsin-Madison, 475 N. Charter St., Madison, WI 53706 USA}

\author[0000-0002-9136-8876]{Claudia Scarlata}
\affiliation{Minnesota Institute for Astrophysics, University of Minnesota, 116 Church Street SE, Minneapolis, MN 55455, USA}

\author[0000-0003-0605-8732]{Evan D. Skillman}
\affiliation{Minnesota Institute for Astrophysics, University of Minnesota, 116 Church Street SE, Minneapolis, MN 55455, USA}

\author[0000-0003-3903-6935]{Stephen M. Wilkins}
\affiliation{Astronomy Centre, University of Sussex, Falmer, Brighton BN1 9QH, UK}

\author[0000-0001-8289-3428]{Aida Wofford}
\affiliation{Instituto de Astronomía, Universidad Nacional Autónoma de México, Unidad Académica en Ensenada, Km 103 Carr. Tijuana-Ensenada, Ensenada 22860, México}

\author[0000-0002-9217-7051]{Xinfeng Xu}
\affiliation{Center for Astrophysical Sciences, Department of Physics \& Astronomy, Johns Hopkins University, Baltimore, MD 21218, USA}


\begin{abstract}
Observations of low-ionization state (LIS) metal lines provide crucial insights into the interstellar medium of galaxies, yet, disentangling the physical processes responsible for the emerging line profiles is difficult. This work investigates how mock spectra generated using a single galaxy in a radiation-hydrodynamical simulation can help us interpret observations of a real galaxy. We create 22,500  \cII\ and \siII\ spectra from the virtual galaxy at different times and through multiple lines of sight and compare them with the 45 observations of low-redshift star-forming galaxies from the COS Legacy Spectroscopic SurveY (\classy).  We find that the mock profiles provide accurate replicates to the observations of 38 galaxies with a broad range of stellar masses ($10^6$ to $10^9~M_\odot$) and metallicities (0.02 to 0.55~$Z_\odot$). Additionally, we highlight that aperture losses explain the weakness of the fluorescent emission in several \classy\ spectra and must be accounted for when comparing simulations to observations. Overall, we show that the evolution of a single simulated galaxy can produce a large diversity of spectra whose properties are representative of galaxies of comparable or smaller masses. Building upon these results, we explore the origin of the continuum, residual flux, and fluorescent emission in the simulation. We find that these different spectral features all emerge from distinct regions in the galaxy's ISM, and their characteristics can vary as a function of the viewing angle. While these outcomes challenge simplified interpretations of down-the-barrel spectra, our results indicate that high-resolution simulations provide an optimal framework to interpret these observations.
\end{abstract}

\section{Introduction} 
\label{sec:intro}

The interstellar medium (ISM) of star-forming galaxies is an intricate environment, constantly transformed by complex processes such as star formation and evolution, galactic outflows, and gas accretion. Understanding what regulates the ISM transformation through time is crucial for establishing accurate galaxy evolution models. Yet this effort relies on our ability to extract robust and accurate insights into their properties. In this context, low-ionization states (LIS) of metals such as \cII\ (ionization potential between 11.26 and 24.38 eV), \siII\ (ionization potential between 8.15 and 16.34 eV), or \ion{Fe}{2} (ionization potential between 7.90 and 16.19 eV) are powerful tracers of the neutral and low-ionization ISM: their associated transitions, most of which span the ultraviolet (UV) and far-UV wavelength regions, provide us with direct insights into the gas properties such as its metallicity, its geometry, or its kinematics. 

The typical approach to extracting insights from LIS metal lines is to analyze the ``down-the-barrel" absorption lines. The properties and shape of the absorption features are tightly connected to the gas velocity, column density, and covering fraction, which all describe different aspects of the galaxy's ISM. For example, the observation of broad blue-shifted absorption lines suggests the presence of outflowing gas \citep[e.g.,][]{heckman2000, heckman2015, shapley2003, erb2012, rubin2011, Wofford2013, chisholm2015,chisholm2016, chisholm2017mass,  finley2017wind, steidel2018, wang2020, Xu2022}. The analysis of the outflow properties (e.g., speed, structure, and ejected mass) can help us understand the baryon cycle of galaxies, \citep[i.e., the interplay between the outflow rates, gas accretion rate, and their impact on the star formation rate, see, e.g.,][]{Oppenheimer2010}. By combining column densities of LIS lines with neutral hydrogen, one can measure the metal content of the neutral gas, where a large proportion of the baryonic component is thought to lie. Also, by comparing these line properties with the ionized gas abundances, one can decipher mixing timescales between the two phases and ultimately, the metal enrichment of the ISM \citep[e.g.,][]{James2014, James2018, chisholm2018}. 

LIS absorption lines are relevant in the context of reionization studies. The observation of absorption features with large residual flux (i.e., the amount of flux at the maximum depth of the line) may probe the presence of low-column density channels through which ionizing photons can escape the ISM of the galaxy \citep[e.g.,][]{heckman2001, alexandroff2015, reddy2016stack, gazagnes2018, chisholm2018, steidel2018, mauerhofer2021, saldana2022}. This aspect makes LIS absorption lines promising diagnostics for predicting the escape fraction of ionizing photons ($f_{\rm esc}^{\rm LyC}$) at high redshift where it cannot be directly measured due to the opacity of the intergalactic medium (IGM).

Though UV LIS resonant lines are a common approach to exploring the ISM properties of star-forming galaxies, one can also explore the UV LIS ``fluorescent" line properties. Many LIS metal ions are not entirely resonant: their ground state is split into two spin levels (the level just above the ground state is referred to as the fine-structure level), providing an additional channel of photon (de-)excitation. Since the energy level difference of this fluorescent channel is smaller than for the resonant channel, we observe an additional spectral feature redder than the LIS absorption line\footnote{Fluorescent transitions are commonly indicated with a star, e.g., \siII*, \cII*, or Fe II*}, which provide distinct but complementary insights into the physical conditions within the galaxies. 

Over the past decades, multiple works have reported the detection of fluorescent emission in star-forming galaxies from $z \sim 0$ to 5.5  \citep{shapley2003,erb2010, rubin2011, erb2012, martin2012, jaskot2014, James2014, jaskot2019, bosman2019, wang2020}. Simultaneously, several studies used idealized outflow/inflow models or radiation-hydrodynamic simulations to investigate the factors regulating the observed strength of the fluorescent emission \citep{prochaska2011, scarlata2015, mauerhofer2021, carr2022}. Both simulations and observations highlighted that the fluorescent spectral features can trace the presence of outflowing gas, inform about the gas column density along the line of sight, and, importantly, provide clues into the spatial distribution of the absorbing gas \citep[see, e.g., discussion in][]{jaskot2014, rivera2015}. 

Although UV LIS absorption and emission lines have a high and valuable informative potential for constraining the properties of a galaxy's ISM, LIS metal line spectra remain difficult to interpret. The emerging profiles can exhibit complex, asymmetric shapes which result from the contributions of multiple gas clouds. Such clumpy configuration challenges simple interpretations of the line profiles. For example, it impacts the residual flux measurements so that one can only infer from them an upper limit to the geometric covering fraction \citep[see, e.g.,][]{vasei2016, rivera2017, gazagnes2018}. Additionally, commonly studied LIS lines (e.g., \ion{Si}{2} 1260 \AA, \ion{C}{2} 1334 \AA) are resonance lines that may be affected by “emission filling”  \citep[scattered line photons re-emitted in the line-of-sight direction, see, e.g.,][]{prochaska2011, erb2012, scarlata2015} which affects both the shape and properties of the absorption line. Finally, the analysis of the fluorescent emission also suffers some caveats. In particular, the fluorescent emission is typically more spatially extended than the continuum emission \citep{erb2012,finley2017wind, wang2020}, which makes its strength particularly sensitive to aperture losses \citep{prochaska2011, erb2012, finley2017VII, wang2020, mauerhofer2021}. Therefore, while the absorption and emission line analysis often relies on simple spectral diagnostics (e.g., the equivalent width or the residual flux) or idealized models, these approaches might only capture a fraction of the full information content available. 

The maturity of recent simulation techniques has enabled a significant step forward in disentangling the physical processes responsible for the absorption and emission line shapes. In particular, \citet[][hereafter, M21]{mauerhofer2021} provided a thorough analysis of the physics that govern the production of neutral and LIS line spectra using a high-resolution radiation hydrodynamics (RHD) simulation post-processed with radiative-transfer techniques. Their work provided crucial lessons for interpreting rest-frame UV and FUV spectra, uncovering major findings for the era of high-redshift observations with the James Webb Space Telescope (JWST) or with future 40m-class telescopes. Importantly, such a framework now enables us to thoroughly compare simulated spectra constructed from realistic galaxy environments to actual observations. For example, \citet{blaizot2023simulating} recently highlighted the remarkable consistency between the Lyman-$\alpha$ profiles seen in the simulation of \citetalias{mauerhofer2021} and the Lyman-$\alpha$ observations of low-$z$ star-forming galaxies.

In the present paper, we investigate whether the \citetalias{mauerhofer2021} simulation is able to reproduce the \cII\ and \siII\ spectra of 45 low-redshift star-formation galaxies from the COS Legacy Spectroscopic SurveY \citep[\classy,][]{berg2022} whose properties are similar to high-$z$ objects. Showing that it can, we further use the simulation to develop an interpretation of these complex observations. In particular, we explore the effects of limited aperture size, the time and sight-line variations of the spectral properties, and the origin of the different spectral features that compose a LIS line spectrum.

 This paper is organized as follows: Section~\ref{sec:data} details the simulation setup and the observations used in this work. Section~\ref{sec:global} compares the \siII\ and \cII\ properties and line profiles in the simulated and observed spectra, and Section~\ref{sec:apersec} analyzes the impact of aperture loss on the fluorescence emission properties. Section~\ref{sect:disc} discusses the variety of LIS line profiles generated using a single virtual galaxy and explores the spatial properties  of the stellar continuum, fluorescent emission, and residual flux in the simulation. Finally, Section~\ref{sec:conc} summarizes our results.

\section{Data} \label{sec:data}
The goal of this work is to compare the \cIIl+\cIIlstar\ and \siIIl+\siIIlstar\ line profiles from the  \classy\ observations to physically realistic models. To do so, we construct 22,500 mock UV spectra using the simulated galaxy in the high-resolution RHD simulation from \citetalias{mauerhofer2021} and explore their resemblance with the \classy\ spectra. Section~\ref{sec:dataobs} describes the \classy\ sample and Section~\ref{sec:datasim} details the simulation.

\subsection{Observations}
\label{sec:dataobs}
\classy\  \citep{berg2022} is an HST/COS treasury program that provided the first high-resolution, high signal-to-noise FUV spectral catalog of 45 local star-forming galaxies spanning similar properties as reionization-era objects in terms of masses (log(M$_\star$/M$_\odot$) $\approx$ 6 to 10), gas metallicities (0.02 to 1.2 $Z_\odot$), star formation rates (log(SFR) $\approx$ $-2.0$ to $+$1.5), and ionization parameters (log(U) $\approx$ $-3.3$ to $-2.0$). Figure~\ref{fig:simprop} provides more details into the distributions of stellar masses, SFR, and gas metallicities for the 45 \classy\ objects. 
The \classy\ spectra cover a rest-frame wavelength range from $\sim$~1200 to 2000 \AA, utilizing the full potential of the G130M+G160M+G185M gratings on HST/COS. The high data quality makes these observations ideal laboratories for determining the most promising FUV diagnostics for interpreting high-redshift observations. 

All the \classy\ data, including 135 new and 177 archival orbits,  were reduced consistently using \textsc{CalCOS} v.3.3.101. A complete description of the data reduction procedure is detailed in \citet{berg2022}, and a thorough technical overview of the different challenges is exposed in \citet{james2022}.  All observational data presented in this paper were obtained from the Mikulski Archive for Space Telescopes (MAST) at the Space Telescope Science Institute. The observations can be accessed via \dataset[10.17909/m3fq-jj25]{http://dx.doi.org/10.17909/m3fq-jj25}.


Because of detector gaps or Milky Way absorption line contamination, the wavelength region around 1260 \AA\ or 1334 \AA\ is not always usable for our analysis. Of the 45 galaxies, 39 and 37 have ``non-contaminated" \cIIl\ and \siIIl\ observations, respectively, and 32 have both wavelength regions observed simultaneously. All 45 galaxies have at least one of the two wavelength ranges covered, and we, therefore, consider the whole sample in our analysis. The next section details the RHD simulation setup.

\subsection{Simulation}
\label{sec:datasim}
In \citetalias{mauerhofer2021}, the authors use one galaxy from a zoom-in simulation to investigate the processes responsible for the shape of the Lyman-$\beta$ and \cIIl+\cIIlstar\ lines and discuss their results in the context of the escape of ionizing photons. In this work, we use the same framework to build a large dataset of 22,500 mock spectra that we compare to the \classy\ observations. Section~\ref{sec:zoomin} presents the zoom-in simulation and Section~\ref{sec:mocks} details the steps used to generate the mock spectra, building upon the strategy from \citetalias{mauerhofer2021} and \citet{blaizot2023simulating}. 

\subsubsection{Zoom-in simulation}
\label{sec:zoomin}

\begin{figure*}
    \centering
    \includegraphics[width = \hsize]{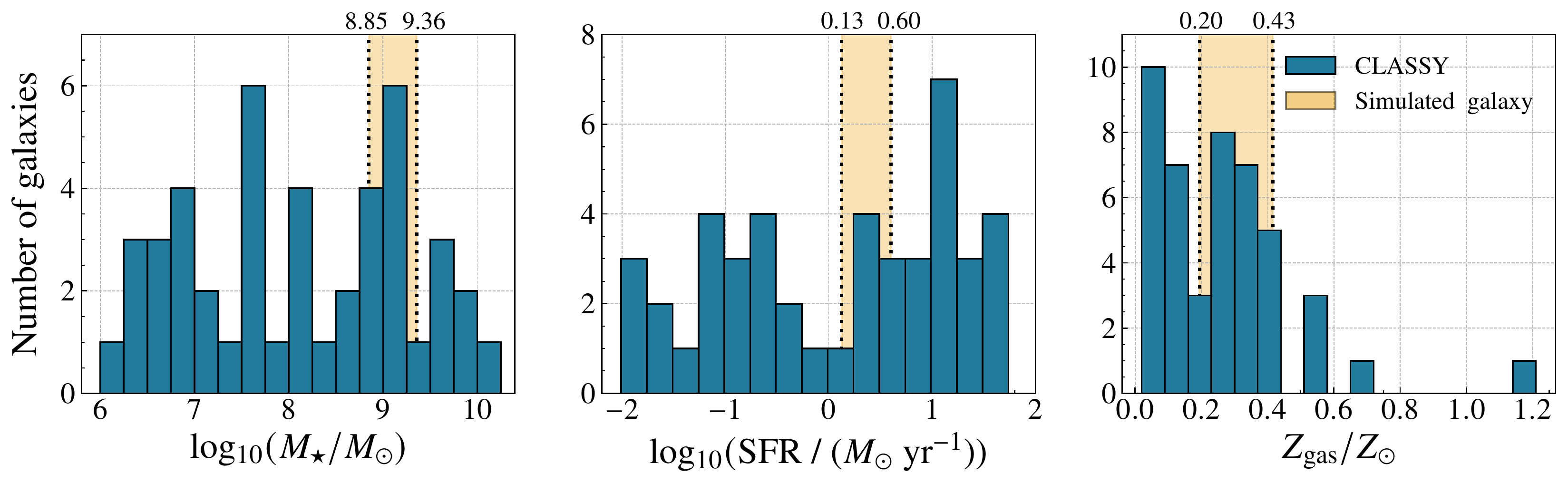}
    \caption{The range of stellar masses ($M_\star$), SFRs,  and ionized gas metallicity ($Z_{\rm gas}$) of the \classy\ sample (blue histogram) and of the virtual galaxy (orange stripe) in the zoom-in simulation during the 750 Myrs ($z = 4.19$ to $z = 3.00$) period considered in this work. }
    \label{fig:simprop}
\end{figure*}

The \citetalias{mauerhofer2021} simulation run was obtained with the adaptive mesh refinement code \textsc{Ramses-RT} \citep{teyssier2002, rosdahl2013, rosdahl2015}. The initial conditions were generated using \textsc{music} \citep{Hahn2013} and chosen such that the resulting galaxy has a stellar mass of $M_\star\sim10^9$ $M_\odot$ at $z = 3$, with a maximum cell resolution of 14 pc around this redshift.  The simulation builds upon the physics of cooling, supernova feedback, and star-formation from \textsc{sphinx} \citep{rosdahl2018}, and includes a self-consistent propagation of the ionizing photons through radiative transfer providing accurate non-equilibrium ionization states of hydrogen and helium atoms as well as radiative feedback. The ionizing radiation emission from stellar particles follows the \textsc{bpass} 2.0 stellar library models \citep{eldridge2008, stanway2016}, which have a maximum (cut-off) $M_\star$ of 100 $M_{\odot}$, with all stars set in binary systems. An additional ionizing UV background is added to all cells that have a density of hydrogen $n_{\rm H} < 10^{-2}$ cm$^{-3}$ to account for the ionizing radiation produced by external galaxies in a realistic universe \citep{Faucher2009}. 

We use 75 simulation outputs from $z~=~4.19$ to $z~=~3.00$ ($\sim$9.2 Myr time-steps). Over this period, the halo mass evolves from $4.0 \times 10^{10}$ to $5.6 \times 10^{10}$ $M_\odot$, the stellar mass increases from $7.1 \times 10^8$ to $2.3 \times 10^9$~${M_\odot}$, and both the averaged stellar and gas metallicity increase from $0.20Z_\odot$ to $0.42Z_\odot$. Additionally, during these 690 Myrs, the star-formation rate (SFR, measured using all stars whose age is less than 100 Myrs) oscillates between $\sim1.5$ and 4.0 $M_\odot$ yr$^{-1}$, with two main SFR peaks around 1.64 and 2.04 Gyrs. 

Figure~\ref{fig:simprop} compares the range of stellar mass, SFR, and gas metallicities in the virtual galaxy and in the \classy\ sample. While the virtual galaxy covers a narrower range of galaxy properties compared to the \classy\ sample, the 75 outputs considered in this work enable us to generate mock spectra at different phases in the galaxy evolution, which is interesting in the context of understanding the physical processes required to produce realistic LIS metal lines observations. Additionally, \classy\ galaxies are at $z$ between 0.002 and 0.182, hence a significantly lower redshift range than the range considered for the simulated galaxy ($z~=~4.19$ to $z~=~3.00$). However, the \classy\ targets have been selected to span similar properties as seen at high redshift, with, in particular, SFRs that are more typical of galaxies at cosmic noon \citep[see Figure 2 in][]{berg2022}. In addition, absorbing gas near the brightest stars dominates the down-the-barrel spectrum, so cosmological contexts that normally affect the circumgalactic medium gas may not be as significant for our analysis. Finally,  in the 690 Myrs period considered in this work, we note that the virtual galaxy did not experience any major merging events which could have significantly transformed its environment and made more difficult the interpretation of the line profiles. Similarly, in \classy, most of the targets are simple, isolated structures, suggesting that these objects did not undergo any recent major merging events.  We detail in the next section the procedure to create the mock \cII\ and \siII\ profiles.

\subsubsection{Creating the mock observations}
\label{sec:mocks}

In order to produce mock spectra, we first need to compute the number densities of C$^+$ and Si$^+$ ions everywhere in the simulation. We do this in two steps: first, we derive the number density of each element ($n_{\rm C}$ and $n_{\rm Si}$), and then we compute their ionization fractions. We derive the densities of carbon and silicon in each simulation cell assuming solar abundance ratios, such that $n_{\rm C} = n_{\rm H} A_{\rm C} Z_{\rm cell}/Z_\odot$ and $n_{\rm Si} = n_{\rm H} A_{\rm Si} Z_{\rm cell}/Z_\odot$, where $A_{\rm C}$ and $A_{\rm Si}$ are respectively the solar ratio of carbon ($2.69 \times 10^{-4}$) and silicon atoms ($3.24 \times 10^{-5}$) over hydrogen atoms taken from \citet{asplund2021}, $Z_\odot$ is the solar metallicity \citep[0.0134,][]{grevesse2010}, and $Z_{\rm cell}$ is the value of the metal mass fraction in each cell. While not included in \citetalias{mauerhofer2021}, this work also considers the phenomenon of dust depletion: large fractions of metals are incorporated into dust grains and hence ``removed" from the ISM gas. We adopt the depletion fraction of metals as a function of gas-phase metallicity from \citet{decia2016} and \citet{Konstantopoulou2022} ($X^{\rm dust}_{\rm Si} = 1-10^{-0.04-0.72*(1.09 +0.60Z)}$ and $X^{\rm dust}_{\rm C} = 1-10^{-0.1*(1.09 +0.60Z)}$)  and apply the correction formula cell by cell in the simulation. We note that both observations \citep{jenkins2009, Parvathi2012_Cextincction} and simulations \citep{Choban2022} highlighted that dust-depletion fractions also depend on the average gas density along the line of sight. In this work, we do not apply a density-dependent model for the depletion fractions as there currently does not exist simple implementations of such models. Yet, as further discussed below, we tested and compared different sets of mock spectra generated using different models. In the context of dust depletion, while we do observe some slight differences in the distribution of spectral properties that are generated, we do not find that it affects the results related to the comparison of the \classy\ sample, which is the main focus of this work.


The ionization fractions of carbon and silicon are computed using \textsc{krome} \citep{grassi2014} using the recombination rates from \citet{badnell2006}, the collisional ionization rates from \citet{voronov1997}, and the photoionization cross-sections from \citet{verner1996}. The complete details of this procedure are given in \citetalias{mauerhofer2021}. Note that \textsc{krome} assumes chemical equilibrium for both ions. To account for potential effects from non-equilibrium ionization states for \siII\ and \cII, one would have to self-consistently propagate these elements in the \textsc{Ramses-RT} simulation run which is not trivial. Yet, accounting for non-equilibrium ionization states would mostly affect the diffuse gas which, in theory, has little impact on the absorption spectra (emerging from denser gas).

\begin{figure*}[!htbp]
    \centering
    \includegraphics[width = 0.48\textwidth,trim={0 0.83cm 0 0},clip]{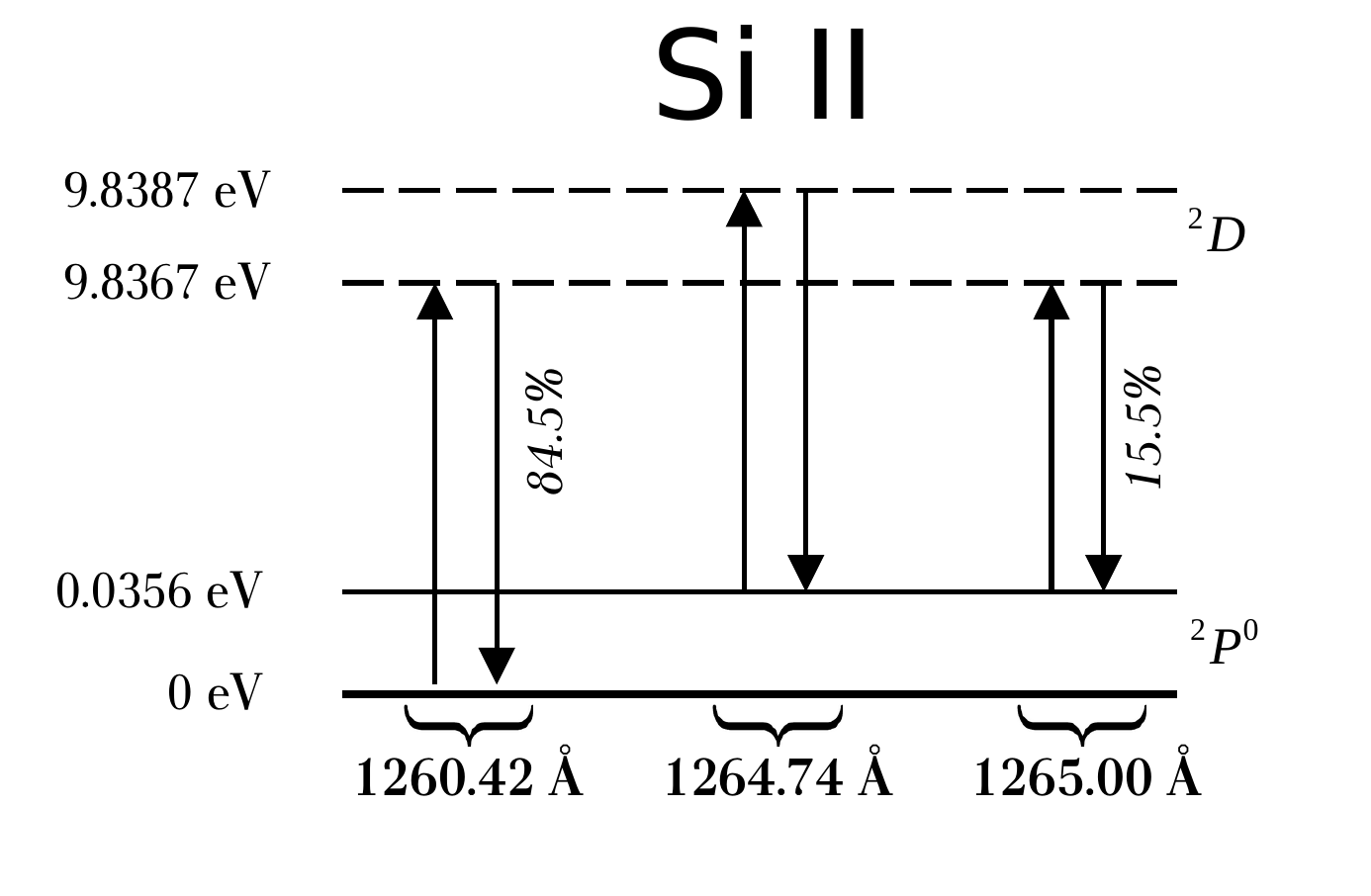}
    \includegraphics[width = 0.48\textwidth,trim={0 1cm 0 0},clip]{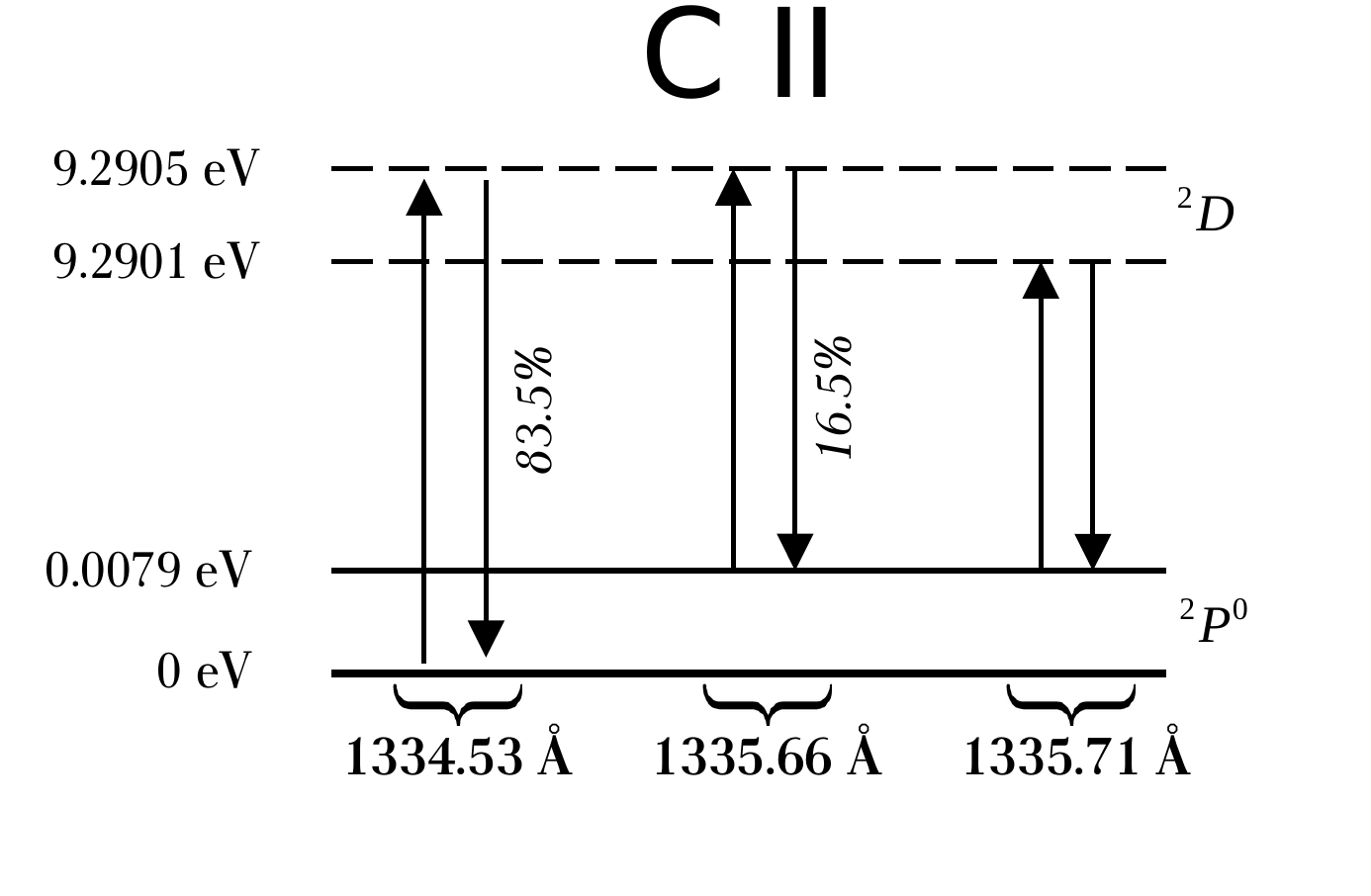}
    \caption{Energy levels of the \siII\ and \cII\ ions, illustrating the different channels of excitation and de-excitation from the ground state or fine-structure level. The percentage values indicate the probabilities that an electron from the excited level is de-excited to the ground state (resonance) or to the fine structure level (fluorescence). A photon absorbed at 1260.42~\AA\ (1334.53~\AA) has a 15.5\% (16.5\%) chance to be re-emitted at the wavelength of the fluorescence for \siII\ (\cII).   Figure adapted from \citetalias{mauerhofer2021}.}
    \label{fig:levels}
\end{figure*}

 As noted in the introduction, the ground-state energy level of the \cII\ and \siII\ ions is split into two spin states having an energy difference of 0.0079 eV for \cII\ and 0.0356 eV for \siII\ (see Figure~\ref{fig:levels}). These fine-structure levels, typically populated through electron collisions, produce two additional channels (referred to as fluorescent channels, noted as \cII* and \siII*) of photon absorption and emission at 1335.66 \AA\ and 1335.71~\AA, respectively, for \cII, and at 1264.73~\AA\ and 1265.02~\AA, respectively, for \siII. Typically, the phenomenon of fluorescence works as follows: a photon absorbed at 1260.42~\AA\ or 1334.53~\AA\ can either be re-emitted at the same wavelength (\textit{resonance}) or at the wavelength of the fine-structure level (\textit{fluorescence}). The probability of either phenomenon depends on the transitions' Einstein coefficients, and the probability of resonance scattering is higher than for the fluorescence (84.5\% and 83.5\% for \siII\ and \cII, respectively).  Typically, fluorescent emission emerges from regions where resonant photons undergo multiple scatterings and are converted to fluorescent emission \citep{scarlata2015}.
 
 Fluorescent lines are usually observed as emission lines \citep[e.g.,][]{shapley2003, erb2010, jones2012, martin2012, finley2017VII, wang2020}, yet, the physics of the fine-structure levels depends on a variety of factors (e.g., density, dust, radiation fields) such that in some cases, absorption features dominate \citep{wolfe2003, lebouteiller2013, jaskot2019, gatkine2019, Xu2023}. This is because, as emphasized by Figure~\ref{fig:levels}, the first energy level is also split in two such that there exist two additional channels of excitation for an electron in the fine-structure level.  Similar to \citetalias{mauerhofer2021}, we consider the physics of all resonant and non-resonant channels of \cII\ and \siII\ when applying the radiative transfer post-processing. The percentage of ions being in the fine-structure level is determined using \textsc{pyneb} \citep{luridiana2015} assuming the default atomic data.


To produce the LIS absorption and emission lines in any direction, we use the radiative transfer post-processing code \textsc{rascas} \citep{rascas2009}. \textsc{rascas} produces the line profiles using the following procedure: one million photon packets sample the spectra of stellar populations contained in each star particle, within a wavelength range of $\pm$ 10 \AA\ around the \cIIl\ and \siIIl\ lines. The photon packets are assigned to stars with a probability that scales with the star particle luminosity. Their position is assigned to the position of the stellar particles in the simulation and their initial direction is drawn from an isotropic distribution. The frequency of the photon packet is drawn randomly from a power law based on the stellar energy distribution of the \textsc{bpass} model with age and metallicity equal to those of the particle.  Photon packets are propagated through the simulation grid and encounter at each cell an optical depth which is the sum of all channels of interaction:

 \begin{equation}
 \begin{split}
     \tau_{\rm cell} = \tau_{\rm C_{II}\ 1334.53}\ +\ \tau_{\rm C_{II}*\ 1335.66}\ +\ \tau_{\rm C_{II}*\ 1335.71}\ +\\ \tau_{\rm Si_{II}\ 1260.42} + \tau_{\rm Si_{II}*\ 1264.73} + \tau_{\rm Si_{II}*\ 1265.02} + \tau_{\rm dust}
     \end{split}
 \end{equation}
 
\noindent The optical depth of each line is the product of the ion column density in each cell and the ion cross section which is defined by Eq~(2) in \citetalias{mauerhofer2021}\footnote{The \cII\ and \siII\ line parameters are taken from the NIST atomic database \url{https://www.nist.gov/pml/atomic-spectra-database}}.
 

We compute the turbulent velocity of the gas ($v_{\rm turb}$) and the optical depth of the dust grains ($\tau_{\rm dust}$) in each cell. We determine $v_{\rm turb}$ using a formula that accounts for the gas density and velocity in the neighboring cells, as in the star-formation recipe of e.g., \citet{trebitsch2021}. This method provides a more physical prescription of $v_{\rm turb}$ than
\citetalias{mauerhofer2021} whose authors assumed a fixed value of 20 km s$^{-1}$. This model results in a broader range of $v_{\rm turb}$ which varies from a few km s$^{-1}$ to hundreds of km s$^{-1}$. For $\tau_{\rm dust}$, we follow the implementation of \citet{katz2022}, inspired by the results of \citet{remyruyer2014}, which adopts a broken power-law function of the metallicity and is set to 0 in cells with temperatures above 10$^5$ K. This temperature cut is not applied when calculating the dust-depletion fractions. Nevertheless,  because the number densities of C+ and Si+ in cells with T $>10^5$ K are relatively small \citep[3\% of C and $\sim0.1\%$ of Si, see][]{mauerhofer2021thesis} in this simulation, we assume that applying this cut would not impact our results. 

\citetalias{mauerhofer2021} shows that the choice of turbulence and dust models can significantly impact the shape and properties of the absorption and emission lines produced. We, therefore, tested how our results are affected by using different radiative-transfer ``recipes" to produce the line profiles (e.g., varying the dust model or $v_{\rm turb}$ parameter, accounting for dust depletion). We observed slight changes in the distribution of the spectral line shapes and properties depending on the model used.  As a result of these preliminary tests, we found that including the dust depletion fractions and calculating $v_{\rm turb}$ and $\tau_{\rm dust}$ using the approaches described in the current work seem to provide a more accurate description of the environments of the \classy\ galaxies (i.e., the best similarity between the simulated and observed line shapes and properties). Yet, importantly, we also noticed that using any of these models would still provide similar outcomes so that the main conclusions drawn in this work would not be affected. Because determining the impact of different recipes on the properties of the simulated spectra is an important aspect of providing physically realistic models of the environments of star-forming galaxies, an upcoming paper (Mauerhofer et al, in prep.) will detail the results of the additional tests  performed in the context of this work.

We compute the radiative transfer of the UV continuum through the simulated galaxy and its ISM/CGM in post-processing. In practice, we emit photon packets from all the star particles located within the virial radius (R$_{\rm vir} =$ 28 kpc) and compute their propagation and scattering until they reach 3 $\times$ R$_{\rm vir}$ or are absorbed by dust. Along the process, we use the peeling-off technique to construct spectra in 300 directions (corresponding to a regular sampling of the sphere obtained with HealPix) and within an aperture of 1'' diameter ($\sim$7 kpc at $z = 4.2$ and 7.9 kpc  at $z=3$), which is large enough to capture the full ISM extent of the simulated galaxy, plus some of the light scattered in the circum-galactic medium (CGM)\footnote{Because the separation between ISM and CGM is not well-defined, we tend to refer only to the ISM when discussing our results.}. In total, we construct 22,500 \cII\ and \siII\ mock spectra based on 300 different lines of sight observations of each of the 75 simulation outputs. In all this work, we use the noiseless, synthetic spectra, but adapt their resolution to that of the \classy\ spectra (see details in Section~\ref{sec:global}).  In the next section, we detail the comparison between these simulated mock profiles and the \classy\ observations.

\section{Comparing the C II and Si II line profiles}
\label{sec:global}
In this section, we compare the 22,500 mock LIS spectra constructed using the simulated galaxy with the \classy\ observations. Importantly, we simultaneously compare  the \siII\ and \cII\ properties and profiles. This exercise is more difficult than simply looking at how well each ion property/profile is reproduced individually but provides additional insights into the accuracy of the physical models used in the simulation. Section~\ref{sec:lineprop} first compares the lines' properties and Section~\ref{sec:fitting} determines the best-matching mock spectrum for each \classy\ observation. 

\subsection{Comparison of line properties}
\label{sec:lineprop}

Here we measure and compare (i) the equivalent widths (EW) of the \cIIl\ and \siIIl\ absorption lines, (ii) the equivalent widths of the \cIIlstar\ and \siIIlstar\ fluorescent emission, (iii) the residual fluxes ($R_f$) of the absorption lines, and  (iv) the absorption lines' central velocities ($v_{\rm cen}$). In this work, the $v_{\rm cen}$ values for each line are relative to the systemic velocity of the objects, which is defined as the velocity of the center of mass for the virtual galaxy, and as the offset velocity of the strong emission lines (e.g., H$\alpha$, H$\beta$, \ion{O}{3}) for the \classy\ galaxies.

\begin{figure}[!htbp]
    \centering
    \includegraphics[width = \hsize]{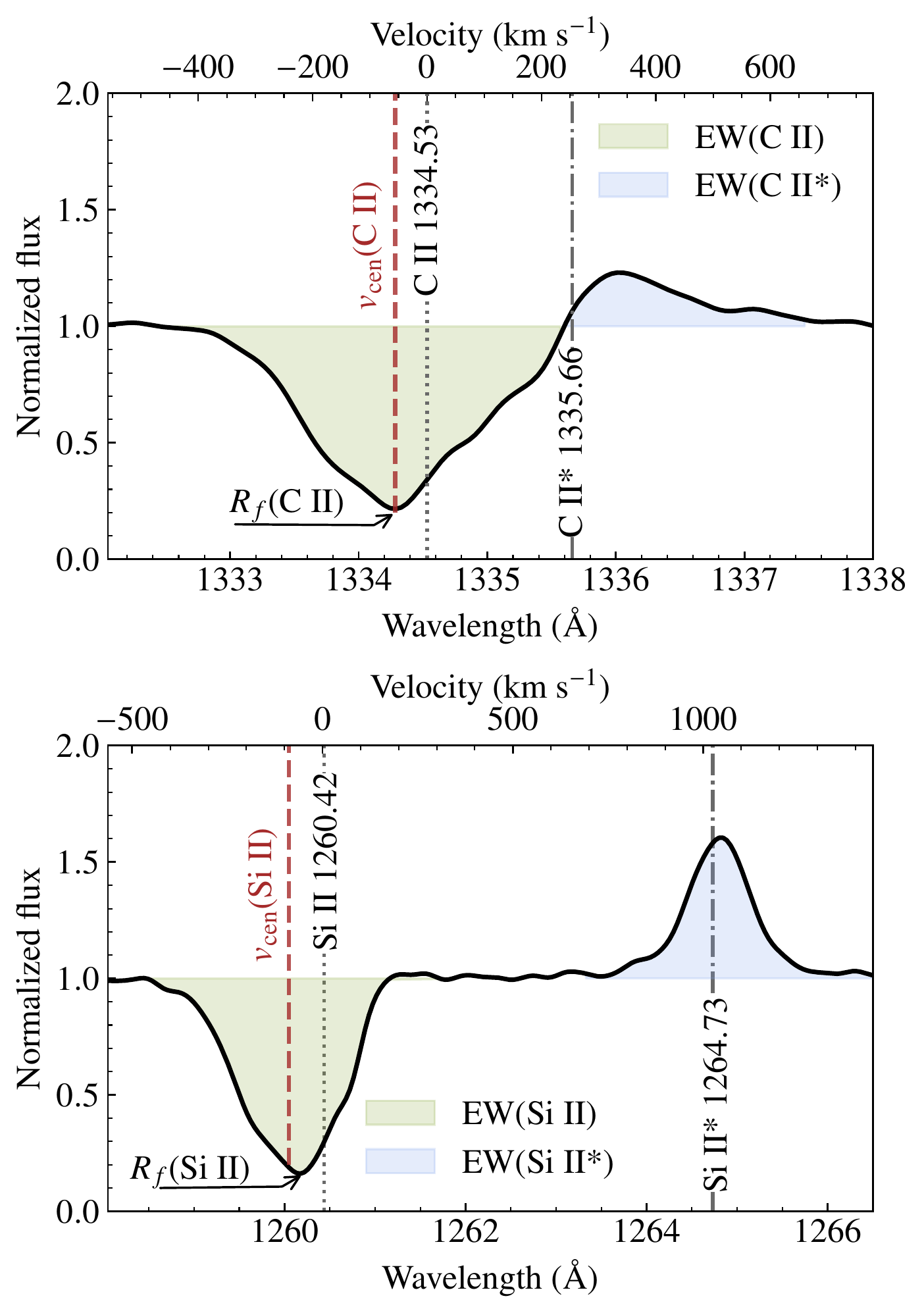}
    \caption{An example of \cII\ (top panel) and \siII\ (bottom) simulated spectrum with the characteristic measurements annotated. Each spectrum has been normalized using the median of the flux in intervals around 1256~\AA\ and 1331~\AA\ for \siII\ and \cII, respectively. } 
    \label{fig:measurements}
\end{figure}

We used a simple and consistent procedure to derive these properties for the 45 \classy\ observations and for the 22,500 mock profiles. First, the \siII\  and \cII\ spectra are normalized using the median of the flux in 1\AA\ wide feature-free intervals around 1256 \AA\ and 1331 \AA, respectively. Then, we use a Gaussian Line Spread Function to smooth all simulated spectra to a resolution of 60~km~s$^{-1}$, matching the median resolution for the \classy\ spectra \citep[see Table 3 in][]{berg2022}. Finally, we measure each property as follows:

\begin{itemize}
    \item EW(\cII) and EW(\siII) are derived by integrating the flux below the normalized continuum in an interval of $\pm$2.5~\AA\ ($\approx$600~km~s$^{-1}$) around $\lambda = 1334.53$~\AA\ and $\lambda = 1260.42$~\AA, respectively.\footnote{In this work, EWs are calculated such that absorption features have positive values and emission features have negative values.}
    \item EW(\cII*) and EW(\siII*) are derived by integrating the flux above the normalized continuum in an interval of $\pm$2~\AA\ ($\approx$400~km~s$^{-1}$) around  $\lambda = 1335.71$~\AA\ and $\lambda = 1265.02$~\AA, respectively. 
    \item $R_f$(\cII) and $R_f$(\siII) are obtained by taking the flux value at minimum depth of the absorption line in an interval of $\pm$2.5~\AA\ ($\sim$600~km~s$^{-1}$)  around $\lambda = 1334.53$~\AA\ and $\lambda = 1260.42$~\AA, respectively.
    \item The central velocities, $v_{\rm cen}$(\cII) and  $v_{\rm cen}$(\siII), are derived by taking the velocity at which the \cIIl\ and \siIIl\ absorption lines can be separated in two components holding each half the total equivalent width. These values are relative to the systemic velocity.
\end{itemize} 

\noindent  Figure~\ref{fig:measurements} presents an example of \siII\ and \cII  simulated spectrum with the characteristic measurements annotated. It is important to keep in mind that the EW measurements only provide information on the strength of the resonant absorption and of the fluorescent emission in the different spectra. By construction, our measurement approach always yields EW(\cII) and EW(\siII) values $\geq$~0, and EW(\cII*) and EW(\siII*) $\leq$~0. Consequently, spectra exhibiting pure emission around \cIIl\ and \siIIl\ (due to resonant scattering) will have EW(\cII) and EW(\siII) equal to 0. Similarly, spectra showing pure absorption around the \cIIlstar\ and \siIIlstar\ will have EW(\cII*) and EW(\siII*) equal to 0. 
 
\begin{figure*}[!htbp]
    \centering
    \includegraphics[width = \textwidth]{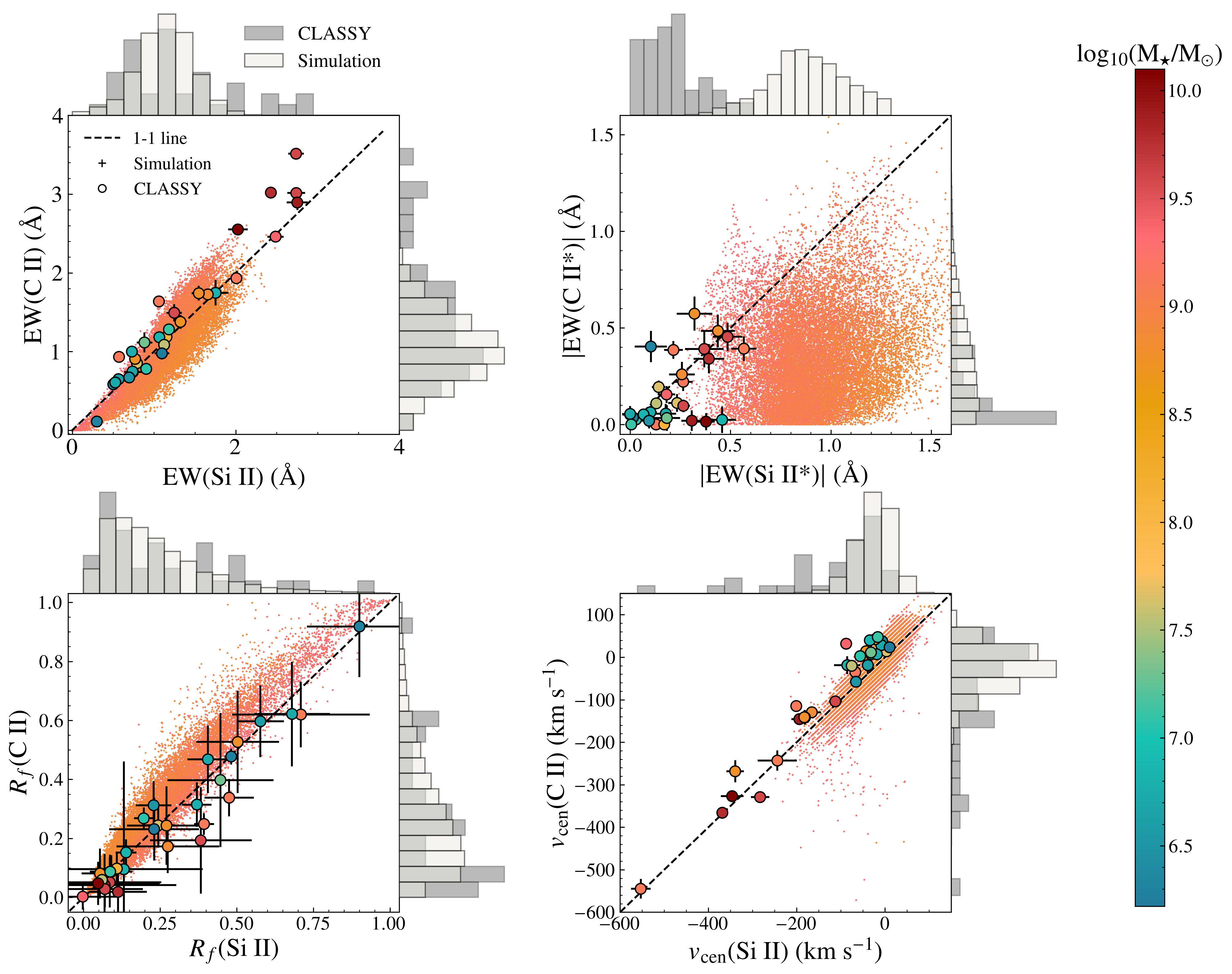}
    \caption{The comparison of the EW(\cII) and EW(\siII) (top left), $|$EW(\cII*)$|$ and $|$EW(\siII*)$|$ (top right), $R_f$(\cII) and $R_f$(\siII) (bottom left), and $v_{\rm cen}$(\cII)  and $v_{\rm cen}$(\siII) (bottom right) for the 22,500 mock spectra in the simulations (small points) and the \classy\ observations (circles). The procedure used to derive each measurement is given in the text. Each point is color-coded by its stellar mass (for the simulation points, it corresponds to the stellar mass of the simulation output used to construct the spectra). The dotted lines indicate the 1-1 relations. On the side of these panels, we show histograms presenting the 1-D distribution of the measurements. Except for \siII*, the simulated spectra properties overlap with 87\% of the \classy\ sample.}.
    \label{fig:glob}
\end{figure*}

The different velocity intervals used to measure each property have been chosen to capture the maximal extent of the absorption and emission line profiles observed in the \classy\ and simulated spectra. Nevertheless, our approach has two main caveats. First, we use intervals of 600 and 400~km~s$^{-1}$ to measure EW(\cII) and EW(\cII*), yet, the \cIIl\ and \cIIlstar\ transitions are separated by only $\sim250$ km~s$^{-1}$. Hence, absorption at the fluorescent wavelength will contribute to the value of EW(\cII), and resonantly scattered emission around 1334.54~\AA\ will contribute to the value of EW(\cII*). The blending of both resonant and fluorescent contributions does not impact the \siII\ measurements since the \siIIl\ and \siIIlstar\ transitions are separated by $\sim$1,100 km~s$^{-1}$. 

Second, there exists a \ion{S}{2} absorption line at 1259.52~\AA, resolved in some \classy\ spectra, which is often blended with the \siIIl\ line. While it is typically much weaker than the \siII\ absorption feature, its contribution sometimes impacts the EW(\siII) and $v_{\rm cen}$(\siII) measurements from the \classy\ spectra. This line is not included in our radiative transfer model. 


The impact of these caveats is further discussed in the context of the joint \siII\ and \cII\ properties analysis below. Overall, they do not affect the main results presented in this paper which go beyond the simple comparison of the line spectral properties. An upcoming paper will detail the refined measurements of the LIS properties in the \classy\ spectra (Parker et al., in prep). 


\subsubsection{Line properties ranges}
\label{sec:1D}

Here we discuss the histograms of the \cII\ and \siII\ properties, shown in the top and right parts of each panel of Figure~\ref{fig:glob}, for \classy\ and the simulation.
The range of EWs in both absorption lines (upper left panel) is typically between 0 and $2.5$~\AA\ in the simulated spectra and overlaps with most of the \classy\ measurements. We denote five \classy\ objects, with stellar masses larger than the virtual galaxy ($M_{\star} > 10^{9.5}$~$M_\odot$), which have EWs $>$ 2.5\AA. The range of residual fluxes (lower left panel) in both lines covers a similar interval for the observed and mock spectra, from 0 to 1. The mock line profiles have $v_{\rm cen}$ values between $\sim-300$ to 150 km~s$^{-1}$ (lower right panel). Most \classy\ galaxies have $v_{\rm cen}$ values within that range, and four objects have $v_{\rm cen}$(\siII) $<-300$ km~s$^{-1}$. Despite a few \classy\ outliers having either larger EWs or $v_{\rm cen}$ values, the overlap between the ranges of the resonant line properties is remarkable.\\

Regarding the fluorescent emission properties (upper right panel), the range of $|$EW(\cII*)$|$ is between 0 and 1 for both samples. However, the simulated spectra typically have $|$EW(\siII*)$|$ between 0.4 and 1.5 \AA\ while \classy\ galaxies have  $|$EW(\siII*)$|$ between 0 and 0.6~\AA. There are several possible factors that could explain the fact that we observe an overlap for $|$EW(\cII*)$|$ but not for $|$EW(\siII*)$|$. First, one must keep in mind that the separation between the \cII* fluorescent and \cII\ resonant wavelengths is small. Consequently, \cII* photons emitted close to the resonance (i.e., without a large velocity offset) could be re-absorbed in the resonant channel as they travel in the ISM. This makes the \cII* emission more difficult to interpret because it suggests that we only observe a certain fraction of the total \cII* emission that is produced. Second, these differences might also suggest that either the physics of the simulation or of our radiative transfer recipe cannot fully reproduce the fluorescent emission properties of the \classy\ objects. In particular, differences in the abundance ratios, assumed solar in this work, or in the number of ions populating the fine-structure level for C$^+$ and Si$^+$, would lead to a different EW(\cII*) to EW(\siII*) relation.  \citetalias{mauerhofer2021} also demonstrate that the turbulent velocity and/or dust opacity model can considerably influence the shape and properties of the fluorescent emission. Finally, aperture losses also largely affect the observed \cII* and \siII* emission, and their impact may vary for \cII* and \siII*. We discuss this latter aspect in Section~\ref{sec:apersec}, and a separate analysis, focused on the simulation and its post-processing will further investigate the physical mechanisms that can further explain the strength of the fluorescent emission in the simulated spectra.



\subsubsection{Si II versus C II}
\label{sec:2d}

Figure~\ref{fig:glob} also reveals some interesting trends connecting the resonant \cII\ and \siII\ properties. In the simulated spectra, EW(\cII) $<$ EW(\siII) for weak absorption lines (EWs $<1$~\AA), and $v_{\rm cen}$(\cII) more negative (positive) than  $v_{\rm cen}$(\siII) for blueshifted (redshifted) absorption lines. These trends can be explained by the impact of the \cIIlstar\ features on the \cIIl\  absorption line. Spectra with significant \cII* fluorescent emission will have lower EW(\cII), and more negative $v_{\rm cen}$(\cIIl) because the emission component fills a significant portion of the absorption line. On the other hand, spectra with broad \cIIl\ absorption lines also exhibit a significant absorption component at the fluorescent wavelength. The combined resonant and fluorescent absorption feature yields larger EW(\cII) and more positive $v_{\rm cen}$(\cII) compared to \siII.

The lower left panel of Figure~\ref{fig:glob}  shows that $R_f$(\cII) is typically larger than $R_f$(\siII) in the simulation. This is because the \cIIl\ lines are typically more affected by infilling due to resonant scattering than the \siIIl\ lines.

It is interesting to note that the \classy\ measurements exhibit slightly different behaviors compared to the simulated data.  In particular, we find that the \classy\ objects typically have EW(\siII)~$>$~EW(\cII), $v_{\rm cen}$(\siII)~$>$~$v_{\rm cen}$(\cII), and $R_f$(\cII) $\approx$ $R_f$(\siII). The two former trends can be explained by the presence of the \ion{S}{2} $\lambda1259$ absorption line blended with the \siIIl\ line, which is not modeled in the simulated spectra. This additional absorption on the blue side of the \siII\ line biases the EW and $v_{\rm cen}$ measurements towards larger and lower values, respectively. Appendix~\ref{app:globaltrends} presents examples of spectra that clearly highlight the impact of the \ion{S}{2} absorption line. 

Still, the impact of the \ion{S}{2} absorption line cannot explain why the \classy\ spectra have $R_f$(\cII) $\approx$ $R_f$(\siII) while the simulated spectra yield $R_f$(\cII) $\geq$ $R_f$(\siII). Interestingly, residual flux discrepancies between different ions (or different transitions of the same ion) have been highlighted in several observational works that also found evidence that the LIS absorption lines were saturated \citep{reddy2016stack, gazagnes2018, du2021,saldana2022}.  In general, we expect low-ionized metallic ions to co-exist within the same gas phase, hence to have the same residual flux assuming the transitions are saturated. Yet, the \cII\ ionization potential is slightly higher than \siII, such that it can have a higher contribution from the ionized gas. The different causes of the residual flux disparities can either be differences in the saturation of the transition, in the amount of infilling due to resonant scattering, or also systematic differences in the ionized and neutral fractions, or Si to C abundances between the \classy\ galaxies and the simulation. In Section~\ref{disc:spatial}, we explore the spatial distribution of the residual flux and discuss further the most probable origin of this discrepancy.


Finally, we note that the \classy\ fluorescent EW values are scattered around the $1-1$ line, while EW(\siII*)~$>$~EW(\cII*) in the simulation. We refer the reader to Section~\ref{sec:1D} for the discussion of the different factors that can explain the discrepancy between the \siII* and \cII* emission in the simulation. Additionally, we further highlight in Section~\ref{sec:apersec} that the fluorescent emission observed in \classy\ is affected by aperture losses, so the comparison of both trends must be considered carefully.

The joint comparison of the \siII\ versus \cII\ properties highlights several differences in the trends observed in the simulation and in the observations, some of which can be explained by the caveats of our measurement approach, while others, related to, e.g., the residual flux or the strength of the \siII* emission, are further discussed in Sections~\ref{sec:apersec} and Section~\ref{disc:spatial} in the context of aperture loss.  Overall, only six out of 45 \classy\ galaxies have spectral properties (in particular their absorption line EWs and central velocities) that are clearly not replicated in the simulation, with five of these objects having $M_\star$ larger than the simulated galaxy. Hence, the significant overlap between simulated and observed line properties is extremely promising as it highlights the potential of RHD/RT simulations in replicating the environment of real star-forming galaxies and producing diverse and physically realistic spectra which compare to observations. We further explore this point in the next section.


\subsection{Finding the best matches for each \classy\ objects}
\label{sec:fitting}

\begin{figure*}[!htbp]
    \centering
    \includegraphics[width = 0.32\hsize]{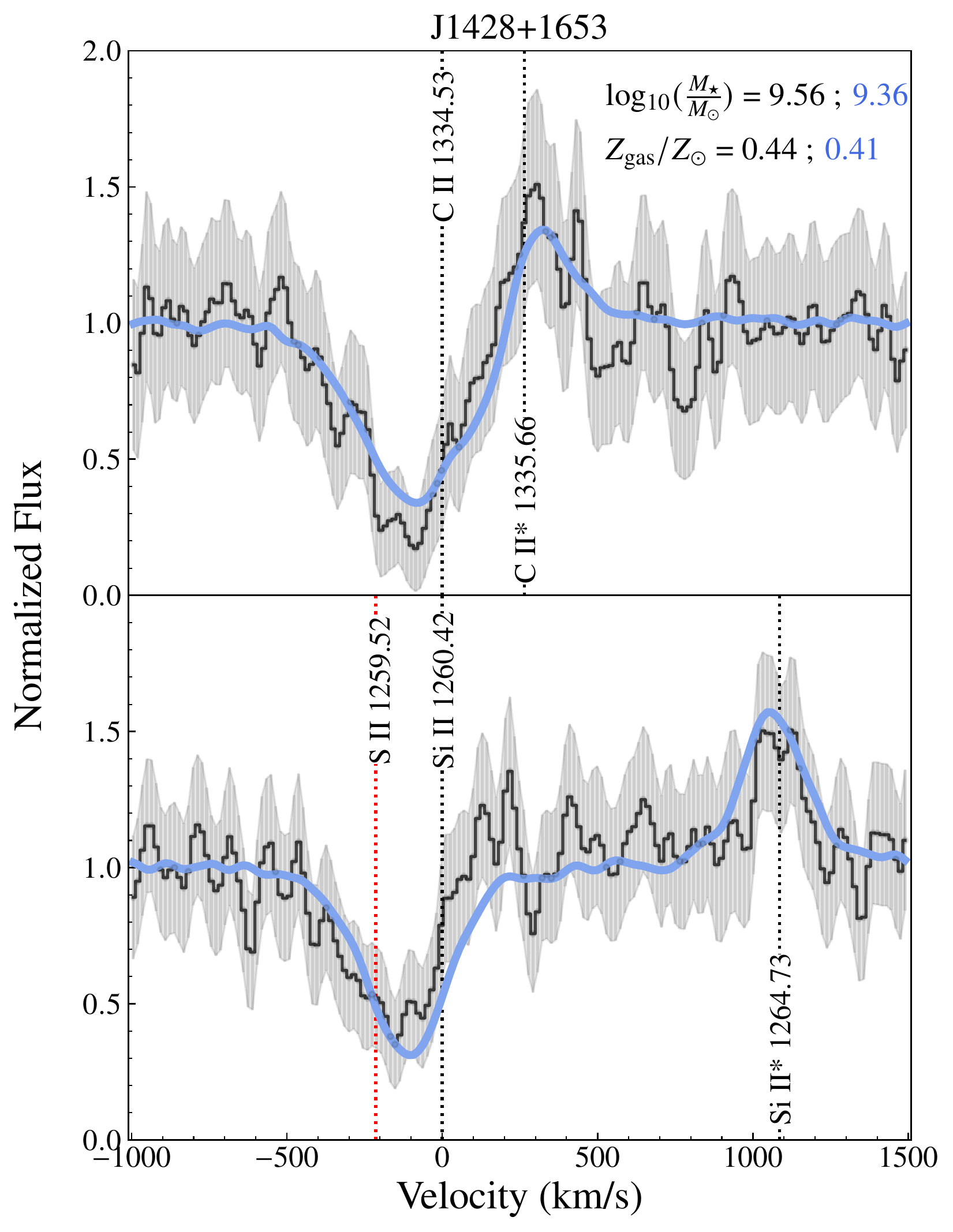}
    \includegraphics[width = 0.32\hsize]{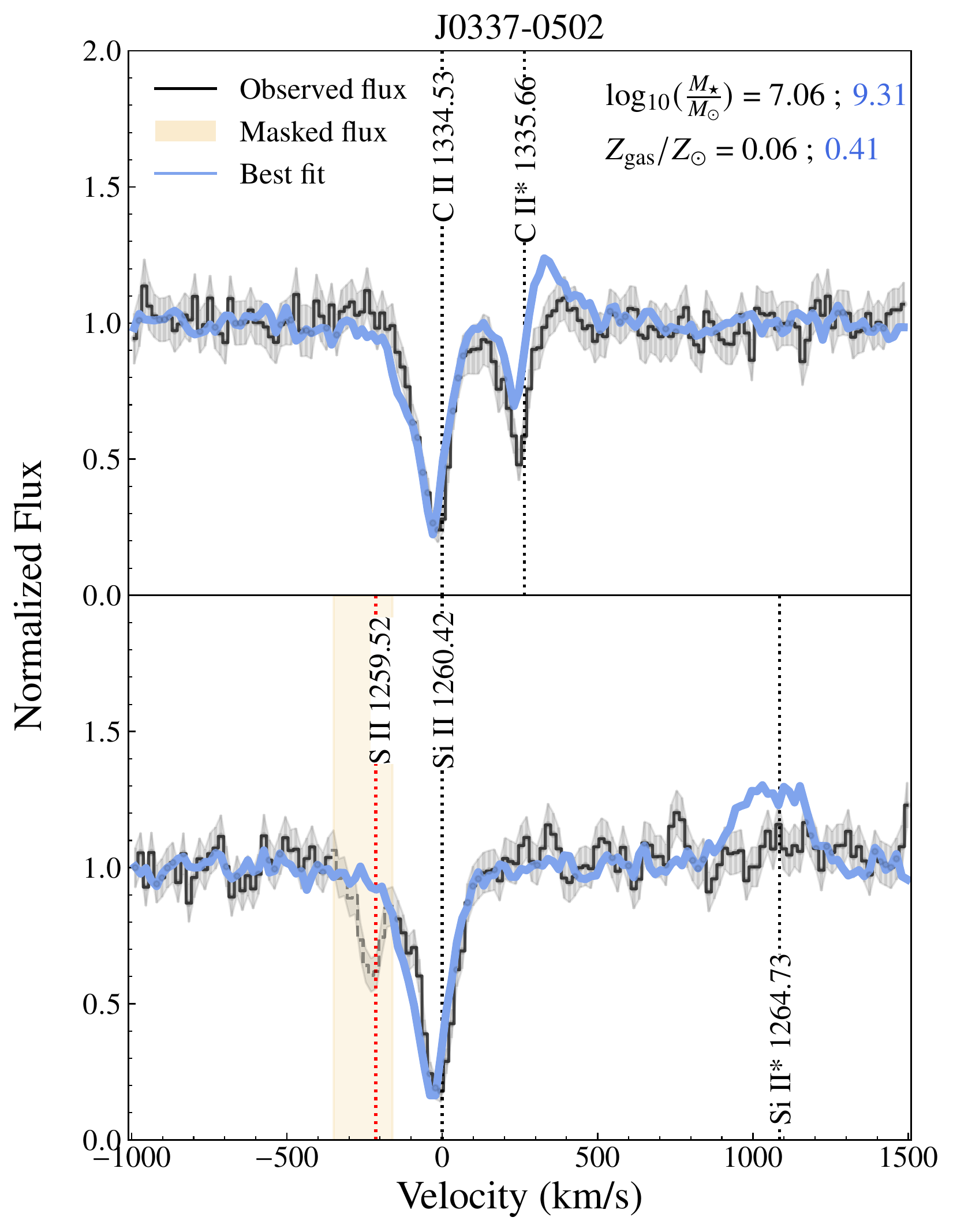}
    \includegraphics[width = 0.32\hsize]{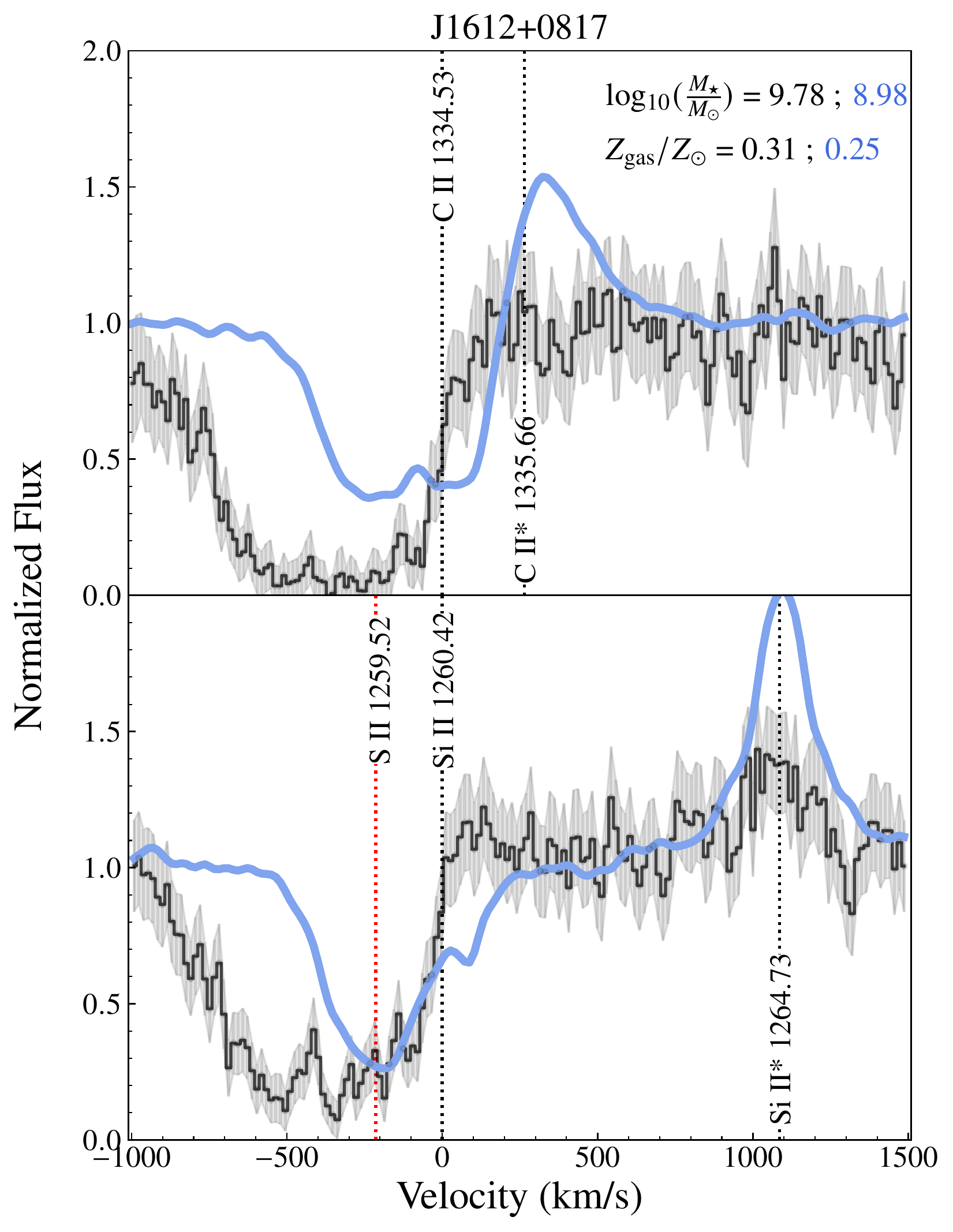}
\caption{The best-matching spectrum of the \cIIl+\cIIlstar\ (top panels) and \siIIl+\siIIlstar\ (bottom panels) line profiles for three \classy\ galaxies  (J1428+1653 left, J0337-0502 middle, and J1612+0817 right) using the procedure detailed in Section~\ref{sec:fitting}. The black line is the observed flux,  the grey shaded area is the flux error, the vertical dotted lines indicate the expected location of each transition, the yellow shaded area shows masked regions and the grey shaded area is the flux error. We only mask the wavelength region around \ion{S}{2} 1259~\AA\ when it is clearly resolved (e.g., only in the middle panel here). The blue line is the mock spectrum that best matches the observation. In the top right corner of the top panels, we detail the stellar mass and gas metallicity of the \classy\ galaxy (in black) and of the virtual galaxy at the time step where the best fit has been found (in blue).  The best-matched spectrum of J1428+1653 accurately reproduces both the absorption and fluorescent emission, and similar matching results are found for 28 \classy\ galaxies in total. The best-matched spectrum of J0337-0502 matches the resonant absorption well but overestimates the fluorescent emission. We observe a similar discrepancy for nine \classy\ galaxies. The best-matched spectrum of J1612+0817 fails to reproduce both the absorption and emission features. We find that the spectra of seven \classy\ galaxies are poorly matched, and five of them (including J1612+0817) have a stellar mass larger than the simulated galaxy. Appendix~\ref{app:fits1} presents the plots for all 45 \classy\ galaxies. }
    \label{fig:fitexamples}
\end{figure*}

In this section, we directly compare the \cII\ and \siII\ line profiles of the 45 \classy\ galaxies to the spectra generated using the simulation. In particular, for each \classy\ object, we look for the mock spectrum that best matches the observed profiles. We implement a simple minimization procedure which is performed as follows: the mock and observed spectra are normalized using the median of the flux in a feature-free interval around 1256~\AA\ and 1331~\AA, respectively. Then, for each observed spectrum, the 22,500 mock spectra are degraded to the resolution of the observation using a Gaussian Line Spread Profile, which full width at half maximum (FWHM) is fixed to the value reported in Table 3 in \citet{berg2022}. We then determine which mock spectrum best minimizes the equation

\begin{equation}
\label{eq:chi2}
\begin{split}
    \tilde{\chi}^2 &= \frac{\chi^2_{\rm C\ II} +\chi^2_{\rm Si\ II}}{n_{\rm C\ II}^{\rm obs} + n_{\rm Si\ II}^{\rm obs}} , \text{where} \\
    \chi^2_{\rm C\ II} &= \sum{\left( \frac{f_{\rm C\ II}^{\rm obs} -  f_{\rm C\ II}^{\rm sim}}{\sigma_{\rm C\ II}^{\rm obs}} \right)^2} \text{ and }  \\  
    \chi^2_{\rm Si\ II} &=\sum{\left(\frac{f_{\rm Si\ II}^{\rm obs} -  f_{\rm Si\ II}^{\rm sim}}{\sigma_{\rm Si\ II}^{\rm obs}} \right)^2}.
\end{split}
\end{equation}

\noindent Here, $f_{\rm ion}^{\rm obs}$ and $f_{\rm ion}^{\rm sim}$ are the observed and simulated flux in the wavelength regions of each ion, respectively, $\sigma_{\rm ion}^{\rm obs}$ is the error on the observed flux, and $n_{\rm ion}^{\rm obs}$ is the number of bins in the spectrum of each ion. If the wavelength range around \siII\ or \cII\ is not observed, the corresponding $\chi^2$ and $n^{\rm obs}$ are set to 0. We mask all the Milky Way absorption and emission lines falling within each spectrum. Regarding the \ion{S}{2} absorption line at 1259.52~\AA, which contribution is not modeled in the mock spectra, we choose to mask a region of $\pm$100 km~s$^{-1}$ around it only if it is clearly resolved.

We compute $\tilde{\chi}^2$ using an interval from $-800$ to 700 km~s$^{-1}$ for \cIIl\ and from $-800$ to 1500 km~s$^{-1}$ for \siIIl. We tested different interval sizes and found that it does not impact the results if the full extent of the absorption and emission features is included for both lines. We also compute a $\tilde{\chi}^2_{\rm abs}$ and $\tilde{\chi}^2_{\rm fluo}$ to provide insights into the fit quality around the absorption and fluorescent features, respectively. $\tilde{\chi}^2_{\rm abs}$ is derived using an interval of $\pm$200~km~s$^{-1}$ around the velocity at the minimum depth of the \cII\ and \siII\ absorption lines. $\tilde{\chi}^2_{\rm fluo}$ is derived using an interval of $\pm$200~km~s$^{-1}$ around the \cII\ and \siII\ fluorescent transition wavelength.

Figure~\ref{fig:fitexamples} presents the matching results for three galaxies, J1428+1653, J0337-0502 (also known as SBS 0335-052E)  and J1612+0817, and Appendix~\ref{app:fits1} shows the matching results for all \classy\ objects. We select these galaxies as examples because (1) they have different absorption and emission line morphologies, suggesting that they have distinct ISM properties, and (2) they are representative of the level of agreement between best-matched mocks and CLASSY spectra over the whole sample. The J0337-0502 spectrum shows narrow, relatively symmetric lines, and lacks fluorescent emission with \cII* seen in absorption. The spectrum of J1429+0643 shows broad, asymmetric \cII\ and \siII\ absorption lines and relatively significant \cII* and \siII* emission. Finally, the spectrum of J1612+0817 presents broad, blueshifted absorption features for both \siII\ and \cII\, with little fluorescent emission.  Importantly, J1612+0817 has a stellar mass of $10^{9.78}\ M_\odot$, larger than the simulated galaxy. Both its EW(\cII) and EW(\siII) values are outside the range seen in the simulation (see Section~\ref{sec:global}) so we do not expect to find a good match for this object.

Figure~\ref{fig:fitexamples} highlights that the best-matching spectrum of J1428+1653 reproduces well the \cII\ and \siII\ line profiles ($\tilde{\chi}^2_{\rm abs}$ and $\tilde{\chi}^2_{\rm fluo}$ $<1.5$), and we find similar matching results for 28 \classy\ galaxies. The best-matching spectrum of J0337-0502 matches the absorption lines well but overpredicts the observed fluorescent features around both lines ($\tilde{\chi}^2_{\rm fluo}$ $>1.5$). Eight other spectra present the same discrepancy. Finally, the best-matching spectrum of J1612+0817 fails to reproduce both the absorption and fluorescent features ($\tilde{\chi}^2 > 3$). In total, we find that the spectra of seven galaxies (J0808+3948, J1112+5523, J1144+4012, J1157+3220, J1416+1223, J1525+0757, J1612+0817) are poorly matched, and five of these objects (J1112+5523, J1144+4012, J1416+1223, J1525+0757, J1612+0817) have $M_\star$ larger than the simulated galaxy. For the two remaining galaxies, J0808+3948 and J1157+3220, the absorption lines are significantly blue-shifted, and such profiles are not well replicated in our simulation. 

Over the whole sample, we find that the median $\tilde{\chi}^2$ is 0.84, the median $\tilde{\chi}^2_{\rm abs} = 0.82$ and the median $\tilde{\chi}^2_{\rm fluo} = 1.37$. In general, the simulated spectrum matches the resonant absorption well but sometimes significantly misrepresents the observed fluorescent emission. This discrepancy is typically larger for \siII* than for \cII*, which one might expect given the results shown in Section~\ref{sec:global}.  In the next section, we show how aperture losses can explain the lack of fluorescent emission observed in some \classy\ spectra. 

Overall, the ability to reproduce such a diversity of LIS spectra with a single simulated galaxy of $M_\star~\sim~10^9~M_\odot$ is remarkable. While the spectra of higher mass galaxies are not well matched, we do not find a significant variation of the median $\tilde{\chi}^2$ for galaxies with $M_\star~<~10^9~M_\odot$. This suggests that the simulated spectra reproduce equally well the spectra of \classy\ galaxies less or equally massive than the virtual galaxy. This might be surprising given the expected tight relations between stellar mass and the LIS line properties \citep[e.g.][]{heckman2015, alexandroff2015, chisholm2017leak} and we explore this aspect further in Section~\ref{disc:onegal}.

\section{Aperture loss effects on the fluorescent emission}
\label{sec:apersec}

In the past decade, several studies have explored the emission properties of LIS metal lines using different approaches such as Monte-Carlo methods in idealized geometries \citep{prochaska2011}, semi-analytic computations \citep{scarlata2015, carr2018, carr2022}, or zoom-in simulation post-processed with radiative transfer techniques \citep{mauerhofer2021}. These different approaches all emphasized that several factors, including dust extinction, gas densities, or outflow geometry \citep[the latter aspect is also discussed in][]{jaskot2014, rivera2015} regulate the shape of the absorption and fluorescent emission profiles. These works, and most if not all observational papers studying fluorescent emission \citep{shapley2003, erb2010, erb2012, finley2017VII, finley2017wind, steidel2018, wang2020}, also mention the impact of aperture loss which must be carefully considered if one aims to tie together simulations and observations.

The aperture size has a strong influence on the observed emission because the spatial extent of the fluorescent emission is larger than for the UV continuum. If the gas extent of the galaxy is significantly larger than the spectrograph aperture size (which sometimes also includes several star clusters), one will capture only a disproportional fraction of the fluorescent emission. Several studies have shown how aperture loss might explain the absence or weakness of the fluorescent lines in certain observations \citep[e.g.][]{shapley2003, schwartz2006, scarlata2015}. Given the relatively low redshifts and larger apparent sizes of the \classy\ galaxies ($z = 0.002$ to $z = 0.182$) and the limited size of the COS aperture (2.5$\arcsec$), these effects might also justify the lack of fluorescent emission in several \classy\ spectra. In Section~\ref{sec:aperprop}, we explore the connection between the aperture sizes and the \cII* and \siII* features in \classy\ and in the simulated spectra. Section~\ref{sec:aperfits} expands the analysis of the best-matching mock spectra by including the constraints of the COS aperture. 

\subsection{Clues from observations and simulations}
\label{sec:aperprop}

We first explore the impact of aperture size on the \classy\ spectra. \citet{Arellano2022} thoroughly investigate how aperture size affects nebular properties derived from optical spectra of the \classy\ sample and found it has a negligible impact on the derived measurements. Their work focuses on properties inferred from ratios of nebular lines such that the absence of significant differences is not unexpected. Here, we are directly looking at (partially) resonant lines and their properties. The latter depends on the spatial extent of the gas phases and their geometry, such that one expects aperture losses to play a much more critical role. To confirm this hypothesis, we explore the relationship between the equivalent width of the observed fluorescent emission as a function of the galaxy size which we estimate using its optical half-light radius  \citep[$r_{50}^{\rm opt}$, estimated from the u-band, see Table 6 from][]{berg2022}.

\begin{figure}
    \centering
    \includegraphics[width = 0.99\hsize]{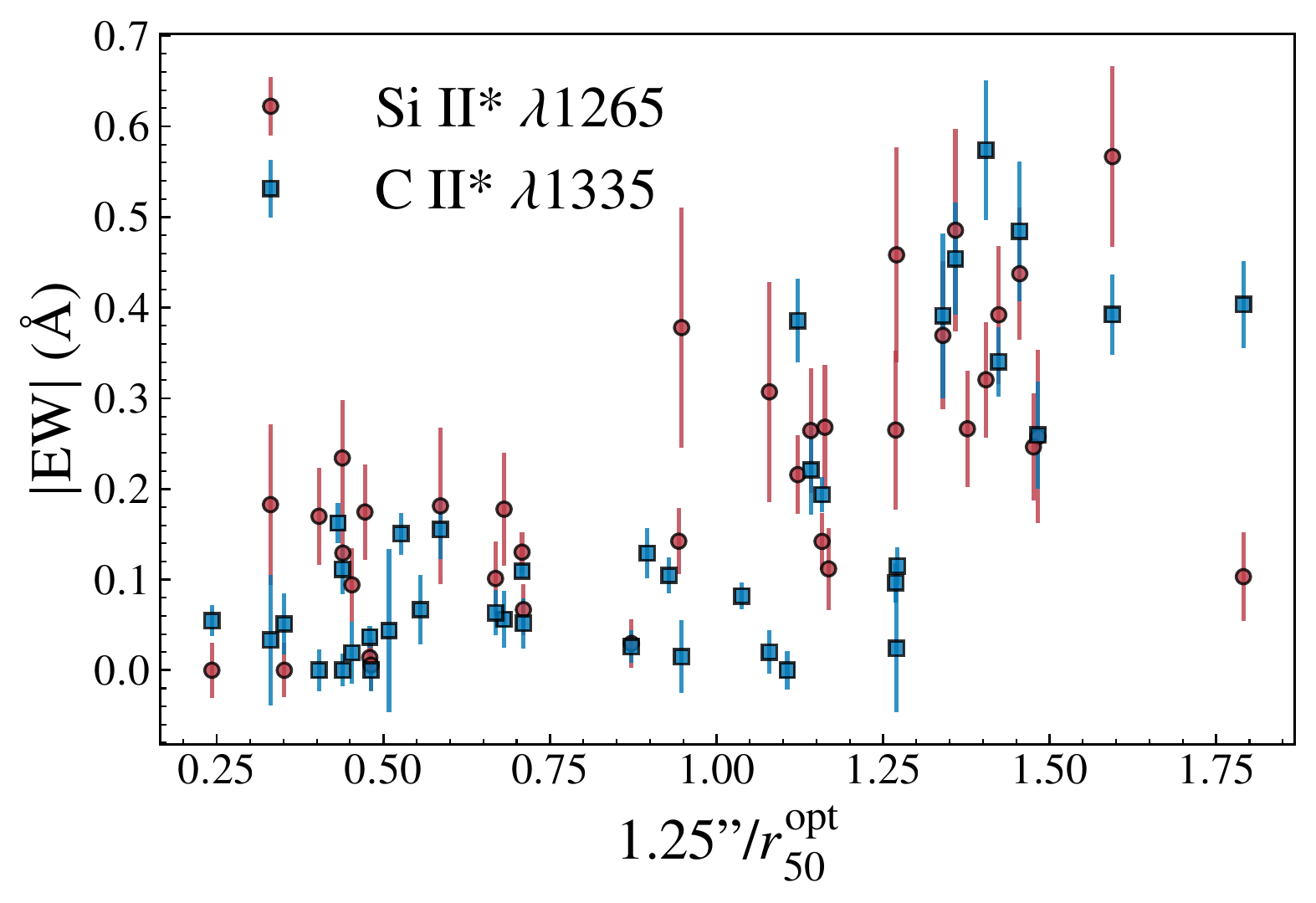}
    \caption{ The measured $|$EW(\cII*)$|$ (blue) and $|$EW(\siII*)$|$ (red) as a function of 1.25$\arcsec$/$r_{50}^{\rm opt}$ (1.25$\arcsec$ being the COS aperture radius) for the \classy\ observations. Galaxies with the largest EWs have the largest 1.25$\arcsec$/$r_{50}^{\rm opt}$ values, with the exception of J1323-0132 (see discussion in the text). }
    \label{fig:classyfluo}
\end{figure}

Figure~\ref{fig:classyfluo} shows the \classy\ $|$EW(\cII*)$|$ and $|$EW(\siII*)$|$ as a function of $r_{\rm \textsc{cos}}$/$r_{50}^{\rm opt}$, where $r_{\rm \textsc{cos}}$ is the radius of the COS aperture, or 1\farcs25. The $r_{\rm \textsc{cos}}$/$r_{50}^{\rm opt}$ ratio provides insights into the fraction of galaxy optical flux that falls within the COS aperture. For example, a ratio of 1 indicates that $\sim$50\%\ of the total optical flux fell within the COS aperture, and larger (lower) values suggest that more (less) light has been captured. 

The comparison of the optical size to the properties of UV lines suffers some caveats. Indeed, the UV light of the \classy\ sample is dominated by young stars and so is sometimes more spatially compact than the optical light. This may mean that a higher fraction of the UV light was actually captured within the COS aperture than assumed here. The UV half-light radii are sometimes estimated using the NUV COS acquisition images, but the COS aperture vignettes the light it collects  (the light at the edges of the aperture has a lower transmission than at the center), resulting in potentially artificially smaller half-light radii. Hence, in the current work, we only use $r_{50}^{\rm opt}$ as a tracer of the galaxy size.

Figure~\ref{fig:classyfluo} highlights that almost all galaxies with EWs~$\gtrsim~0.3$ \AA\ have $r_{\rm \textsc{cos}}$/$r_{50}^{\rm opt} > 1.0$, and the \classy\ spectra with the largest EW(\cII*) and EW(\siII*) have the largest 1.25$\arcsec$/$r_{50}^{\rm opt}$ ratios, with the exception of J1323-0132. For this galaxy, we do not detect any \cII\ or \siII\ absorption, suggesting that this galaxy is largely depleted of low-ionized metal gas and therefore explains the relative weakness of the fluorescent features. These results support that several \classy\ spectra may miss a significant part of fluorescent emission due to aperture losses.

\begin{figure}
    \centering
    \includegraphics[width = 0.99\hsize]{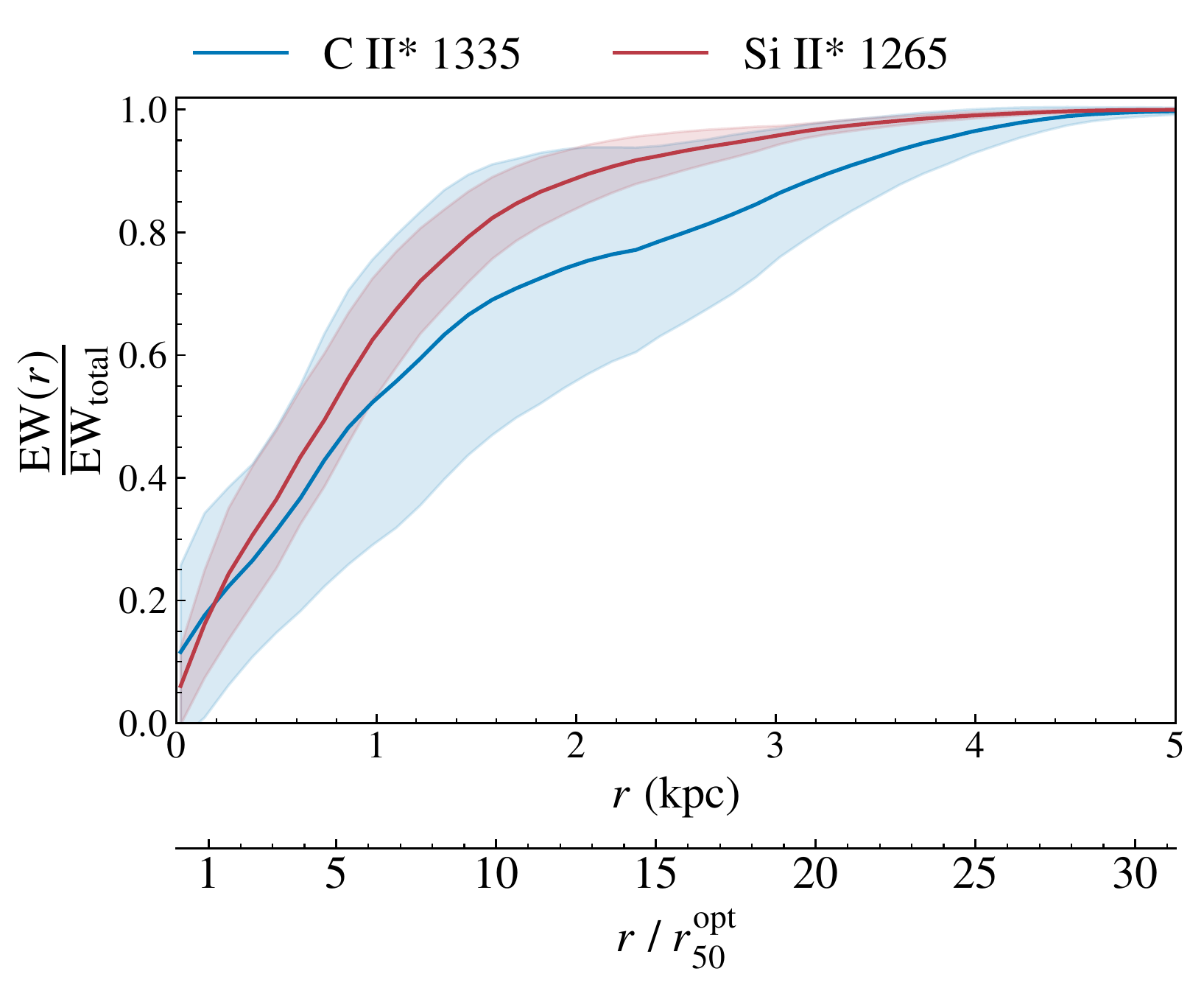}
    \caption{ The evolution of the EW($r$)/EW$_{\rm total}$ for \cIIlstar\ (blue) and \siIIlstar\ (red) as a function of the aperture radius ($r$) for the simulated galaxy. The solid lines and shaded areas are the median and standard deviation over 300 lines of sight, respectively, using a single simulation output of the galaxy at t $=$ 2 Gyrs.  The bottom x-axis shows the ratio between $r$ and $r_{50}^{\rm opt}$ for the virtual galaxy. Simulated spectra with $r < r_{50}^{\rm opt}$ typically exhibit little to no fluorescent emission. $r$ must be at least $\sim$ 20 and 25 times larger than $r_{50}^{\rm opt}$ if one aims to capture all the fluorescent emission for \siII* and \cII*, respectively.}
    \label{fig:simfluo}
\end{figure}


We now investigate how mimicking aperture loss in the simulation impacts the \cII* and \siII* properties of the mock spectra. In Section~\ref{sec:global}, the mock spectra were extracted using a circular fiducial aperture of 1$\arcsec$ diameter which encompasses the galaxy full-extent at $z=3$. To investigate the impact of varying the aperture size, we use a single simulation output of the galaxy at t $=$ 2 Gyrs for which we compute Integral Field Unit (IFU) cubes for each of the 300 lines of sight. Each cube consists of 4,096 individual spectra corresponding to the $64\times64$ spatial pixels. The pixel physical size is $120\times120$ pc, such that the entire cube face is $\sim 7.7$ by 7.7 kpc.

For each of the 300 directions, we extract integrated \cII\ and \siII\ spectra using circular apertures of increasing radius, using a one-pixel step, centered on the brightest pixel. We then measure EW(\cII*) and EW(\siII*) in each spectrum and measure for each ion the evolution of EW($r$)/EW$_{\rm total}$, where EW($r$) is the measurement at a given radius, and EW$_{\rm total}$ is the total EW  obtained by integrating the entire cube face. The EW($r$)/EW$_{\rm total}$ ratio provides insights into the fraction of the total fluorescent flux which is captured as a function of the aperture size.

Figure~\ref{fig:simfluo} presents the \cII* and \siII* EW($r$)/EW$_{\rm total}$ profiles derived over the 300 lines-of-sight measurements, as a function of radii. Using an image of the virtual galaxy in the optical (at 3500~\AA), we also measure the $r_{50}^{\rm opt}$ of the virtual galaxy (see Appendix~\ref{app:uvopt}) to better compare the results from the simulation with the analysis done for the \classy\ spectra in Figure~\ref{fig:classyfluo}. 


The gradual increase of EW(\cII*) and EW(\siII*) with radii demonstrates that the size of the aperture critically impacts the strength of the fluorescent emission observed in the simulated spectra. We note that the \siII* EW($r$)/EW$_{\rm total}$ has a steeper increase with $r$, and the typical variations around the median are smaller compared to \cII*. This is because the \cII\ fluorescent emission is blended with the resonant absorption so that the \cII* emission strength also depends on the absorbing gas properties along the line of sight. 

In general, Figure~\ref{fig:simfluo} highlights that little fluorescent emission is observed in mock spectra extracted using apertures with $r<r_{50}^{\rm opt}$, which is consistent from the analysis of Figure~\ref{fig:classyfluo}. Furthermore, this figure suggests that the aperture size must be five times larger than $r_{50}^{\rm opt}$ to capture at least 50\% of the total fluorescent emission for both ions, and $\sim$ 20 and 25 times larger if one aims to capture all of the \siII* and \cII* emission, respectively. 

While Figure~\ref{fig:classyfluo} indicates that all \classy\ galaxies have $r$/$r_{50}^{\rm opt} < 2$, Figure~\ref{fig:simfluo} indicates that we should capture at best 20\% of the total flux. If this was correct, large discrepancies between the simulated and observed fluorescent emission should be seen for the whole sample, yet, we showed in Section~\ref{sec:fitting} that this is only the case for a few (8) objects. This inconsistency can be explained by several factors. First, in the simulation, both EW($r$)/EW$_{\rm total}$ and the measured $r_{50}^{\rm opt}$ varies as a function of the viewing angle (see the error bar on the median $r_{50}^{\rm opt}$ derived in Appendix~\ref{app:uvopt}) and as a function of the time step considered (and hence of the galaxy property at these ages). Second, the \classy\ galaxies present diverse sizes and morphologies \citep[see, e.g., Figure 2 in][]{berg2022} which might not resemble the simulated galaxy. In particular, several objects seems to present several bright star clusters within the COS aperture, while most of the observed continuum flux in the simulation emerges from a single star cluster (see Section~\ref{disc:spatial}). This means that the spatial extent of fluorescent emission in the simulation might strongly differ from these observations, and a relatively lower $r$/$r_{50}^{\rm opt}$ ratio could already capture most of the total fluorescent flux. Hence, because Figure~\ref{fig:simfluo} builds upon a single virtual object, the comparison of the scaling relations in the simulation with the \classy\ measurements should be taken with caution. 

Despite these caveats, Figures~\ref{fig:classyfluo} and~\ref{fig:simfluo} present trends which clearly highlight the importance of constraining aperture losses for analyzing fluorescent emission observations. In the next section, we re-fit some of the \classy\ spectra using an approach that accounts for the COS aperture size. 

\begin{figure*}[!htbp]
    \centering
    \includegraphics[width = \textwidth]{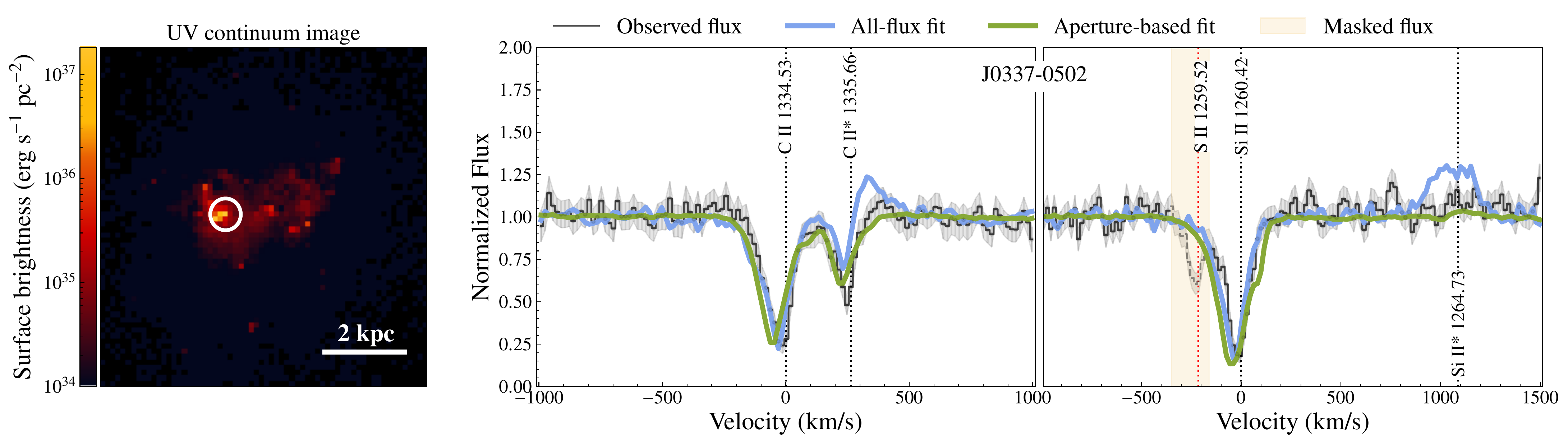}
    \includegraphics[width = \textwidth]{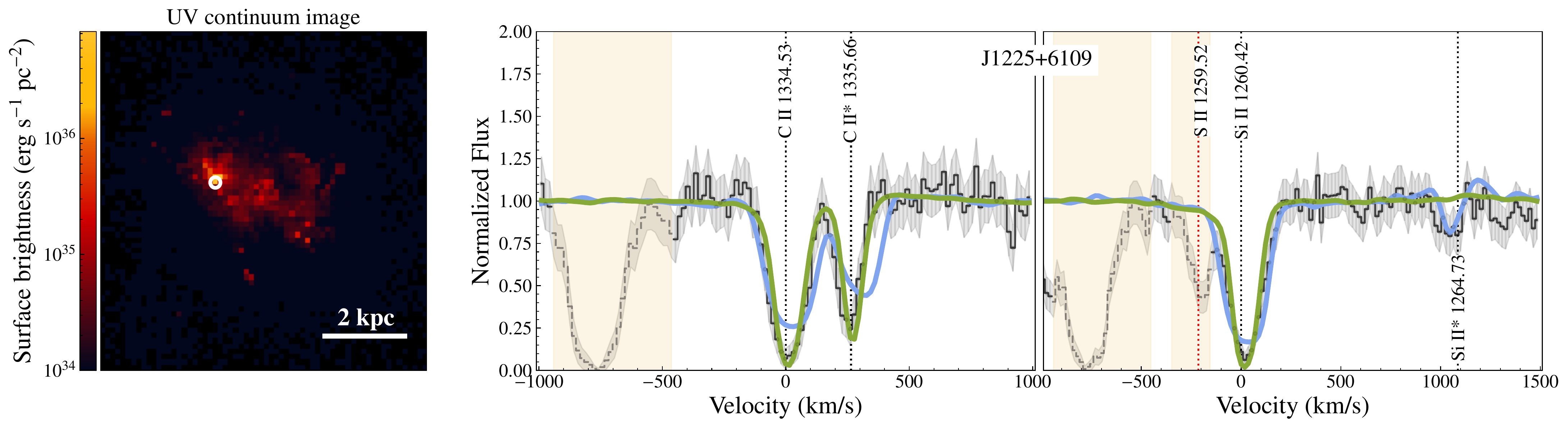}
    \caption{The fitting results based on the aperture-based approach described in Section~\ref{sec:aperfits} for two \classy\ galaxies (J0337-0502 top row and J1225+6109 bottom row).  The left panel presents an image of the simulated galaxy in the direction where the best mock spectrum has been found.  The white circle shows the fiducial aperture used to extract the spectra, centered on the brightest pixel in the region. The size of this region corresponds to the size of the COS aperture at the galaxy redshift. The right panels present the comparison of the previous best-matched mock spectra shown in Section~\ref{sec:fitting} (referred to as ``All-flux fit") and the new ``Aperture-based" fits.  The aperture-based fits reproduce better the line profiles around the \cII* and \siII* fluorescent channels, emphasizing that aperture effects play a critical role in the analysis of the fluorescent emission in the \classy\ galaxies. }
    \label{fig:cosfits}
\end{figure*}

\subsection{Improving the matches using aperture constraints}
\label{sec:aperfits}

Some of the best-matched mock spectra found in Section~\ref{sec:fitting} present fluorescent emission features which exceed the flux observed in the \classy\ spectra. Section~\ref{sec:aperprop} highlighted that aperture effects could be the predominant cause of this discrepancy. In this section, we refine the fitting procedure detailed in Section~\ref{sec:fitting} to explore whether we can improve the matches if we account for aperture effects. For this experiment, we use the simulation output described in Section~\ref{sec:aperprop} (300, 64 by 64 pixels images of the simulated galaxy at t $=$ 2 Gyrs). For each \classy\ spectrum to be matched, we fix the size of the aperture in the simulation to the size of the COS aperture at the galaxy redshift. Then, for each direction, we extract the \siII\ and \cII\  spectra for all regions in which the central pixel luminosity is at least 50\% of the brightest pixel for this direction. This strategy ensures that we follow a procedure consistent with the \classy\ HST/COS acquisition method which automatically centers its aperture on the brightest region.

Here, we use a single simulation output at $t = 2$ Gyrs, while we used 75 different time steps in Section~\ref{sec:fitting}. This is because the production of spectral data cubes is computationally expensive. Fortunately, the output at $t = 2$ Gyrs is best situated to analyze the \classy\ observations: most (21 out of 45) of the best-matched mock spectra found in Section~\ref{sec:fitting} originate from this time step. This output corresponds to the time step A in \citetalias{mauerhofer2021}. 

\begin{figure}[!htbp]
    \centering
    \includegraphics[width = \hsize]{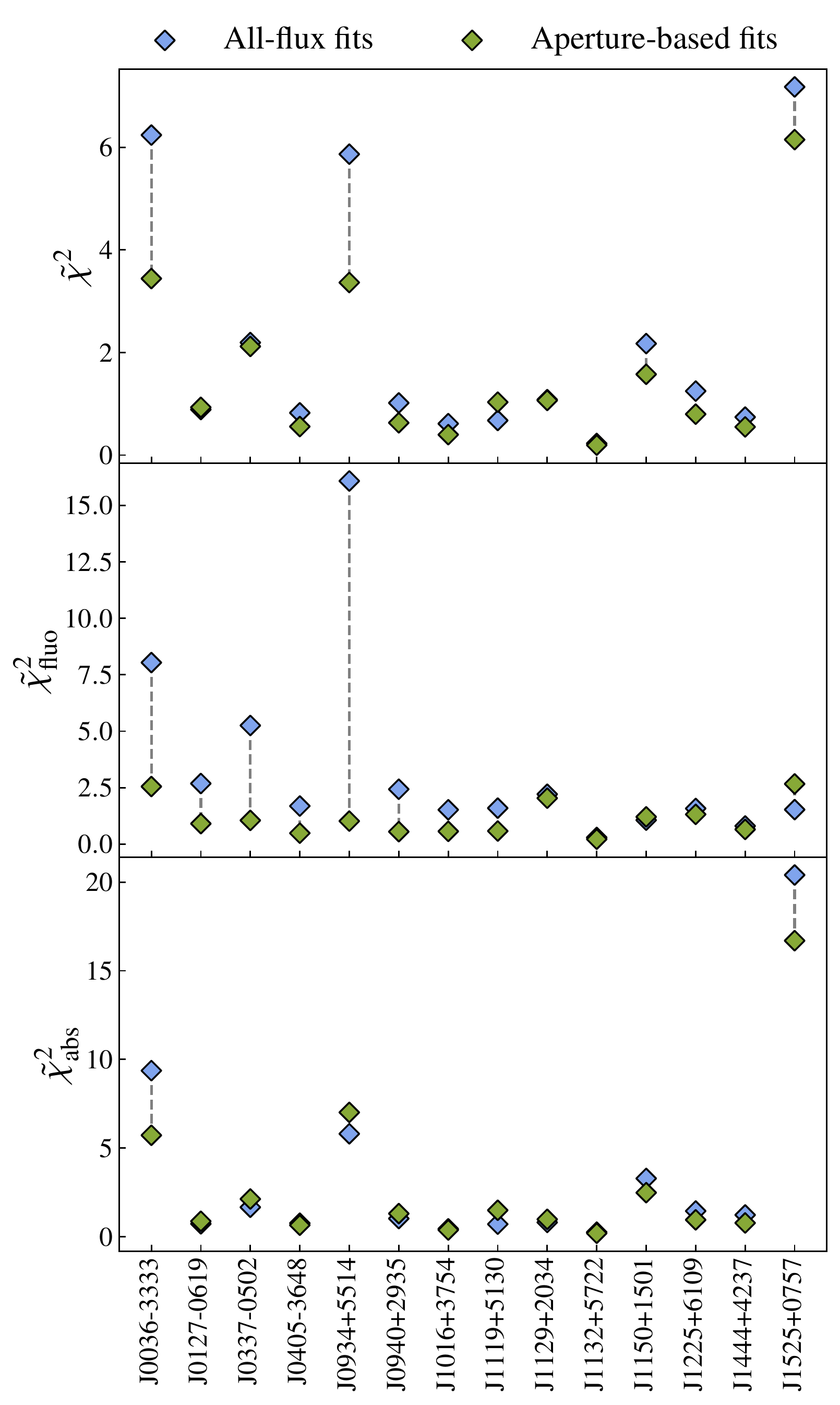}
    \caption{The comparison of the $\chi^2$ values for the all-flux fits (shown in Section~\ref{sec:fitting}) and the aperture-based fits for the 14 \classy\ galaxies re-matched in Section~\ref{sec:aperfits}. The top panel presents the evolution of the $\tilde{\chi}^2$ values derived using Eq~\ref{eq:chi2}. The middle and bottom panels show the evolution of the $\tilde{\chi}^2_{\rm fluo}$ and $\tilde{\chi}^2_{\rm abs}$ values, respectively. Most notably, the aperture-based approach enables significant improvements of the $\tilde{\chi}^2_{\rm fluo}$ for eight galaxies, emphasizing that we need to account for aperture losses when comparing the fluorescent emission properties in simulations and observations. On the other hand, aperture losses may have a lesser impact on the properties and shape of the resonant absorption feature. } 
    \label{fig:chis}
\end{figure}



We use the ``aperture-based" procedure described above to fit 14 \classy\ galaxies  that have (1) $r_{50}/1.25 > 1$ (i.e., the galaxies most affected by aperture losses), (2) observations for both \siII\ and \cII, enabling us to  assess both LIS lines simultaneously, and (3), had their best-match found in the simulation output at $t = 2$ Gyrs.  We compare in Figure~\ref{fig:cosfits} the \cII\ and \siII\ matches obtained using the approach in Section~\ref{sec:fitting} (referred to as ``all-flux fits") and the fits obtained using the new approach detailed in this section (referred to as ``aperture-based fits") for J0337-0502 and J1225+6109. Appendix~\ref{app:fits2} shows this comparison for all 14 galaxies. 

For both J0337-0502 and J1225+6109, the aperture-based fits better replicate the multi-component profile. For J0337-0502, the fluorescent feature fits are significantly improved, while the quality of the resonant absorption fits remains unchanged. For J1225+6109, the  best-matched mock spectrum for J1225+6109 better replicates the absorption features for both lines, the \cII* absorption is also better matched, but, interestingly, the all-flux match seems to better reproduce the absorption around \siII*. Observing absorption around the fluorescent wavelength suggests that the overall gas is too opaque on this line of sight such that fluorescent photons emitted in that direction are either absorbed or scattered back in different directions where they can escape. This absorption would emerge from thick regions, located close to and around the brightest stars. For this galaxy, the COS-based aperture radius is only two pixels wide such that it is not large enough to capture these regions. Overall, this narrow absorption feature is relatively weak, within 1.5$\sigma$ of the flux error, so the differences between both fits are not significant.

To quantify the global improvements enabled by the aperture-based approach, we look at the evolution of the best-match $\tilde{\chi}^2$, $\tilde{\chi}^2_{\rm abs}$, and $\tilde{\chi}^2_{\rm fluo}$ in Figure~\ref{fig:chis}. We define a difference of 1 in the average $\tilde{\chi}^2$ as a significant improvement or deterioration of the fit quality (i.e., the deviation to the observation increases/decreases by 1 $\sigma$ on average). The evolution of the $\chi^2$ values highlights that the aperture-based fits improve the overall fit quality for three of 14 galaxies, while no significant changes are found for the other spectra. 

Most notably, we find significant improvements of the $\tilde{\chi}2_{\rm fluo}$ for eight galaxies, with two galaxies improved by a factor larger than 5, and no substantial changes for five objects. Four of these five objects already had good matches as shown by their low $\tilde{\chi}^2_{\rm fluo}$ values. Hence, using a smaller aperture does not impact the good fits.

Finally, when considering the reduced $\tilde{\chi}^2_{\rm abs}$ values, we only find significant improvements for two galaxies. For one of these, J1525+0757 ($r_{50}$ $\sim$ 1.32$\arcsec$), the best-matched mock still poorly matches the observed spectrum. This galaxy is one of the high-mass objects, and its spectrum exhibits very large absorption lines which none of the available mock spectra can reproduce. This suggests that accounting for aperture losses does not improve the fit quality for the \classy\ galaxies with $M_\star$ larger than the virtual galaxy.

Overall, the approach that mimics the aperture effects for the \classy\ objects enables us to significantly improve the results of Section~\ref{sec:fitting} by extracting mock spectra that better match the \classy\ fluorescent features. This indicates that aperture-biased observations limit the interpretation of fluorescent line spectra because they underestimate the strength of the fluorescent emission. On the other hand, aperture losses may have a lesser impact on the properties and shape of the resonant absorption features. This is because the latter is predominantly produced from the gas around the brightest stars, and is likely affected by the ratio between the angular size of the continuum source and the instrument aperture.  We discuss this aspect further in Section~\ref{disc:spatial}.

\section{Discussion}
\label{sect:disc}

Here, we discuss the outcomes of Sections~\ref{sec:global} and \ref{sec:apersec}. Section~\ref{disc:onegal} discusses the origin of the diversity of the LIS metal line profiles that can be produced using a single simulated galaxy. In Section~\ref{disc:spatial}, we use the simulation to explore the physical and spatial origin of the different spectral features that compose a LIS metal line spectrum. 





\subsection{The feasibility of using a single galaxy simulation to interpret a range of physical conditions}
\label{disc:onegal}

\begin{figure*}
    \centering
    \includegraphics[width = 0.7\hsize]{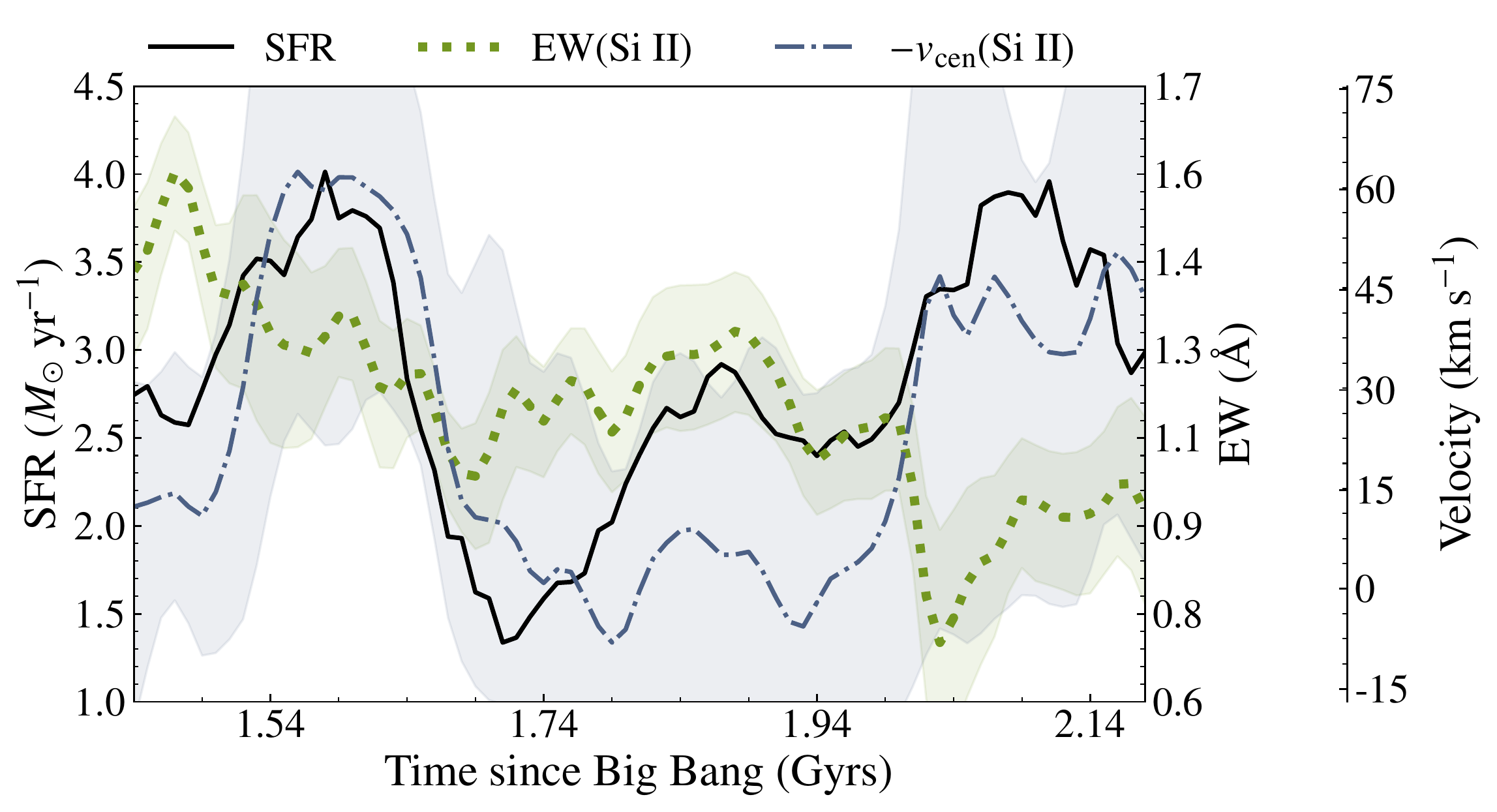}
    \caption{The evolution of the SFR of the galaxy in the zoom-in simulation over the time range of the outputs used to produce the mock \cII\ and \siII\ line profiles (690 Myrs). We show the evolution of the median EW(\siII) (green dotted line) and $-v_{\rm cen}$(\siII) (blue dashed-dotted line) derived from the 300 line-of-sight measurements at each time step. The shaded areas represent the variation per output and is derived using the 16$^{\rm th}$ and 84$^{\rm th}$ percentiles of the 300 measurements. This figure  highlights that both $-v_{\rm cen}$(\siII) and EW(\siII) significantly vary as a function of time and as a function of the viewing angle. Additionally, the variations of $v_{\rm cen}$(\siII) highlight that blue-shifted absorption lines, suggesting the presence of outflows, are commonly found in spectra generated during star-formation events. Overall, the variations of the LIS properties suggest that there exists a significant time and line-of-sight variability of the ISM/CGM properties which can explain how a single galaxy can produce such a diversity of spectra.   }
    \label{fig:sfrsim}
\end{figure*}

Sections~\ref{sec:fitting} and \ref{sec:aperfits} highlighted that the mock profiles generated in this work provide accurate replicates of the HST/COS observations of 38 \classy\ galaxies with a broad range of stellar masses ($10^6$ to $10^{9}$ ${\rm M_\odot}$) and metallicities (0.02 to 0.55~$Z_\odot$). The ability to trace such a diversity of galaxy environments using a single simulated object at $z\sim 3$ with limited stellar mass ($7.1 \times 10^8$ to $2.3 \times 10^9$~${\rm M_\odot}$) and metallicity variations  ($0.24$ to $0.43$ $Z_\odot$) is remarkable. In particular, it supports that the observation of a certain line profile is mostly determined by the ISM properties (geometry, densities, and kinematics) along the line of sight. Yet, this result might come as surprising since one expects the LIS line properties to be tightly  connected to galaxy properties such as stellar mass or metallicity \citep[e.g.][]{heckman2015, alexandroff2015, chisholm2017leak, Xu2022, Xu2023}. In this section, we examine the origin of the diversity of LIS line spectra that can be produced using a single galaxy.


In Figure~\ref{fig:sfrsim}, we compare the variation of the EW(\siII) and $-v_{\rm cen}$(\siII) properties to the variation of the galaxy SFR over the 75 time steps (690 Myrs). Note that we use the \siII\ properties because, for \cII, the contributions of the resonant absorption and fluorescent emission features are often blended, complicating their analysis. 


Figure~\ref{fig:sfrsim} highlights that the evolution of $-v_{\rm cen}$(\siII) is tightly related to the temporal variations of the SFR (Kendall-tau correlation coefficient of 0.67, p-value of 4.92$\times10^{-13}$). We find spectra with blue-shifted absorption lines (suggesting dominant outflows) in time ranges where the galaxy experiences intense star-formation bursts (around 1.6 and 2 Gyrs). We observe a slight offset between the gas kinematics traced by $-v_{\rm cen}$ and the SFR. This behavior was already discussed in e.g. \citet{Trebitsch2017, rosdahl2018} and indicates that outflows happen a short time after the SFR peak. 

On the other hand, between bursts, the virtual galaxy experiences a period of relative quiescence (we note that it is not entirely quiescent as it still forms stars). During that period, the typical \siII\ spectra exhibit absorption lines close to the systemic velocity or red-shifted, suggesting the gas might be relatively static or with potential inflows. 

While we observe a decrease in the median EW(\siII) values close to SFR peaks, the temporal evolution of EW(\siII) is less tightly related to the SFR than $v_{\rm cen}$ (Kendall-tau correlation coefficient of 0.01, p-value of 0.92), suggesting that other aspects must be accounted for to understand its decline through time. Nevertheless, both the fluctuations of the EW and $v_{\rm cen}$ values during the 690 Myrs period considered highlight that there exists a large variability of the \siII\ profile properties as a function of time.

In addition to the time-variability, we denote a significant line-of-sight variability of the \siII\ properties seen at each time step. This aspect is highlighted by the shaded areas in Figure~\ref{fig:sfrsim} which shows the fluctuations of the EW(\siII) and $-v_{\rm cen}$(\siII) values over the 300 orientations at each time step. This line-of-sight variability may suggest that there is a broad distribution of metal densities and gas kinematics  over the entire ISM. This aspect has already been shown in \citetalias{mauerhofer2021}, whose authors highlighted that there exist significant spatial fluctuations (up to a factor of 10) of the gas metallicity along different lines of sight. The \siII\ properties variations as a function of viewing angle emphasize that ISMs are complex heterogeneous environments.

The overall decrease of EW(\siII) through time and the tight connection between the galaxy SFR and $-v_{\rm cen}$(\siII) might indicate that the physical processes involved during star-formation bursts have a significant impact on the structure of the galaxy's ISM and CGM. In particular, the fragmentation of the ISM into dense star-forming clouds, plus the stellar and/or supernovae feedback triggered during or shortly after these events may significantly transform the ISM and introduce large fluctuations in the typical spectral properties observed through time and per viewing angle. Overall, this time- and line-of-sight variability likely explains how a single galaxy can produce such a diversity of spectra that are representative of galaxies of comparable or smaller masses.

\begin{figure*}
    \centering
    \includegraphics[width = \hsize]{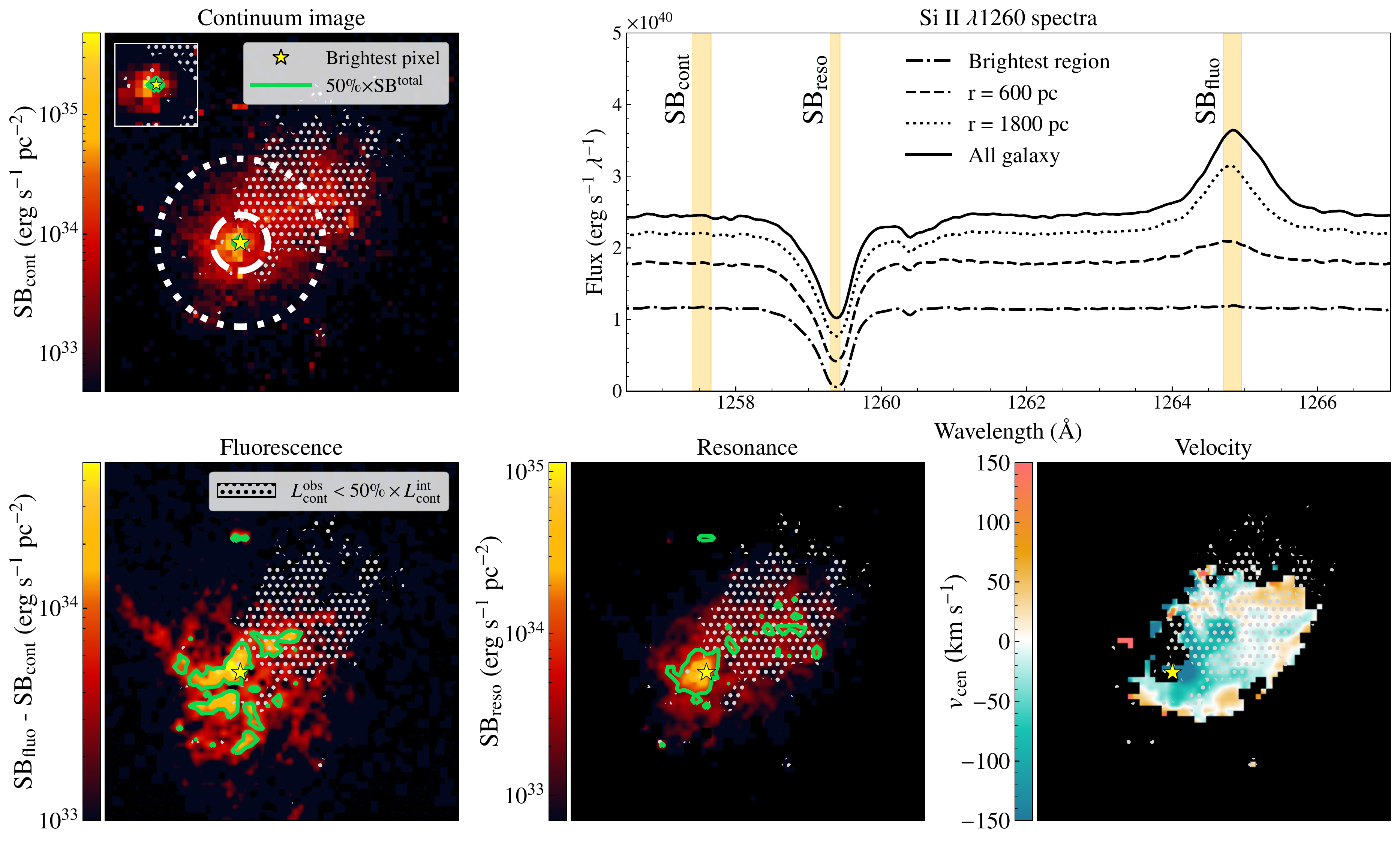}
    \caption{\siII\ images and spectra of the virtual galaxy for the line of sight that provides the best-matched mock spectrum for J0926+4427. The top left panel shows the integrated surface brightness from the simulation corresponding to the continuum emission (SB$_{\rm cont}$), and the bottom left and middle panels show the spatial distribution of the \siII\ fluorescence emission (measured on continuum-subtracted images, i.e., SB$_{\rm fluo}$ $-$ SB$_{\rm cont}$) and residual flux (SB$_{\rm reso}$, measured in an interval around the absorption line maximal depth), respectively. In each of these images, a yellow star represents the brightest continuum pixel, green contour lines show the 50\%\ level of the total SB, and the white hashed-dotted areas represent regions where the dust suppresses at least half of the intrinsic emission (the ratio of the observed and intrinsic luminosities is less than 0.5). The bottom right panel shows the distribution of $v_{\rm cen}$ values derived from the \siII\ spectrum in each pixel. Black pixels represent cases where the absorption component at 1260~\AA\ is too faint or non-existent such that $v_{\rm cen}$ values cannot be robustly measured. All images are 64 by 64 pixels, with a pixel size of 120 pc.
    Finally, the top right panel presents the \siII\ spectra extracted based on (i) the brightest pixel (dot-dashed line), (ii) a circular aperture of radius 600 pc (dashed), (iii) a circular aperture of radius 1800 pc (dotted line) and (iv) the entire image (solid line). The r = 600 pc and r = 1800 pc apertures are shown with matching line styles on the top left panel. The orange areas show the wavelength regions integrated to produce the continuum, fluorescence, and resonance images. }
    \label{fig:contoursJ0926}
\end{figure*}
\begin{figure*}
    \centering
    \includegraphics[width = \hsize]{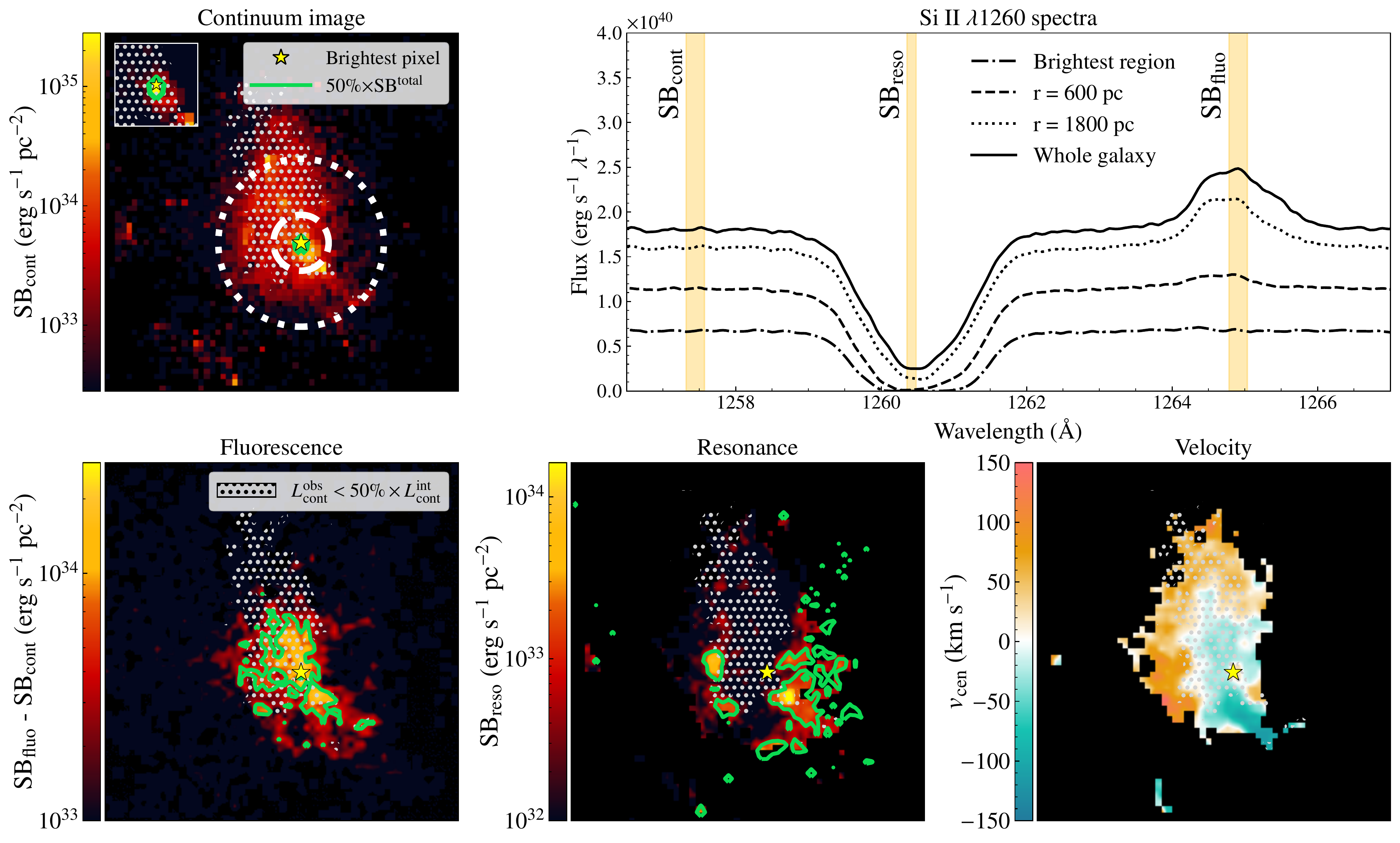}
    \caption{Same as Figure~\ref{fig:contoursJ0926} for the line of sight that provides the best-matched mock spectrum for J0938+5428}
    \label{fig:contoursJ0938}
\end{figure*}

These results may suggest that there only exists a limited dependence on, e.g., $M_\star$ and $Z$ for observing specific LIS spectral properties. Nevertheless, we highlighted in Section~\ref{sec:fitting} that the spectra of the \classy\ objects with $M_\star$ larger than the virtual galaxy are not well reproduced using this simulation. As already noted for the Lyman-$\alpha$ line by \citet{blaizot2023simulating}, this indicates that this simulation can't help us interpret the spectra of more massive objects because the latter might have too complex ISM environments. Additionally, the 1-D histograms shown in Figure~\ref{fig:glob} (Section~\ref{sec:1D}) highlight that the range of simulated LIS properties follows a normal distribution function, which means that there does exist a set of ``typical" LIS properties for this $10^{9}$ ${\rm M_\odot}$ virtual galaxy at $z\sim 3$. In particular, it also suggests that the good agreement between spectra from very low mass \classy\ galaxies and the simulated spectra do not necessarily mean that the virtual galaxy (and its simulation physics) accurately depicts all these objects. It is possible that these spectra are rare in the simulation, arising from time steps where the star-formation rate is low for the virtual galaxy but would be considered intense for these low-mass objects.  Therefore, $M_{\star}$ and $Z$ may still play an important role in understanding the typical LIS spectral properties observed in certain galaxy populations because the likelihood of observing certain spectra is determined by these parameters. To test this hypothesis, one could perform the same experiment done in this work building upon virtual galaxies within a broader range of properties, using e.g. satellite galaxies from the current simulations or galaxies from the SPHINX suite of cosmological RHD simulations \citep{rosdahl2018}. In turn, the ability of lower and larger mass virtual galaxies in reproducing observations in the same regime will determine the exact influence of $M_\star$ on the LIS spectral properties. This effort, although necessary to constrain the range of spectral properties single-galaxy simulation can reproduce as a function of their properties, is beyond the current analysis.

\subsection{Using the simulation to disentangle the spatial origin of the spectral features}
\label{disc:spatial}

This section uses the simulation to interpret the spatial and physical origin of the different spectral features (continuum emission, residual flux, or fluorescent emission) that compose LIS line spectra. We build upon the same data set we used in Section~\ref{sec:apersec} which contains 64 by 64 images of the simulated galaxy in 300 directions at t $=$ 2 Gyrs. We explore the spatial properties of the \siII\ features\footnote{We chose \siII\ because the resonant and fluorescent features are not blended, which simplifies the spatial analysis.}  in  two directions corresponding to the lines of sight that provided the best-matched mock spectrum for the galaxies J0926+4427 and J0938+5428. These two galaxies show distinct line profiles: in J0926+4427, the \siIIl\ line exhibits a relatively narrow, blue-shifted absorption with significant residual flux ($v_{\rm cen}\approx -$300 km~s$^{-1}$ and $R_f\approx$0.5). On the other hand, in J0938+5428, the line profile is broader, centered around the systemic velocity, and shows little residual flux ($R_f\sim$0.1). Additionally, the EW of the fluorescent emission in J0926+4427 is typically two times larger than in J0938+5428. These line profile differences make these two galaxies optimal toy examples to investigate the origin of their spectral features using the simulation.

Figures~\ref{fig:contoursJ0926} and \ref{fig:contoursJ0938}  show images of the \siII\ continuum, fluorescence, and resonance surface brightness in the orientation and time step that best matches the J0926+4427 and J0938+5428 spectra, respectively. Each panel presents the location of the brightest continuum pixel (yellow star), the contours enclosing 50\% of the total flux (green lines), and the regions where dust attenuates by a factor of 2 or more the intrinsic emission (hash-dotted areas).  We also construct maps representing the spatial distribution of $v_{\rm cen}$(\siII) values per pixel. We show the \siII\ spectra extracted from (i) the brightest pixel, (ii) a circular aperture of radius 600 pc, (iii) a circular aperture of radius 1800 pc, and (iv) the whole image. We discuss below the origin of the stellar continuum, fluorescent emission, and residual flux contributions (bottom panels in each figure) in the two directions. To simplify the discussion, we refer to Figure~\ref{fig:contoursJ0926} as ``the J0926 direction" and to Figure~\ref{fig:contoursJ0938} as ``the J0938 direction". 


\paragraph{Stellar continuum:}\
 A single, compact region encompassing two and three pixels dominates half of the observed stellar continuum in the J0926 direction and in the J0938 direction, respectively. This region is more significantly attenuated by dust in the J0938 direction than in the J0926 direction. In both cases, the spectra extracted from the brightest continuum region exhibit fully-covered absorption lines (no residual flux) and no fluorescent emission, indicating that the brightest stars are facing optically thick gas in these two directions. Additionally, in the J0926 direction, the absorption line in the brightest-region spectrum is significantly blue-shifted ($v_{\rm cen}$ is $\leq -150$ km s$^{-1}$), suggesting the presence of strong outflows.  

\paragraph{Fluorescent emission:}\
The fluorescent flux emerges from a larger region than the continuum in both directions, and the 50\% contours either enclose regions with outflowing gas (the median $v_{\rm cen}$ values within the fluorescent contours is $-150$ and $-30$ km~s$^{-1}$ in the J0926 and J0938 direction, respectively), or with optically thin gas (i.e. the black pixels in the $v_{\rm cen}$ images, where no absorption lines are detected). Additionally, the fluorescent emission region is largely dust-depleted in the J0926 direction, but not in the J0934 direction. This may explain why EW(\siII*) is two times larger in the J0926 direction than in the J0934 direction.  Finally, the growth of the fluorescent emission in spectra extracted with increasing aperture radii highlights that most of the emission emerges from regions further away from the brightest stars, which is consistent with the analysis of Section~\ref{sec:aperfits}. 

\paragraph{Residual flux:}\ 
We observe different spatial origins for the residual flux in the J0926 and J0938 directions. For J0926, it emerges from a relatively small and compact region surrounding the brightest stellar cluster. For the J0938 direction, it arises from an extended region further away from the brightest stars. In both directions, the 50\% contours enclose pixels where we do not detect any absorption features, suggesting that some fraction of the residual flux comes from gas-depleted regions. Additionally, we find that the residual flux dominantly emerges from dust-depleted regions in both cases (87\% and 92\% of the pixels contributing to the residual flux are low-dust-attenuation regions for the J0926 and J0938 directions, respectively). The different spectra extracted in both directions highlight that we observe larger residual flux in spectra with increasing aperture radii, yet, the overall shape of the absorption line remains relatively unchanged. This has two important outcomes. First, aperture losses can also affect residual flux measurements and may explain why we observe different \cII\ and \siII\ $R_f$ trends in the simulation and in the \classy\ sample (see Section~\ref{sec:2d}).  Second, the shape of the absorption line feature relates to the properties of the gas surrounding the brightest stars but provides little information about the gas properties in regions where the stellar continuum is strongly attenuated. 


The analysis presented in this section paves the way for follow-up analyses that will further explore (1) the spatial properties of outflows and their impact on the ISM structure, and (2) the origin of the trends that connect the LIS line residual flux and the escape of ionizing photons. In this work, for simplicity, we only consider two directions, which is already enough to highlight that the different spectral features (continuum emission, residual flux, fluorescent emission) have quite distinct spatial origins (little or no overlap in the 50\% contours), which also vary as a function of the galaxy orientation. Importantly, this means that these different spectral features trace the properties of distinct regions in the galaxy's ISM, which makes the interpretation of LIS line spectra complex and delicate. Nevertheless, such a high-resolution simulation provides an optimal laboratory to understand how these features relate to each other, which makes it a very promising framework for interpreting spectroscopic observations.

\section{Conclusion}
\label{sec:conc}

LIS absorption and emission lines provide crucial insights into the ISM of star-forming galaxies, but the complexity of the emerging line profiles makes their interpretation difficult. Building upon the work of \citet{mauerhofer2021}, we used radiative transfer post-processing of a $\sim10^9$ $M_\odot$ dwarf galaxy in a zoom-in RHD simulation to create 22,500 \cIIl+\cIIlstar\ and \siIIl+\siIIlstar\ spectra (corresponding to 300 different outputs in 75 time steps from $z = 4.19$ to $z = 3.00$). We compared the simulated LIS line properties and profiles to the observed spectra of the 45 low-redshift star-forming galaxies from \classy\ \citep{berg2022} to investigate if and how such a simulation setup can be used as a laboratory to interpret spectroscopic observations. We summarize our findings as follows:

\begin{itemize}[leftmargin=*]
    \item The range of the \cII\ and \siII\ resonant line properties (equivalent width, residual flux, and central velocity) seen in the simulated spectra overlaps with 39 out of 45 \classy\ galaxies (Section~\ref{sec:global}, Figure~\ref{fig:glob}). The six remaining objects have either too large EW(\cII) and EW(\siII), or too large $v_{\rm cen}$(\cII) and  $v_{\rm cen}$(\siII) compared to the simulated spectra. Five of them have stellar masses larger than the $\sim10^9$ $M_\odot$ simulated galaxy.
    
    \item We identified a mock spectrum that closely matches the observed absorption and emission line profile for 28 \classy\ galaxies. For nine galaxies, the best-matched mock spectrum  matches well the absorption feature but overpredicts the fluorescent emission. For seven galaxies (including the six objects with either too large EWs or $v_{\rm cen}$), the best-matched mock spectrum poorly reproduces both the absorption and fluorescent features. In general, the mock spectra do not match well the LIS line spectra of the \classy\ galaxies with a stellar mass larger than the simulated galaxy, suggesting that the latter does not represent well more massive galaxies (Section~\ref{sec:fitting}). 
    \item We show that accounting for the COS aperture size faithfully solves the tension between the amount of fluorescent emission in the simulation and in the \classy\ observations. This is because fluorescent emission is more extended than the  UV continuum (Section~\ref{sec:aperprop}) and a small aperture centered on the UV peak will miss a significant fraction of scattered light. Using mock spectra extracted using a fiducial aperture whose size corresponds to the actual size of the COS aperture, we show significant improvements in reproducing the fluorescent emission in eight \classy\ observations (Section~\ref{sec:aperfits}).
    
    \item The typical LIS line shape and properties seen in the simulated galaxy  vary significantly both with time and per viewing angle (Section~\ref{disc:onegal}). In particular, we highlighted the connection between the fluctuations of the \siII\ properties and the evolution of the galaxy SFR, which suggests that the combination of several physical processes (e.g. star formation bursts, feedback mechanisms, gas accretion) may lead to large spatial and temporal variability in the ISM properties. This explains how the evolution of a single galaxy at $z\sim 3$ can produce such a diversity of line profiles  that replicate the spectra of 38 out of the 45 low-redshift \classy\ galaxies spanning a broad range of galaxy properties ($M_\star$ between 10$^6$ to 10$^9$ $M_\odot$ and $Z_{\rm gas}$ between 0.02 to 0.55 $Z_\odot$). 
    
    \item We explored the spatial origin of the  continuum, residual flux, and fluorescent emission in the mock spectra that best match the spectra of J0926+4427 and J0938+5428 (Section~\ref{disc:spatial}). We showed that there is little overlap in the regions contributing to each spectral feature, and the properties of these regions also vary depending on the galaxy orientation. This analysis demonstrates that integrated LIS line spectra are complex combinations of multiple, spatially distinct components, which makes their interpretation difficult. Yet, the present simulation presents an optimal framework for understanding and analyzing spectroscopic observations.
\end{itemize}


Overall, this work challenges simplified interpretations of down-the-barrel spectra but introduces a new methodology where high-resolution simulations can help us understand physically motivated scenarios that explain the observations. In particular, the remarkable success of reproducing real observations with mock spectra from RHD-RT simulations opens diverse and extremely promising perspectives for follow-up analyses. In future work, we will first explore how using different radiative transfer recipes (e.g. diverse dust opacity and turbulent velocity models) impact our results in order to determine the crucial physical ingredients needed to reproduce observations of star-forming galaxies. Second, we will extend the analysis of the physical and spatial origin of the different spectral features in the simulation. This effort will enable us to tighten our understanding of how absorption and emission line diagnostics relate to gas kinematics (in particular the importance of outflows) and to the escape of ionizing photons. This work will look simultaneously at the LIS line and Lyman-$\alpha$ profiles, building upon the work of \citet{blaizot2023simulating}. Finally, we plan to use spatially resolved observations with KCWI and MUSE to further assess the simulation's predictions on the spatial origin of the different spectral features that compose the LIS metal line spectra. 

\begin{acknowledgments}
The authors would like to thank the referee for thoughtful feedback that improved the quality of this manuscript. SG is grateful for the support enabled by the Harlan J. Smith McDonald fellowship which made this project possible. TG and AV are supported by the professorship grant  SNFProf\_PP00P2\_211023 from the Swiss National Foundation for Scientific Research.  VM acknowledges support from the NWO grant 016.VIDI.189.162 (``ODIN”). Observations for this paper come from HST-GO-15840. CLM acknowledges support from NSF under AST-1817125. B.L.J. is thankful for support from the European Space Agency (ESA). We are grateful for the support of this program that was provided by 
NASA through a grant from the Space Telescope Science Institute, 
which is operated by the Associations of Universities for Research in Astronomy, 
Incorporated, under NASA contract NAS5-26555.
\end{acknowledgments}

%
\facilities{HST(COS)}
\software{Astropy \citep{astropy}, NumPy \citep{numpy}, pandas \citep{panda}, SciPy \citep{scipy}, Photutils \citep{photoutils}, RASCAS \citep{rascas2009}}

\vspace{5mm}





\appendix
\section{Further discussion of the trends observed in Figure~3}
\label{app:globaltrends}

\begin{figure*}
    \centering
    \includegraphics[width = 0.49\hsize]{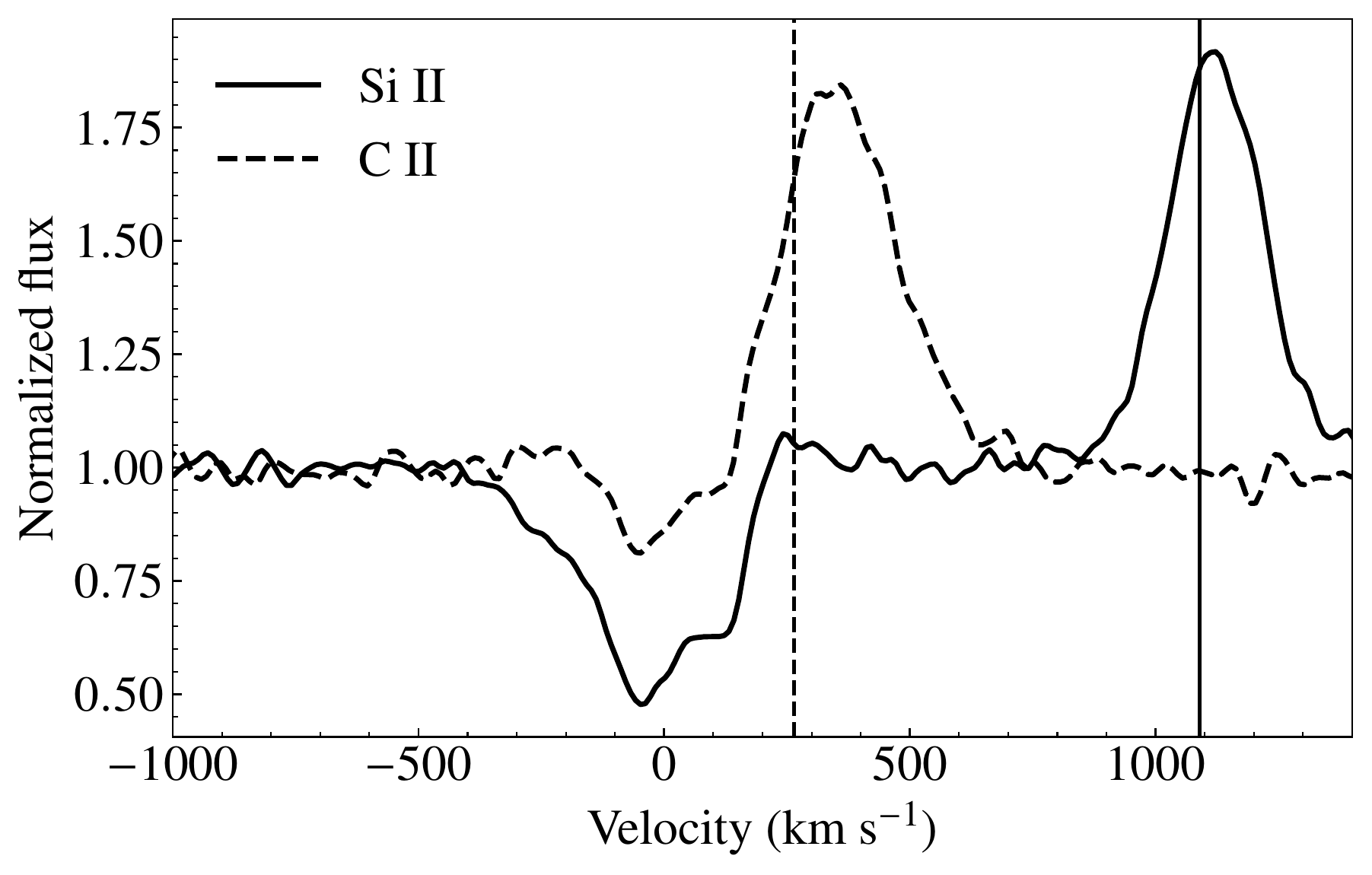}
    \includegraphics[width = 0.49\hsize]{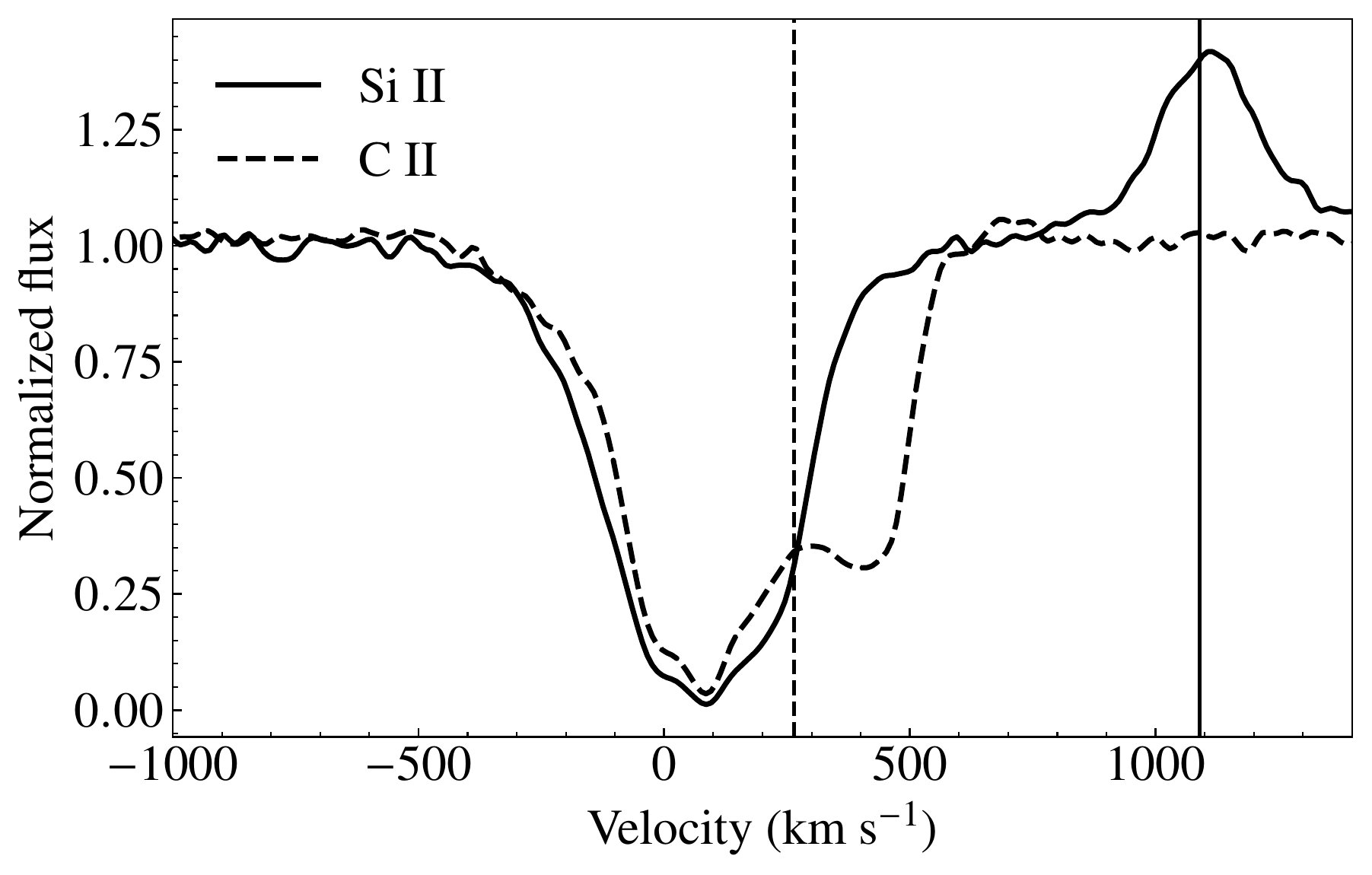}
    \caption{Comparisons of mock \cII\ and \siII\ spectra (dotted and solid lines, respectively, with vertical lines showing the expected wavelength of the fluorescent feature for each ion) for two configurations. In the left panel, both spectra exhibit a considerable fluorescent emission feature. For \cII, because of the small wavelength separation between the resonance and fluorescent channel, this emission fills a large portion of the absorption line. As we note in the text,  resonant scattering infilling, which is stronger for \cII\ than for \siII, also plays a role in producing weaker \cIIl\ lines. Overall, the analysis of the combined spectrum will yield $|$EW(\cIIl)$|$ $<$  $|$EW(\siIIl)$|$ and $R_f$(\cIIl) $<$ $R_f$(\siIIl). The right panel shows a different case where the \cIIlstar\ feature is in absorption. While we observe $R_f$(\cIIl) $\approx$ $R_f$(\siIIl), suggesting the fluorescent absorption has little impact on $R_f$(\cIIl), this scenario will yield $|$EW(\cIIl)$|$ $>$  $|$EW(\siIIl)$|$ and $v_{\rm cen}$(\cIIl) $>$ $v_{\rm cen}$(\siIIl). These different configurations are responsible for the mock spectra trends seen in Figure~\ref{fig:glob}. }
    \label{fig:EWspect}
\end{figure*}

In Section~\ref{sec:global}, we discussed the origin of several trends between the \cII\ and \siII\ properties shown in Figure~\ref{fig:glob}. In particular, we noted that EW(\cII) is typically smaller than EW(\siII) in spectra with weak absorption lines (EWs $<1$\AA). A similar behavior is observed for $v_{\rm cen}$(\cII) versus $v_{\rm cen}$(\siII).  Additionally, we found that $R_f$(\cII) is often found larger than $R_f$(\siII). As we noted in Section~\ref{sec:global}, all these trends can be explained by the impact of the \cIIlstar\ features on the EW(\cIIl) measurements. Spectra with significant \cII* fluorescent emission will have lower EW(\cII), $v_{\rm cen}$(\cII), and $R_f$(\cII) because the emission fills a significant portion of the absorption line. \cIIl\ lines are affected by infilling due to resonant scattering, typically more than for the \siIIl\ lines,  the combination of both effects yielding \cII\ absorption lines considerably weaker in that regime. On the other way, spectra with broad \cIIl\ absorption lines commonly have a significant absorption around the fluorescent wavelength. The combined resonant and fluorescent absorption feature yields large EW(\cII) and ``redder" $v_{\rm cen}$(\cII), sometimes mimicking the presence of significant inflowing gas. This scenario less significantly impacts $R_f$ because the minimum \cIIl\ flux is often deeper for the fluorescent absorption line. Figure~\ref{fig:EWspect} provides two examples of these configurations using mock spectra generated in this work.

The blending of the \cIIl\ and \cIIlstar\ contributions can explain the \cII\ and \siII\ trends observed in the simulation, yet, we also find that the \classy\ measurements do seem to follow slightly different trends. In particular, we find that the classy objects typically have EW(\siII) $>$ EW(\cII), $v_{\rm cen}$(\siII) $>$ $v_{\rm cen}$(\cII), and $R_f$(\cII) $\approx$ $R_f$(\siII). The two former trends can be explained by the presence of the \ion{S}{2} absorption line blended with the \siIIl\ line, which contribution is not modeled in the simulated spectra. This additional absorption on the blue side biases the EW(\siII) and $v_{\rm cen}$(\siII) towards larger and lower values, respectively. The spectra of J1129+2904, shown in Figure~\ref{fig:fits1a}, nicely illustrate the impact of \ion{S}{2} absorption line on the \siIIl\ line.

\section{Best-fit mock spectra}

\subsection{Using the matching approach of Section~\ref{sec:fitting}}
\label{app:fits1}

Figure~\ref{fig:fits1a} presents the spectra that best match the \classy\ \siII\ and \cII\ observations using the approach detailed in Section~\ref{sec:fitting}. As discussed in the latter section, the best fits reproduce accurately the absorption features seen in the observations but overestimate the contribution of the fluorescent emission for 6 of 39 galaxies. 

\begin{figure*}
    \centering
    \includegraphics[width = 0.45\textwidth]{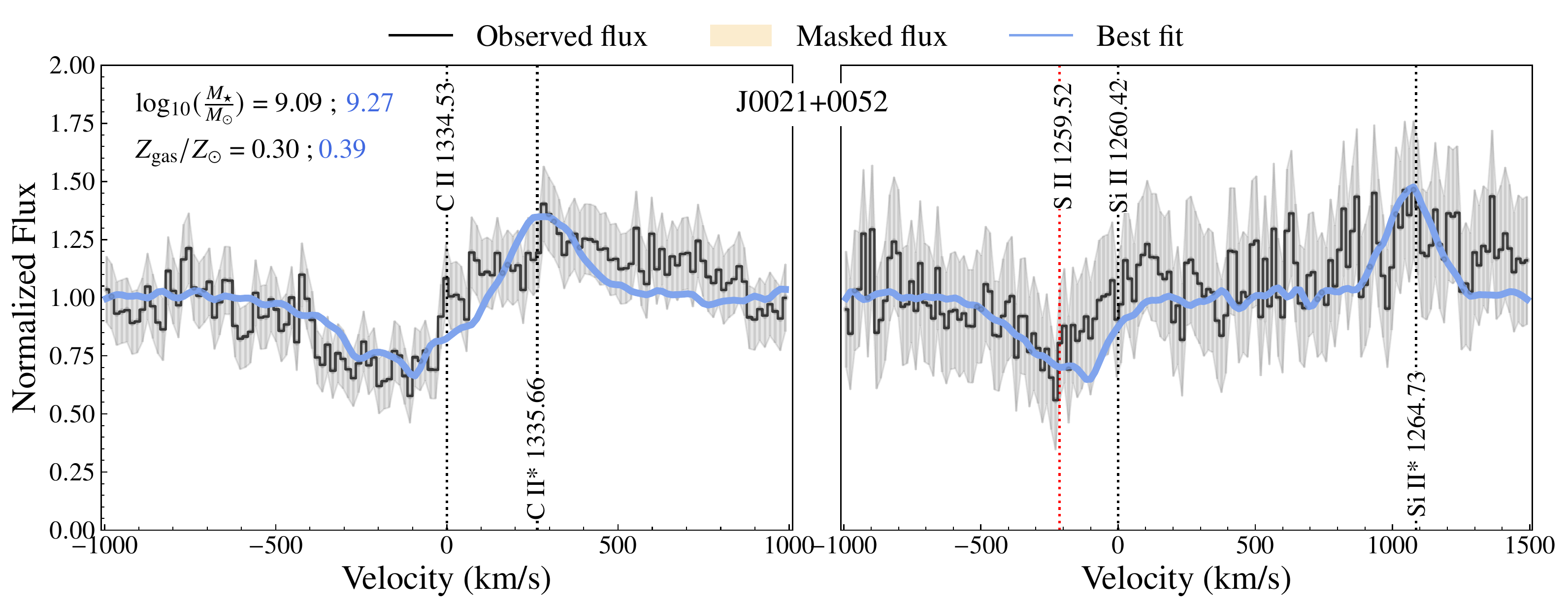}
    \includegraphics[width =0.45\textwidth]{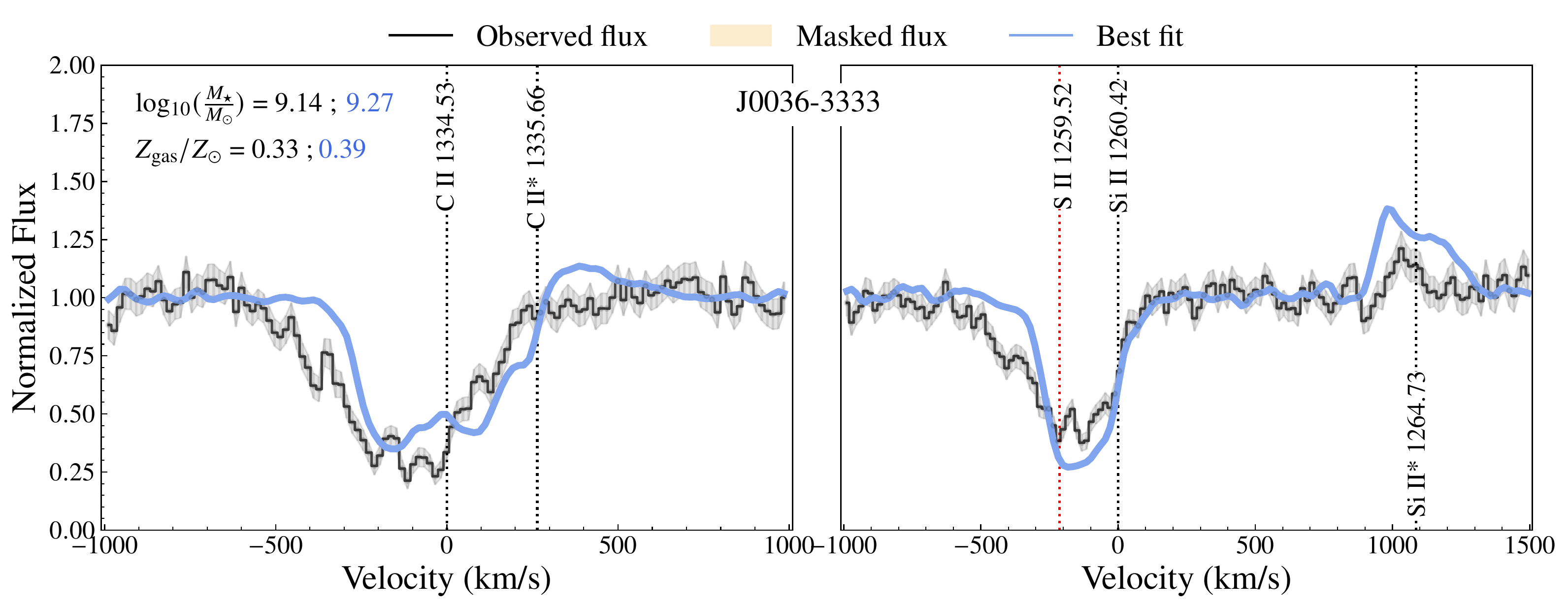}
    \includegraphics[width =0.45\textwidth,trim={0 0 0 1cm},clip]{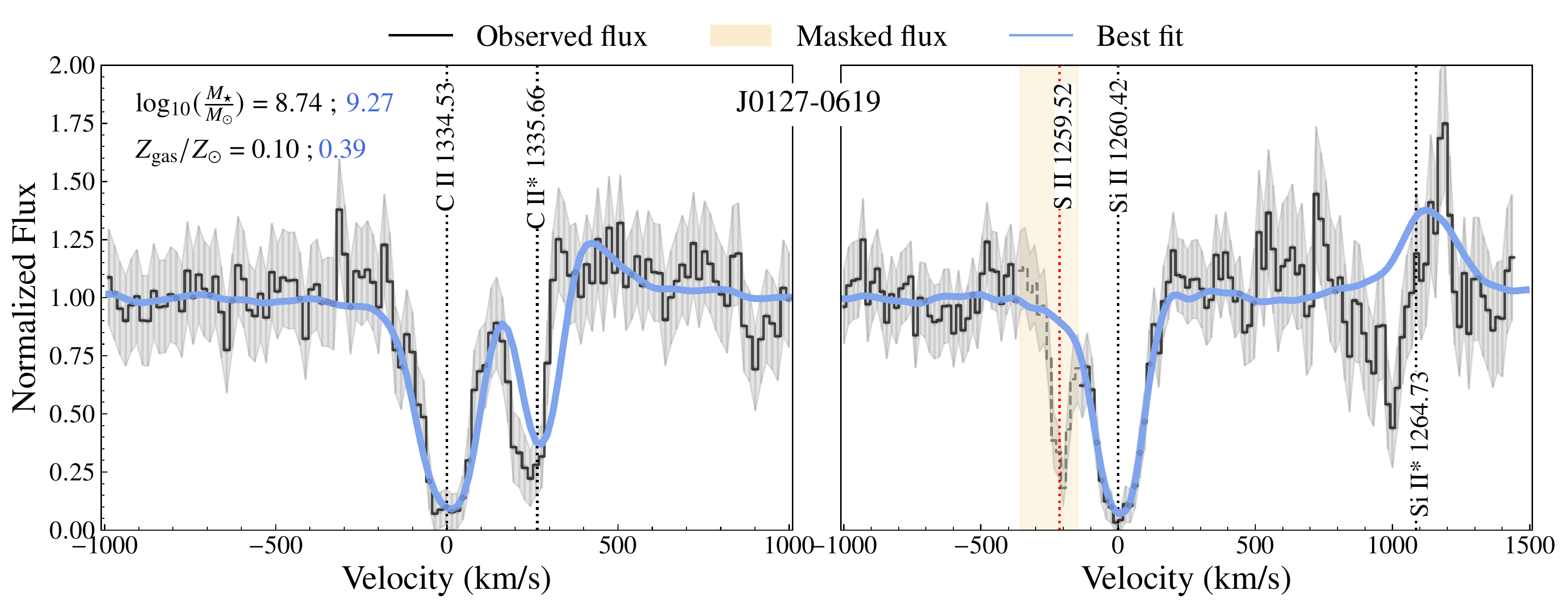}
    \includegraphics[width =0.45\textwidth,trim={0 0 0 1cm},clip]{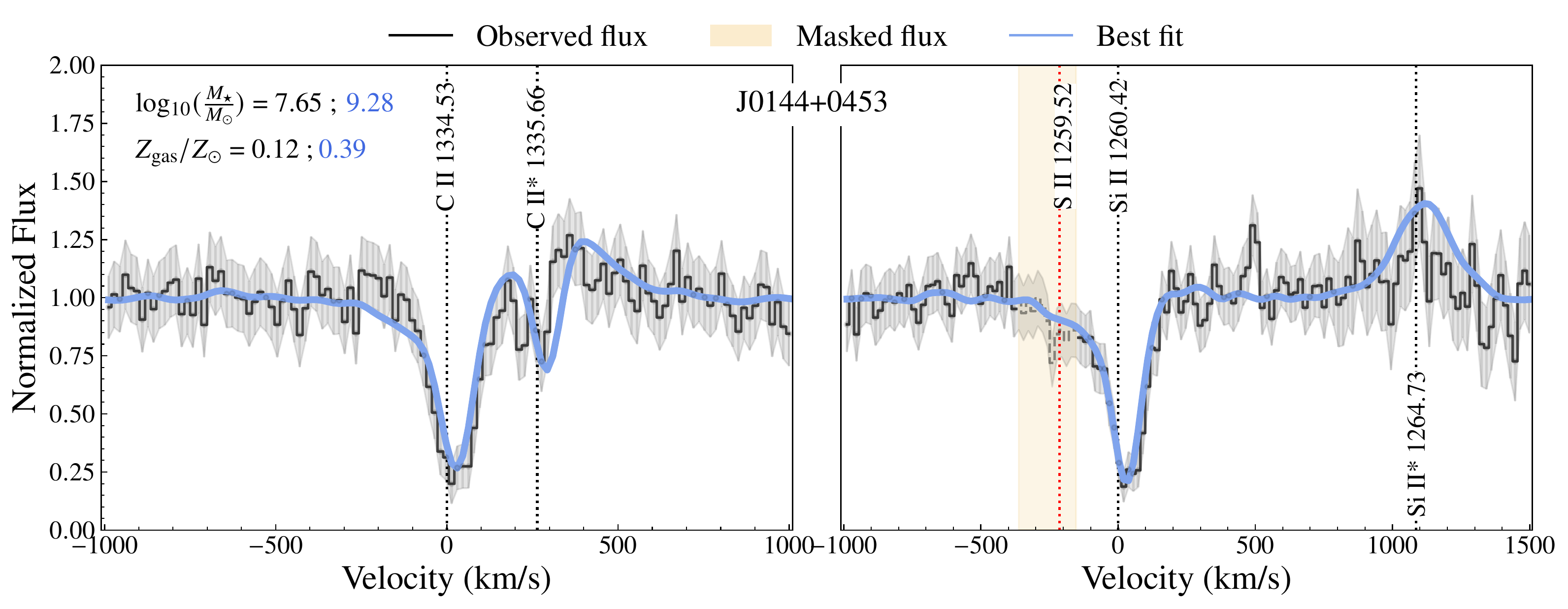}
    \includegraphics[width =0.45\textwidth,trim={0 0 0 1cm},clip]{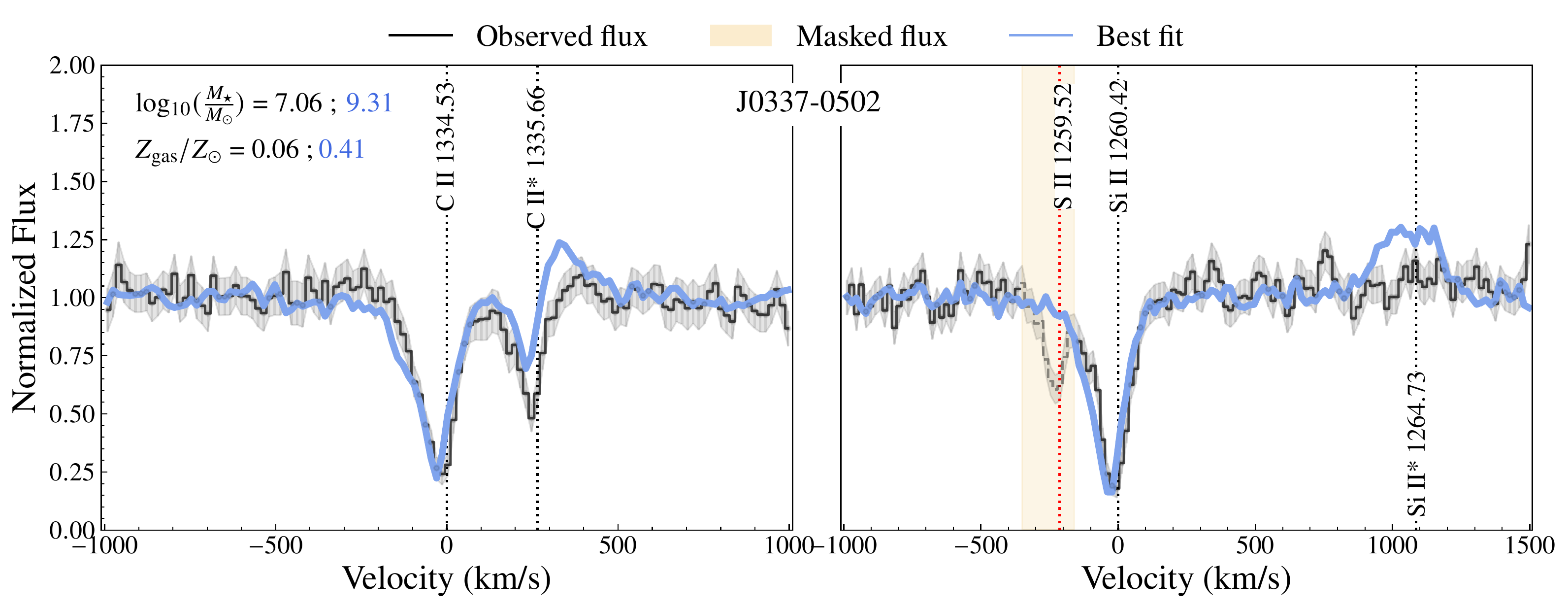}
    \includegraphics[width =0.45\textwidth,trim={0 0 0 1cm},clip]{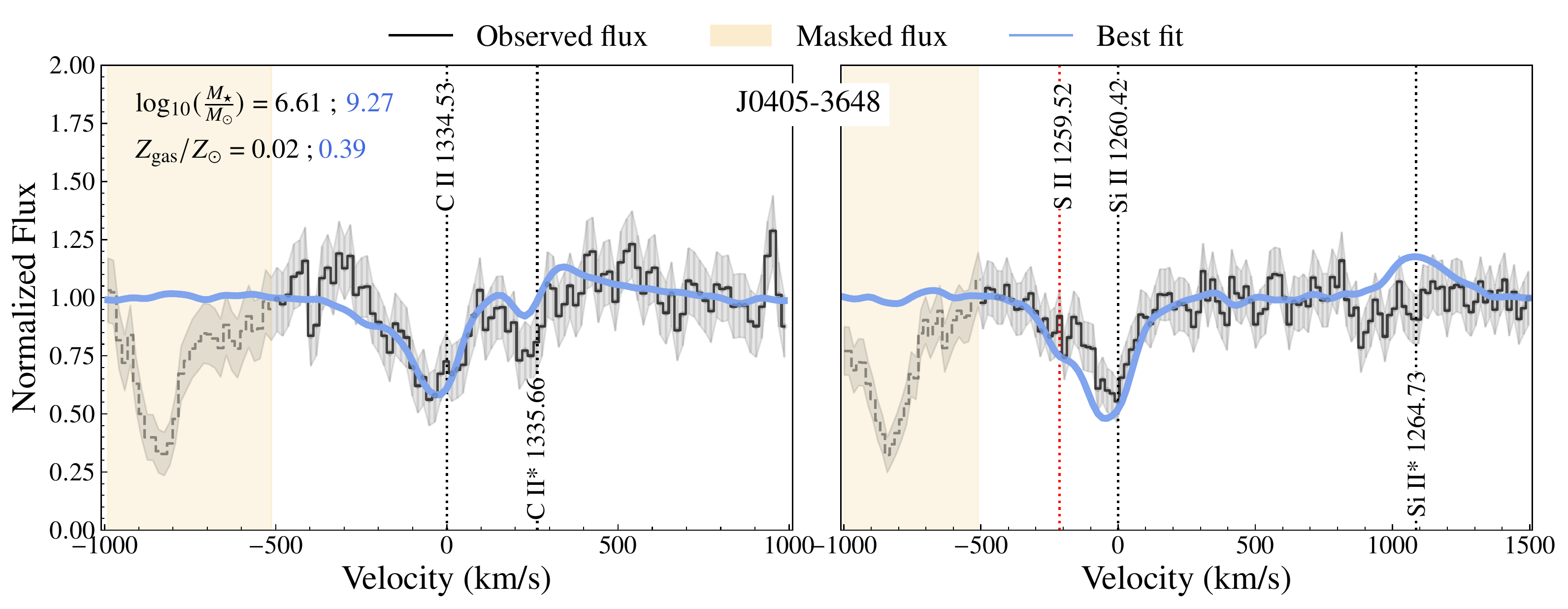}
    \includegraphics[width =0.45\textwidth,trim={0 0 0 1cm},clip]{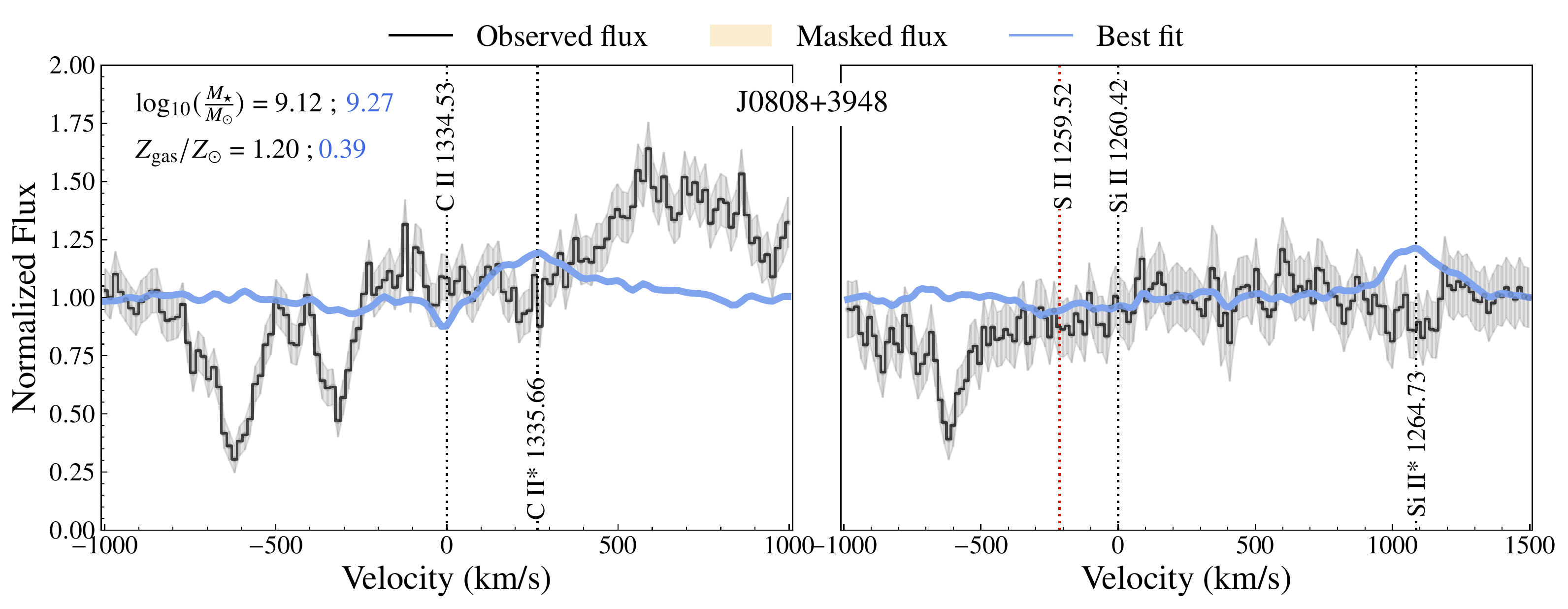}
    \includegraphics[width =0.45\textwidth,trim={0 0 0 1cm},clip]{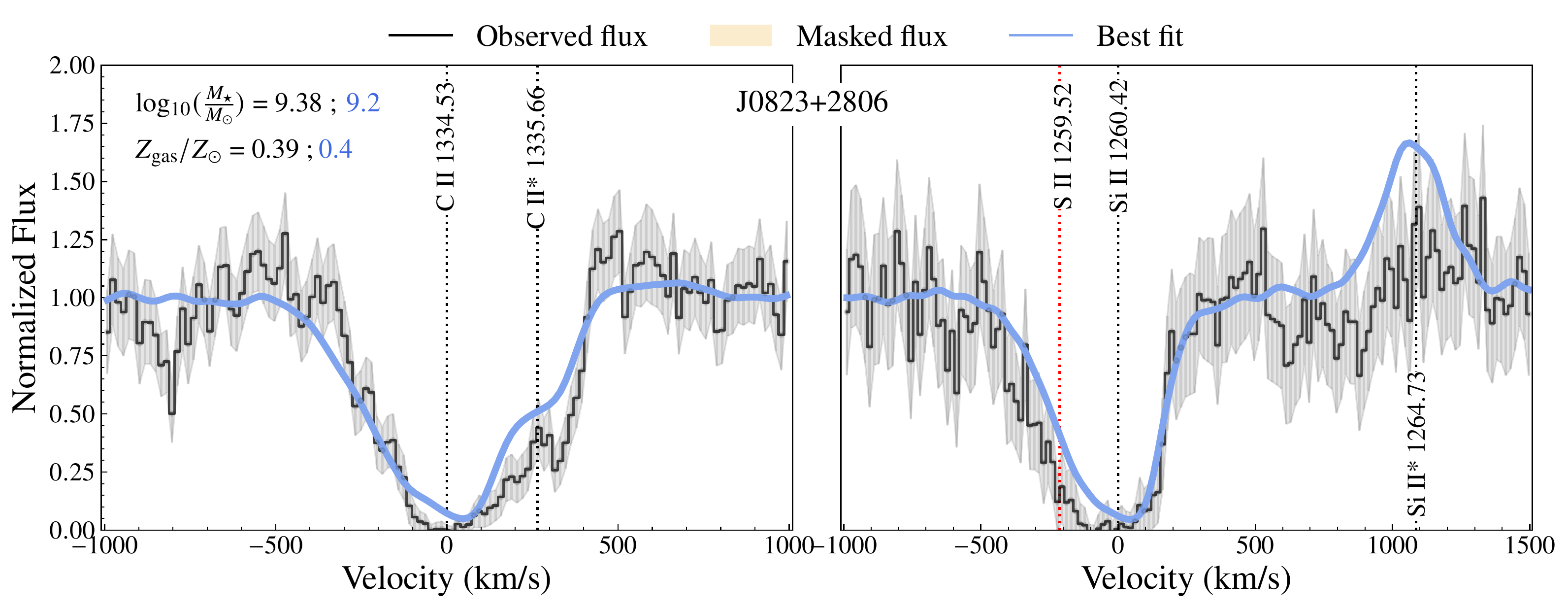}
    \includegraphics[width =0.45\textwidth,trim={0 0 0 1cm},clip]{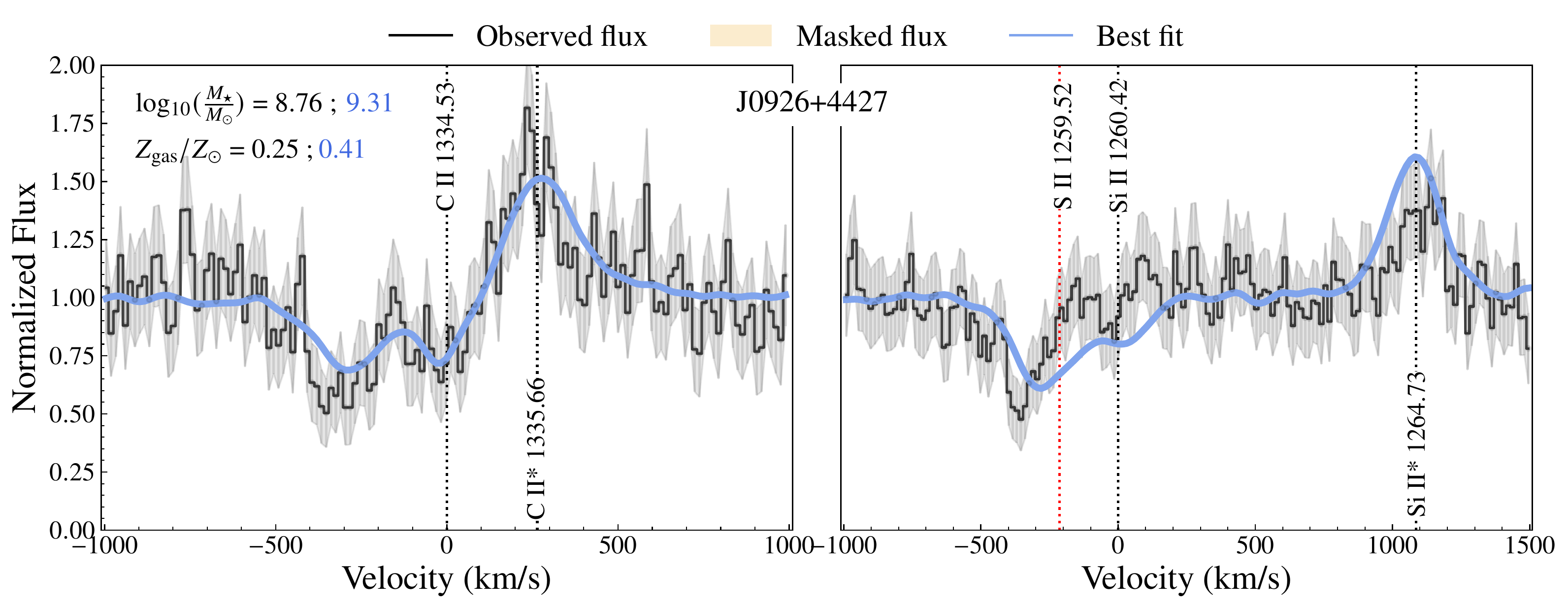}
    \includegraphics[width =0.45\textwidth,trim={0 0 0 1cm},clip]{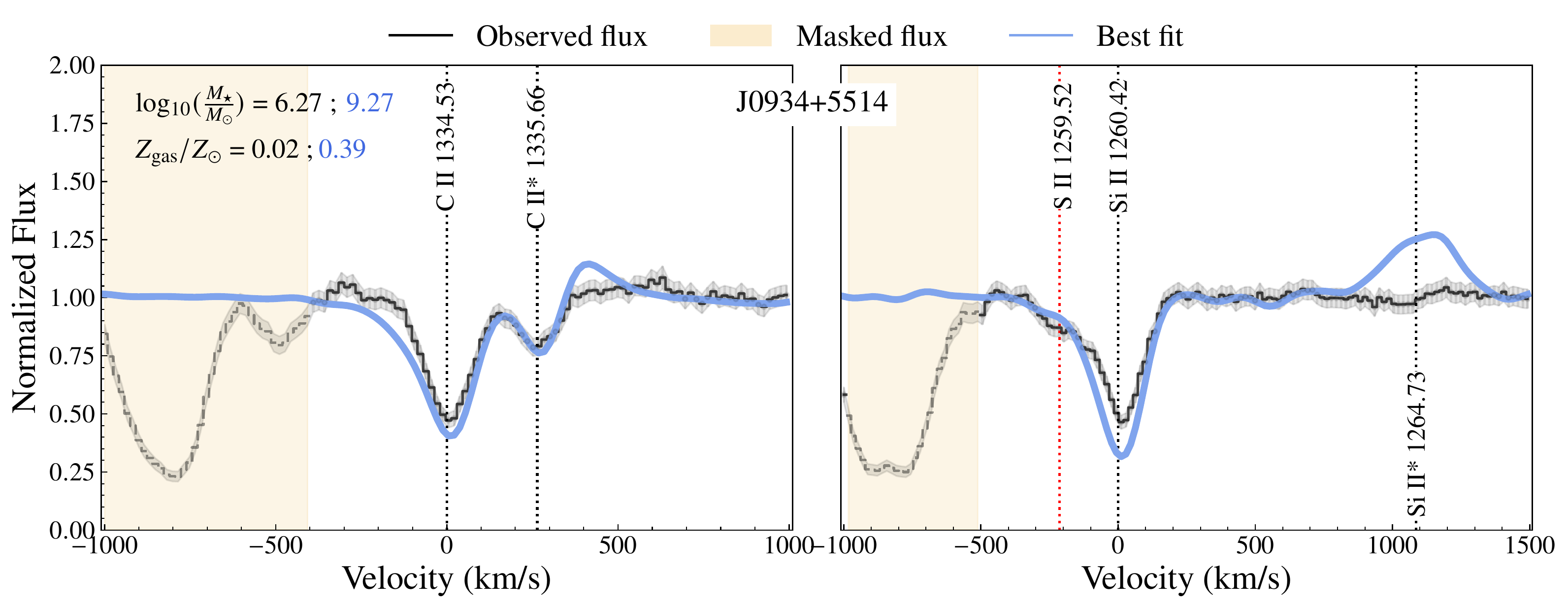}
    \includegraphics[width =0.45\textwidth,trim={0 0 0 1cm},clip]{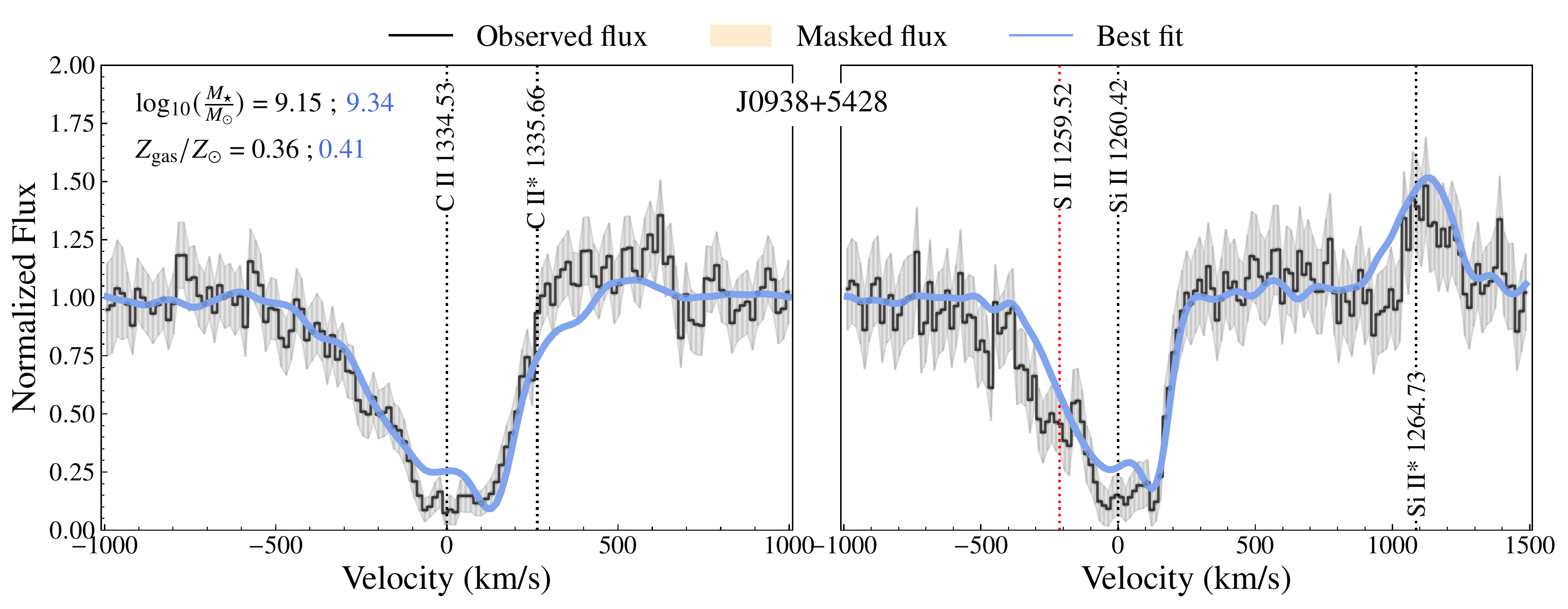}
    \includegraphics[width =0.45\textwidth,trim={0 0 0 1cm},clip]{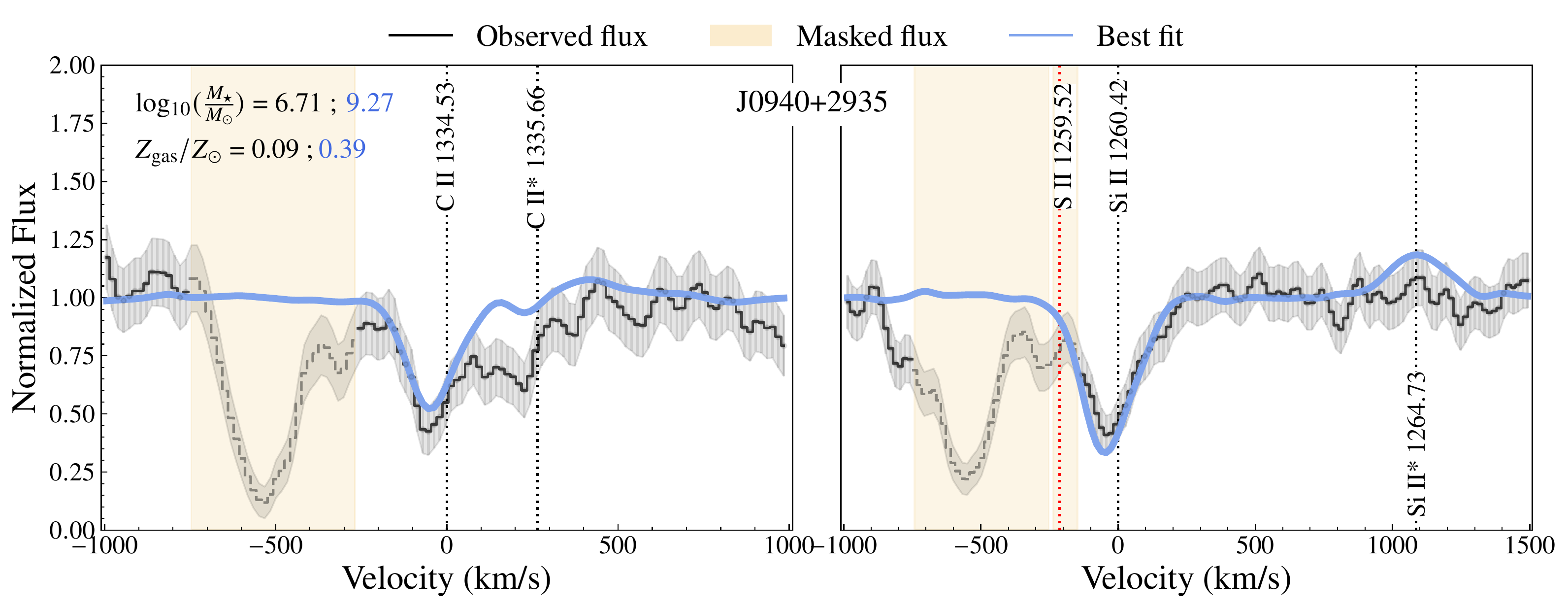}
    \includegraphics[width =0.45\textwidth,trim={0 0 0 1cm},clip]{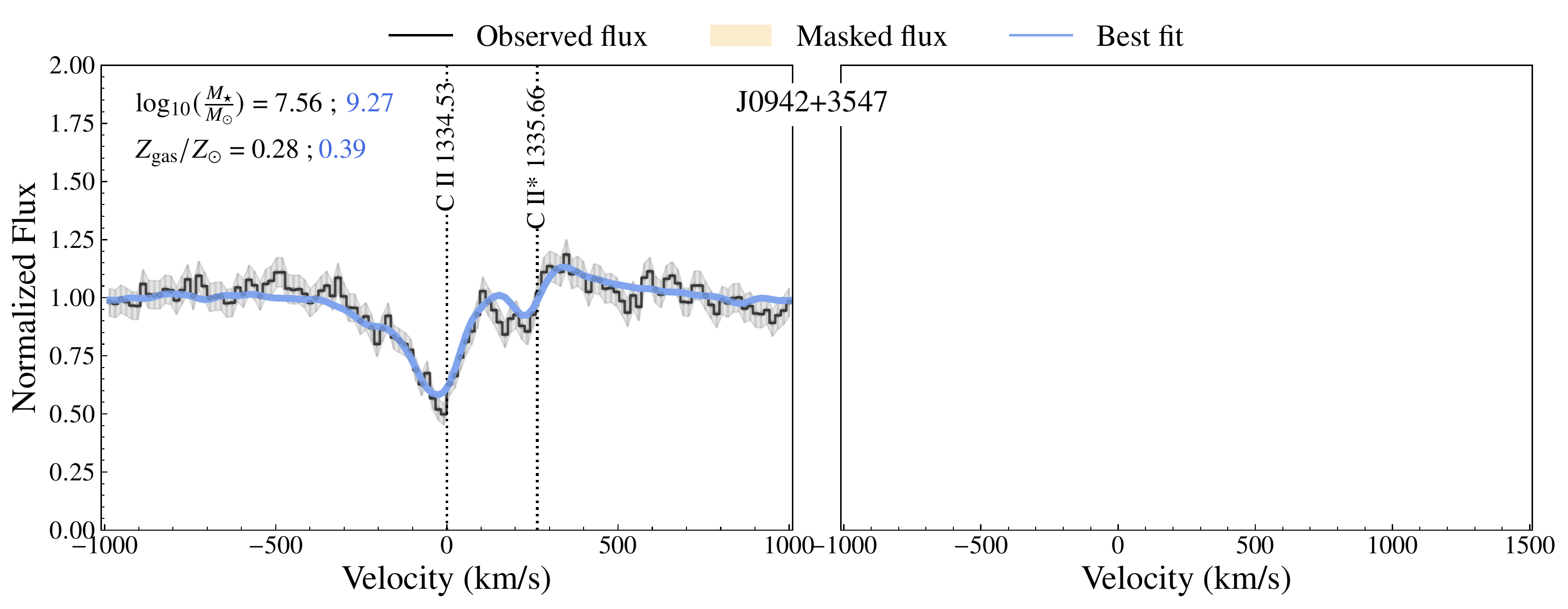}
    \includegraphics[width =0.45\textwidth,trim={0 0 0 1cm},clip]{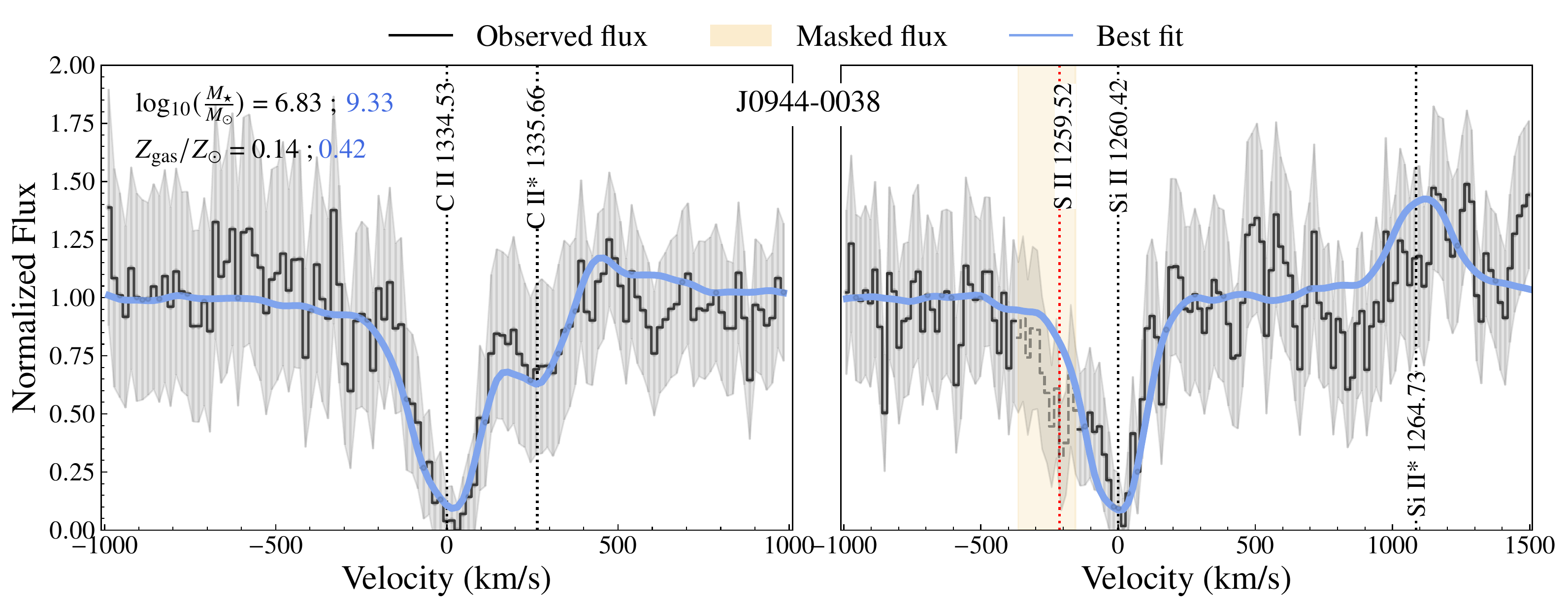}
    \caption{The fits of the \cIIl+\cIIlstar\ (left panels) and \siIIl+\siIIlstar\ (right panels) line profiles for all \classy\ galaxies using the procedure detailed in Section~\ref{sec:fitting}. The black line is the observed flux, the gray dashed line shows regions that are masked, and the grey shaded area is the flux error.  The vertical dotted lines indicate the expected location of each transition. The blue line is the best mock spectrum found, i.e., the mock spectrum with the lower $\chi^2$ based on Eq~\eqref{eq:chi2}. In the top left corner of the left panels, we detail the stellar mass and gas metallicity of the classy galaxy (in black) and of the virtual galaxy at the time step where the best fit has been found (in blue). } 
    \label{fig:fits1a}
\end{figure*}

\begin{figure*}[!htbp]
    \centering
    \includegraphics[width =0.45\textwidth]{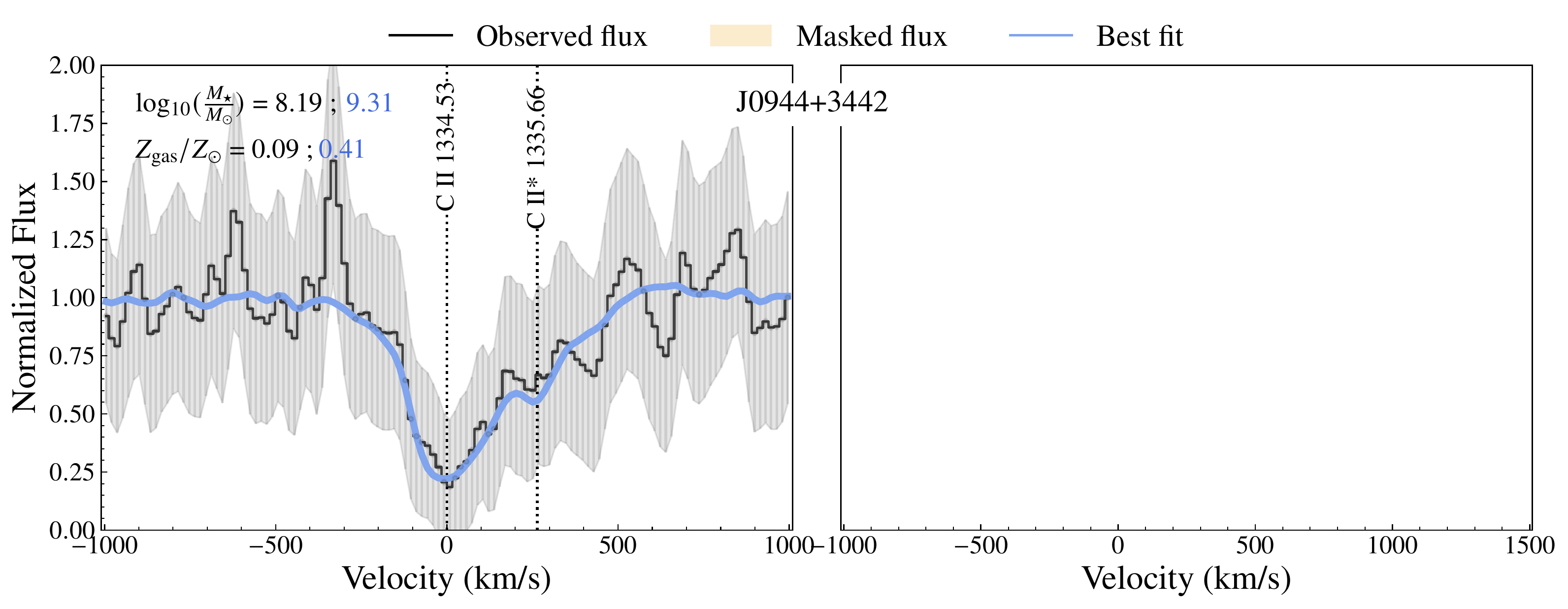}
    \includegraphics[width =0.45\textwidth]{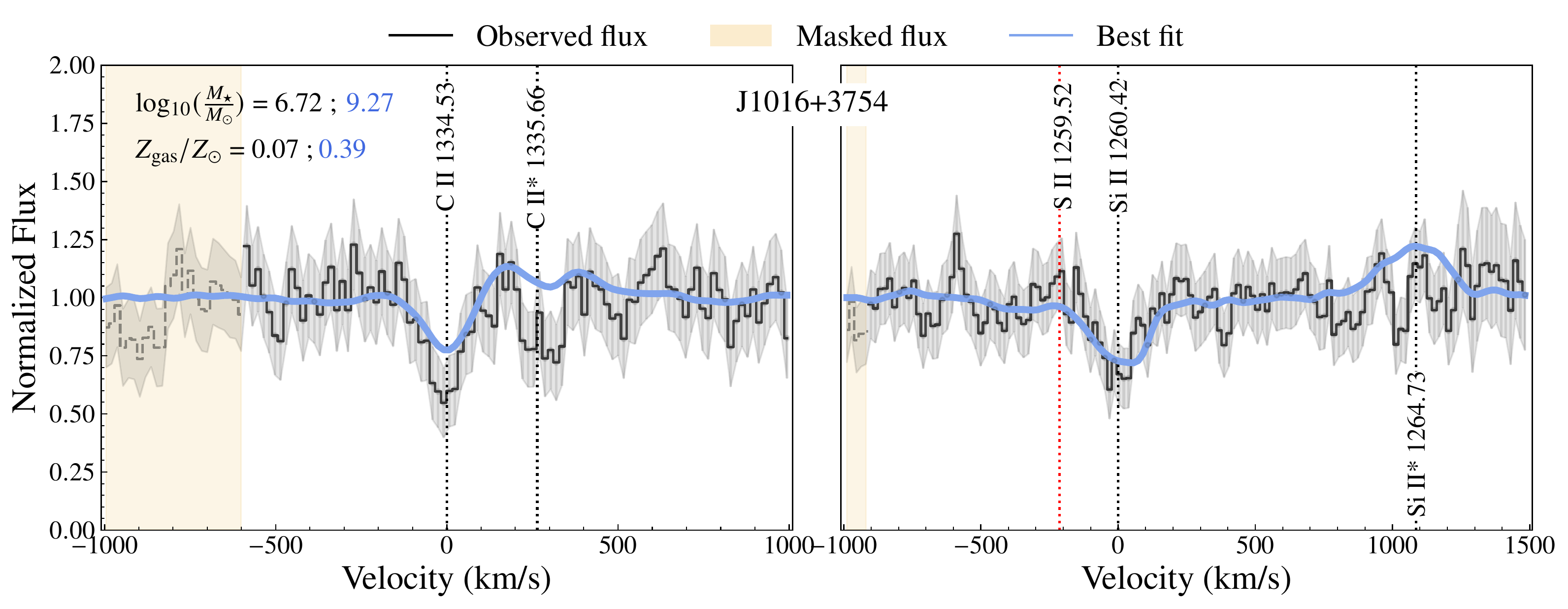}
    \includegraphics[width =0.45\textwidth,trim={0 0 0 1cm},clip]{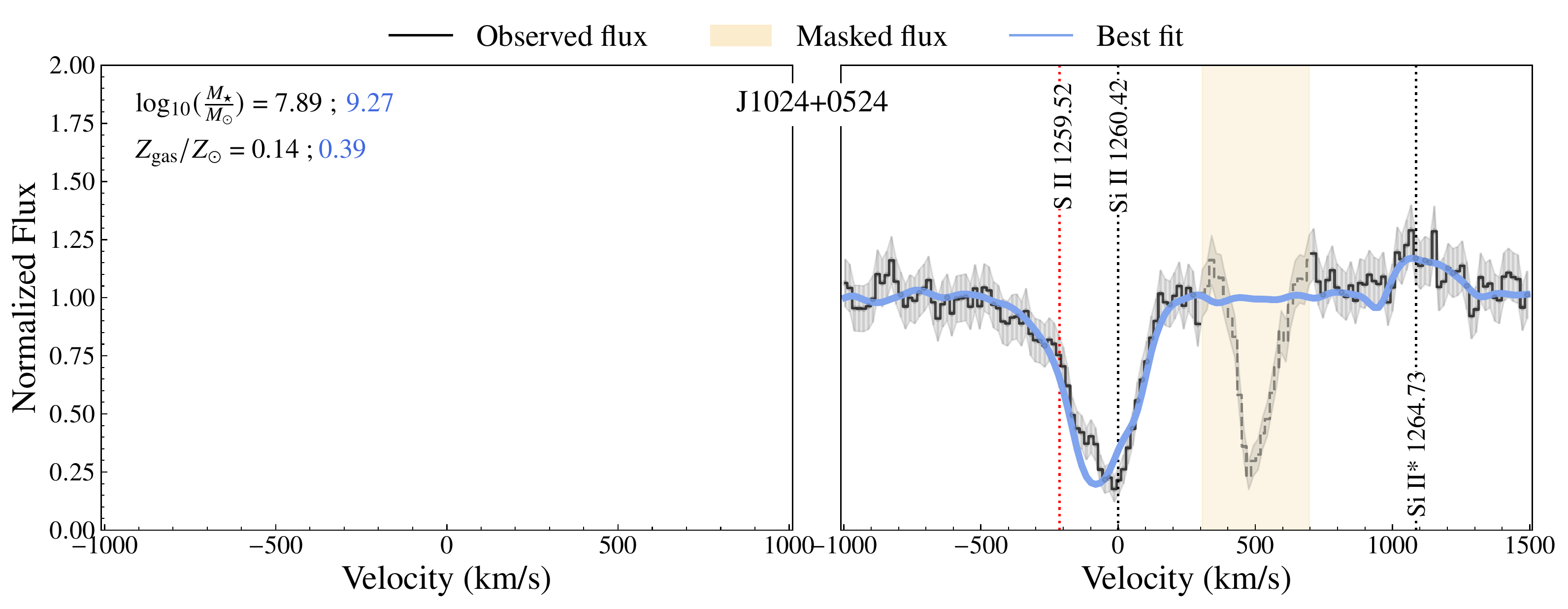}
    \includegraphics[width =0.45\textwidth,trim={0 0 0 1cm},clip]{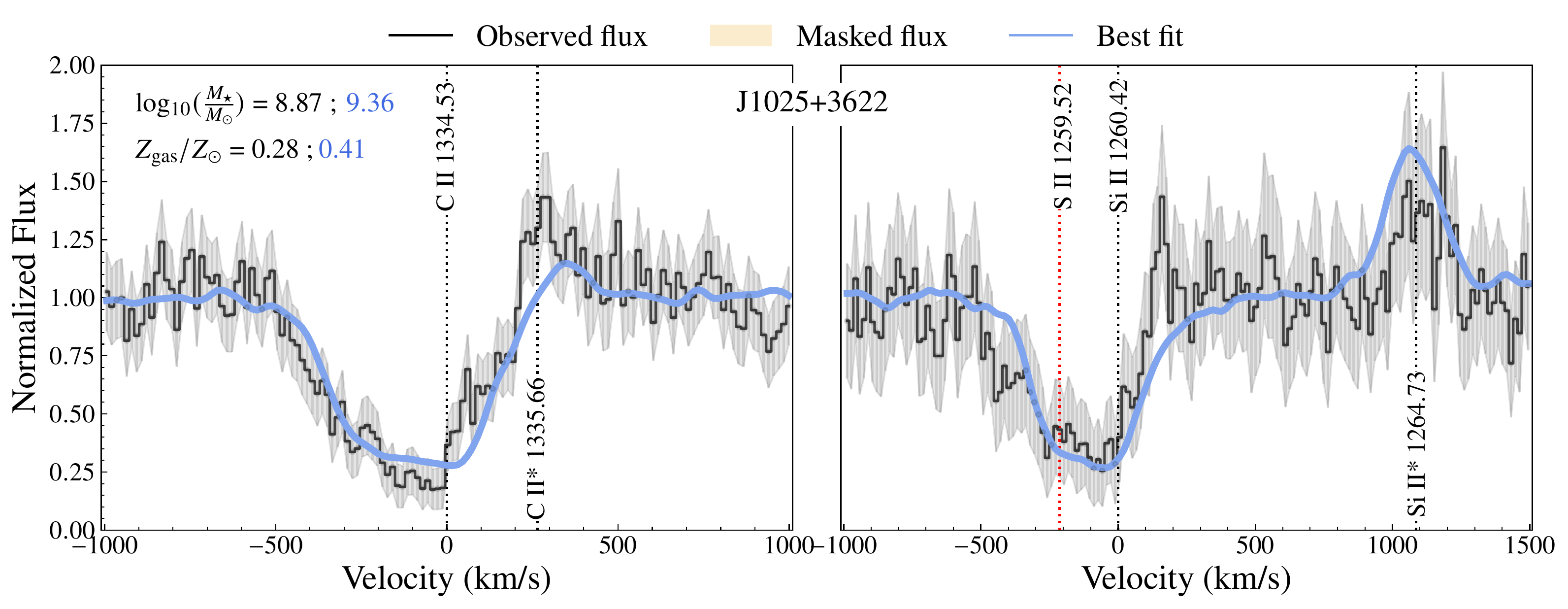}
    \includegraphics[width =0.45\textwidth,trim={0 0 0 1cm},clip]{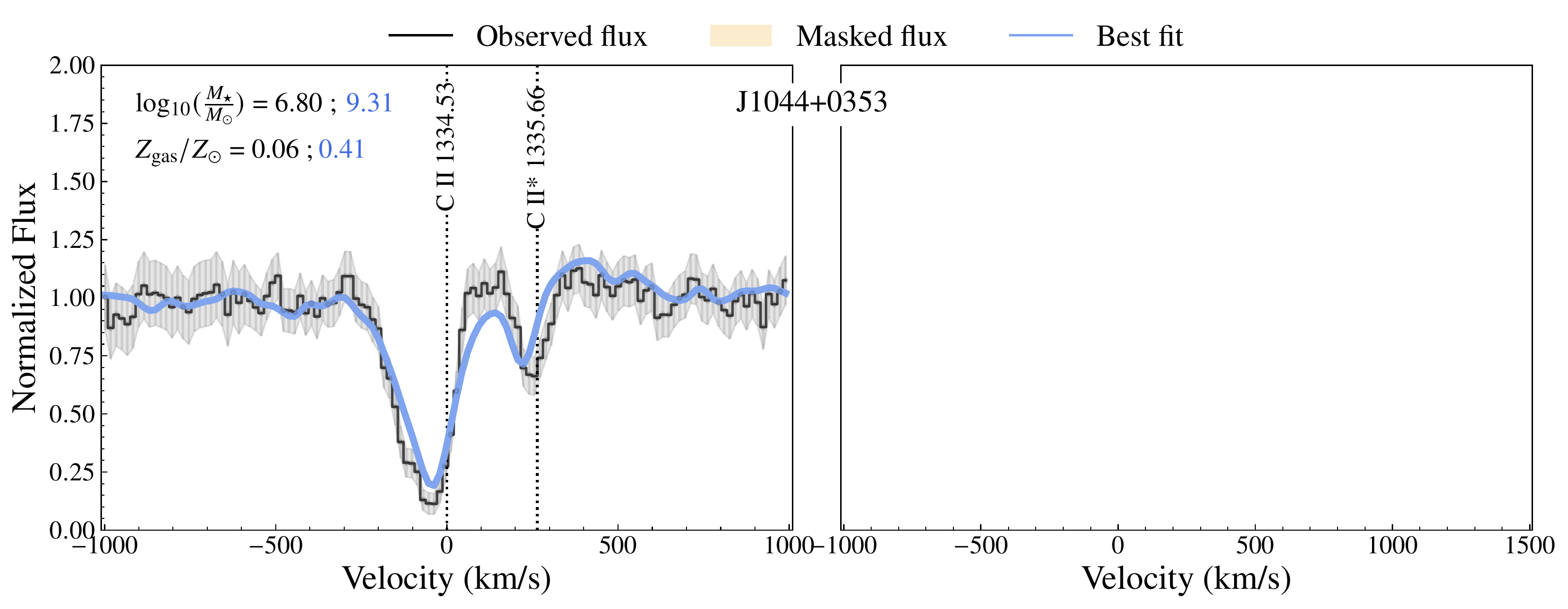}
    \includegraphics[width =0.45\textwidth,trim={0 0 0 1cm},clip]{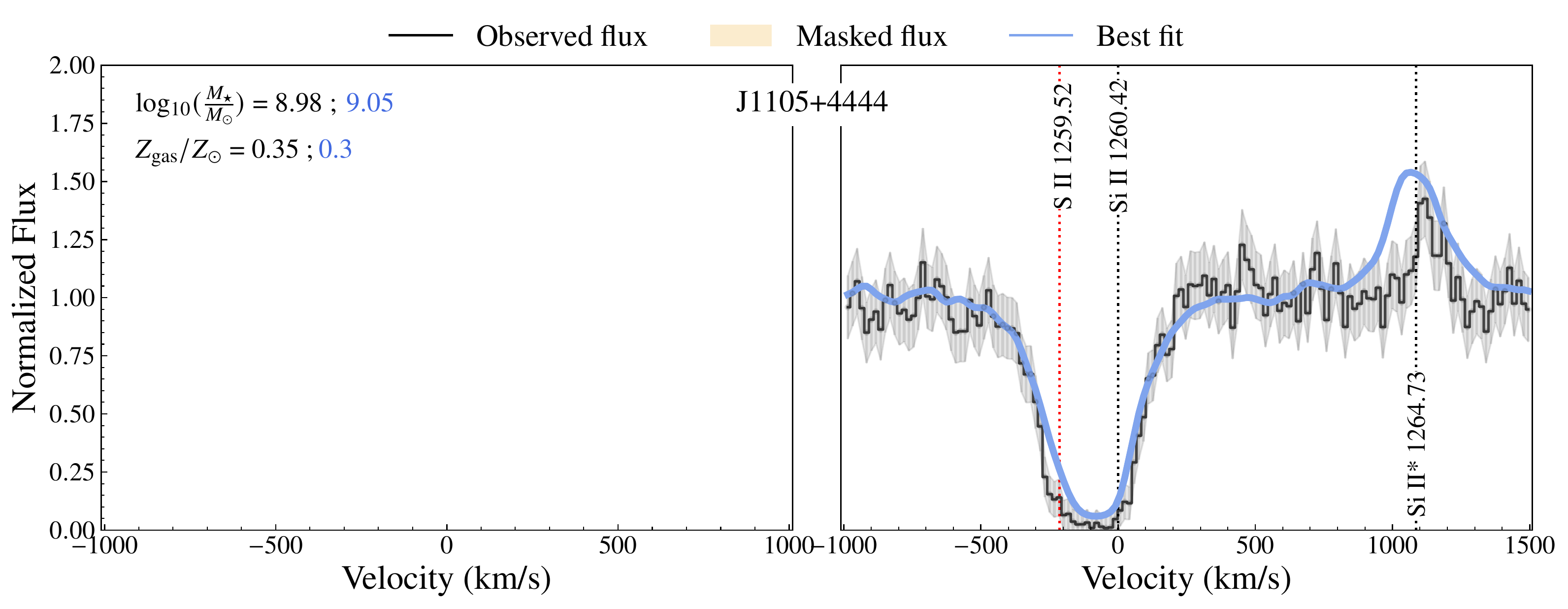}
    \includegraphics[width =0.45\textwidth,trim={0 0 0 1cm},clip]{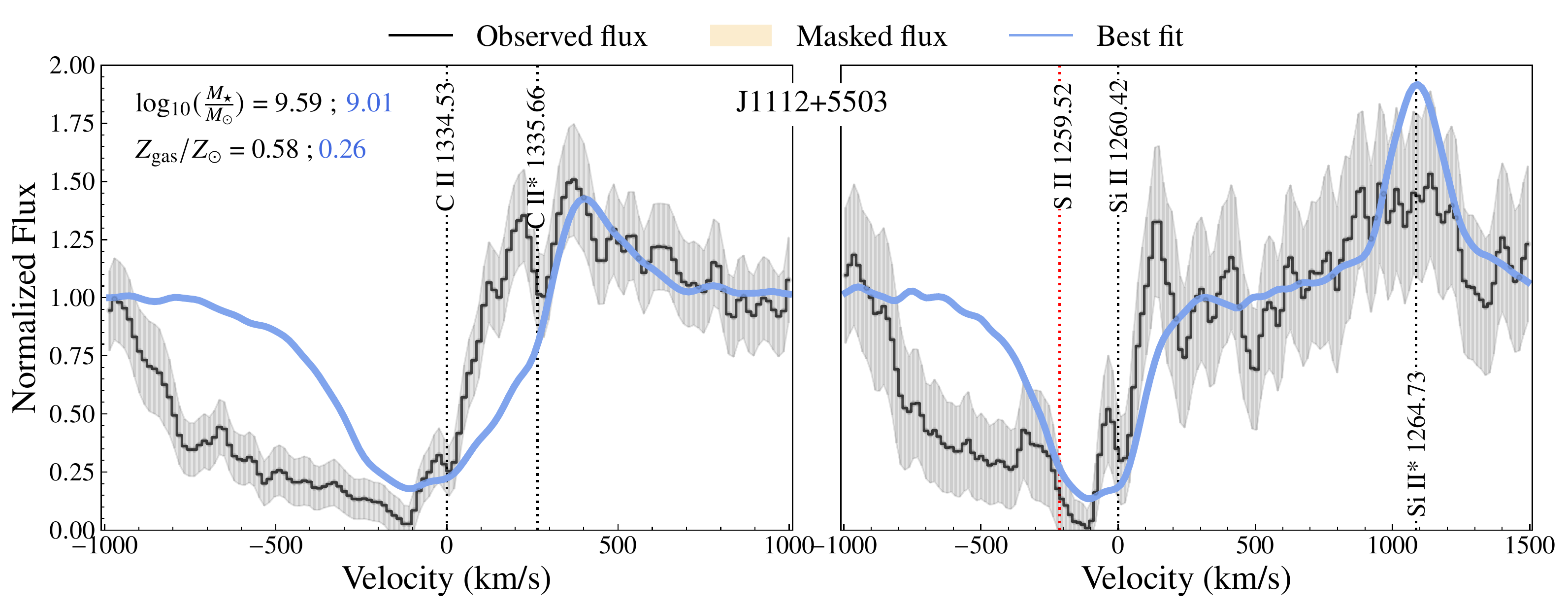}
    \includegraphics[width =0.45\textwidth,trim={0 0 0 1cm},clip]{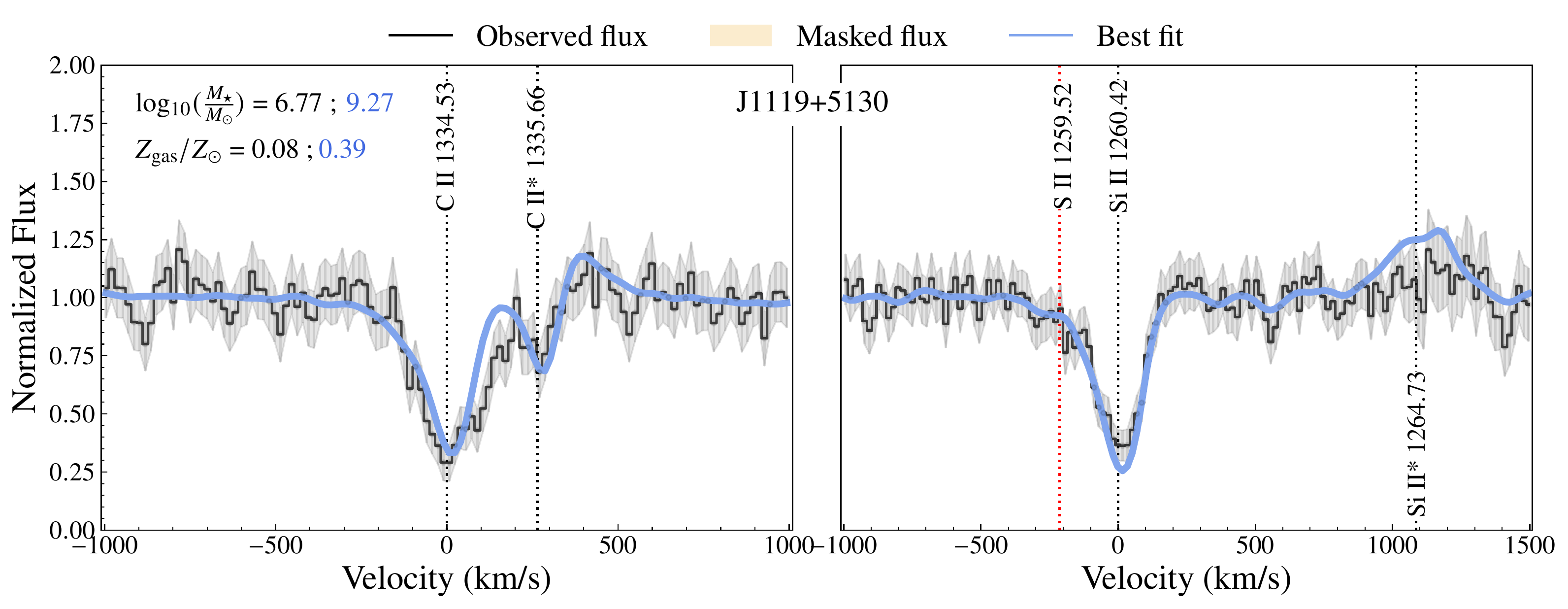}
    \includegraphics[width =0.45\textwidth,trim={0 0 0 1cm},clip]{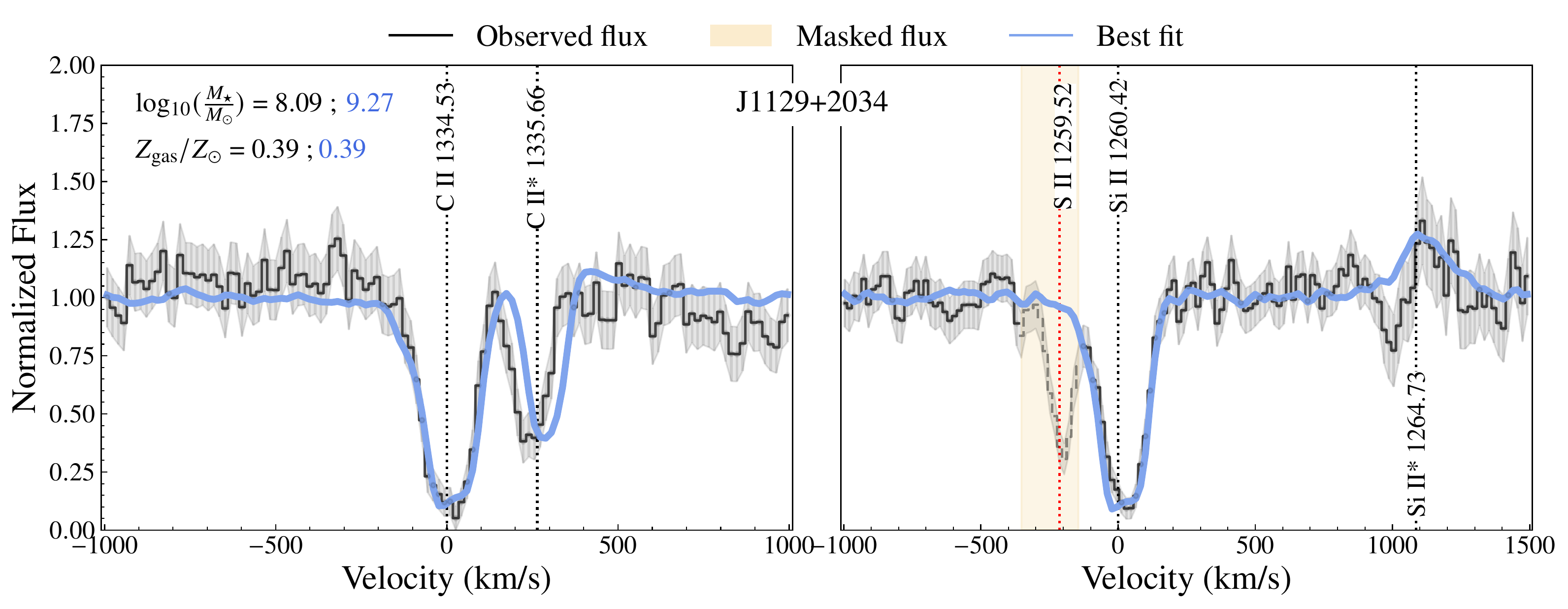}
    \includegraphics[width =0.45\textwidth,trim={0 0 0 1cm},clip]{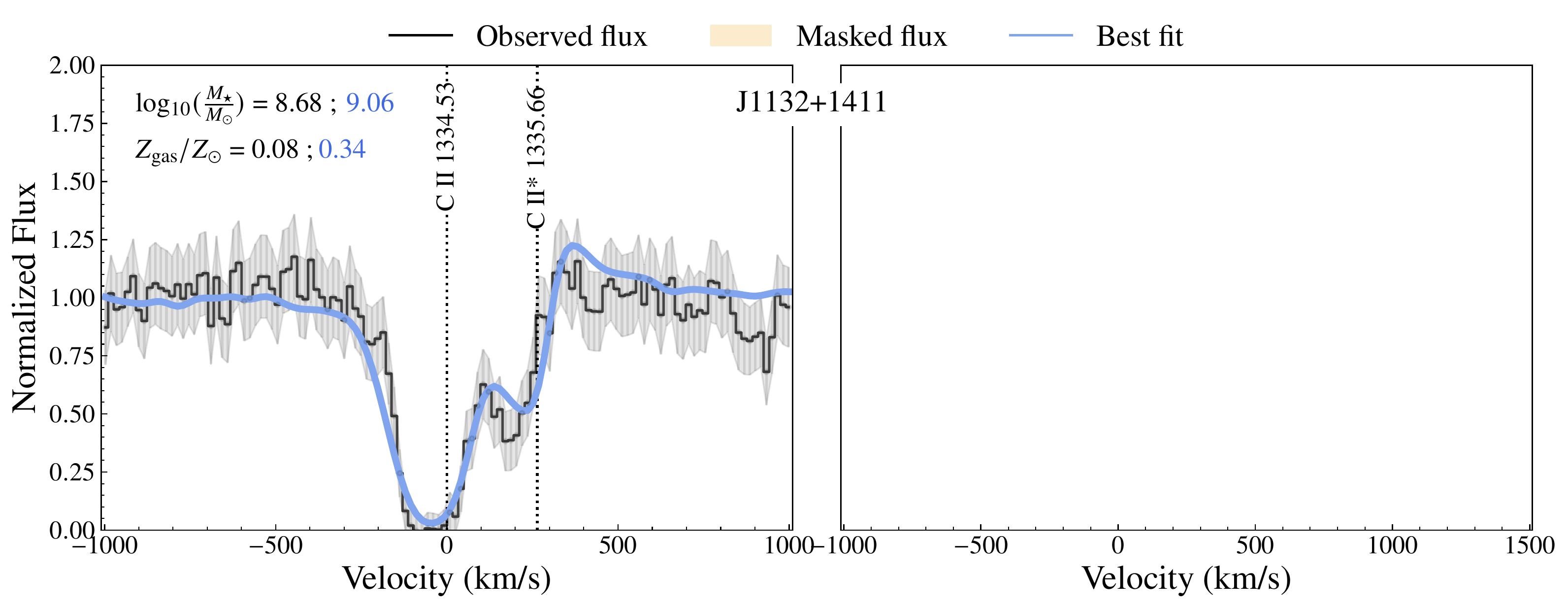}
    \includegraphics[width =0.45\textwidth,trim={0 0 0 1cm},clip]{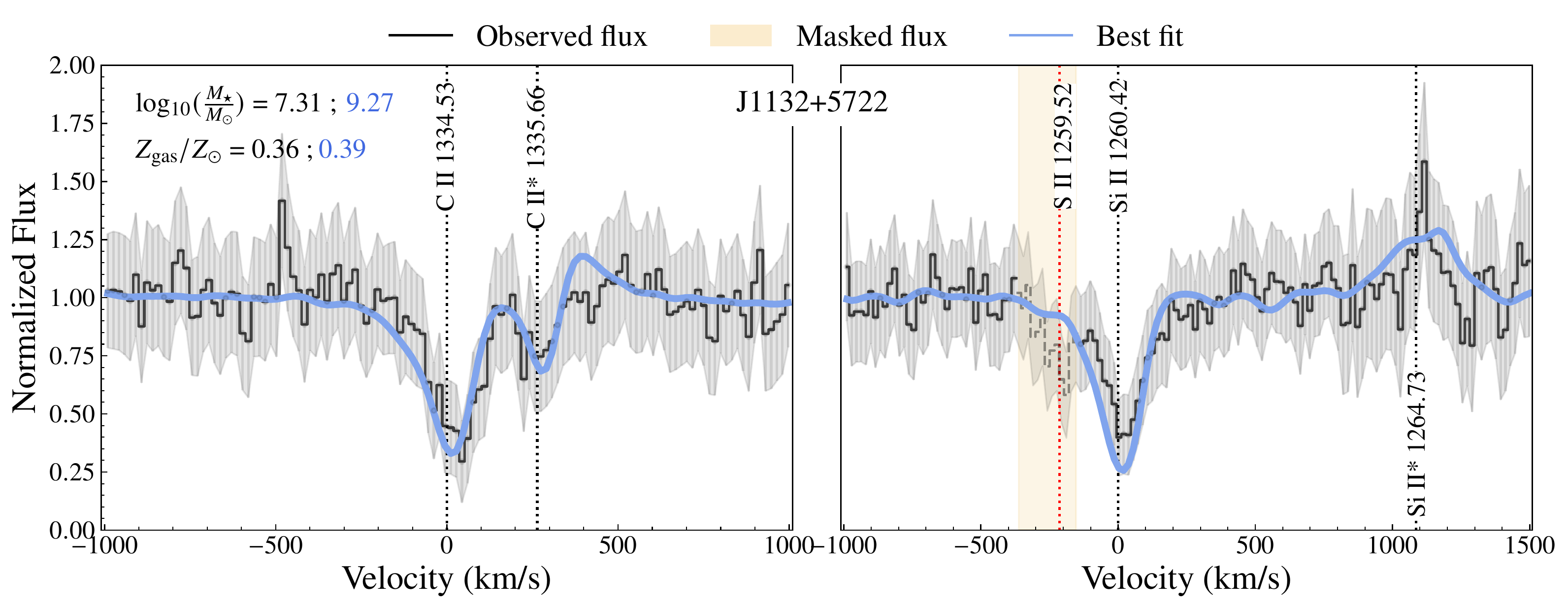}
    \includegraphics[width =0.45\textwidth,trim={0 0 0 1cm},clip]{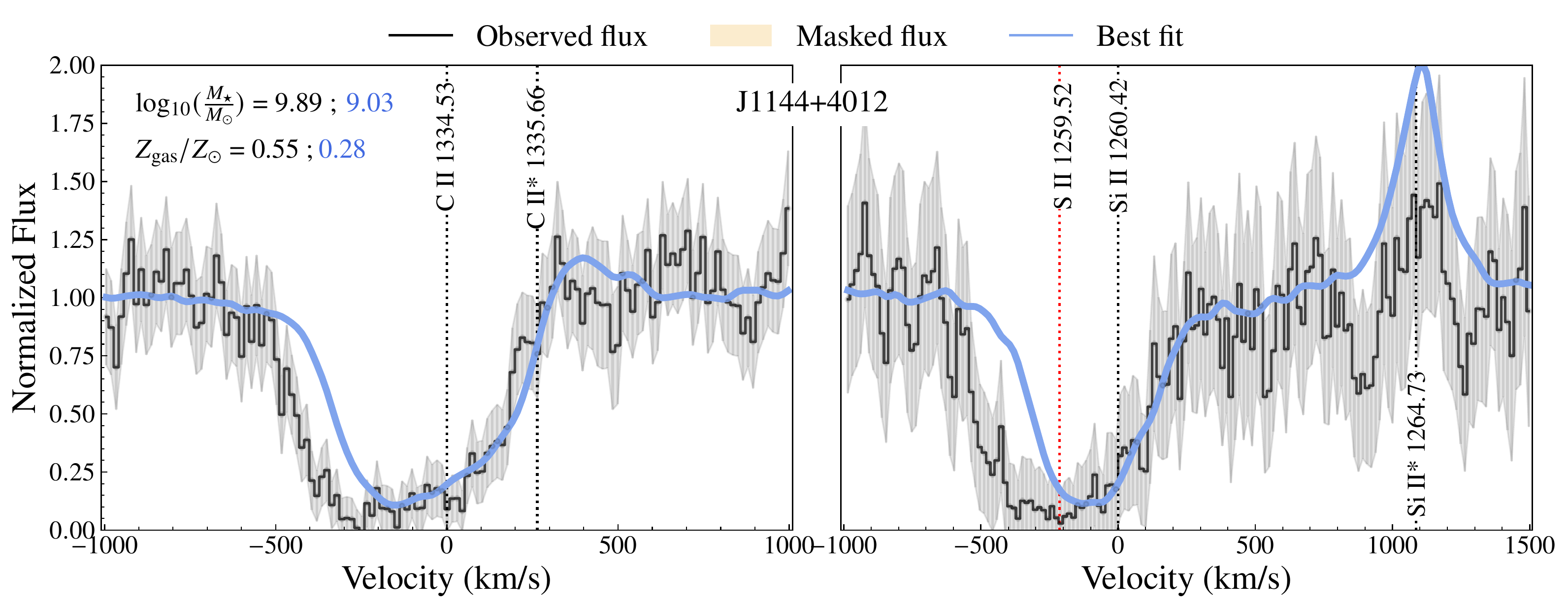}
    \includegraphics[width =0.45\textwidth,trim={0 0 0 1cm},clip]{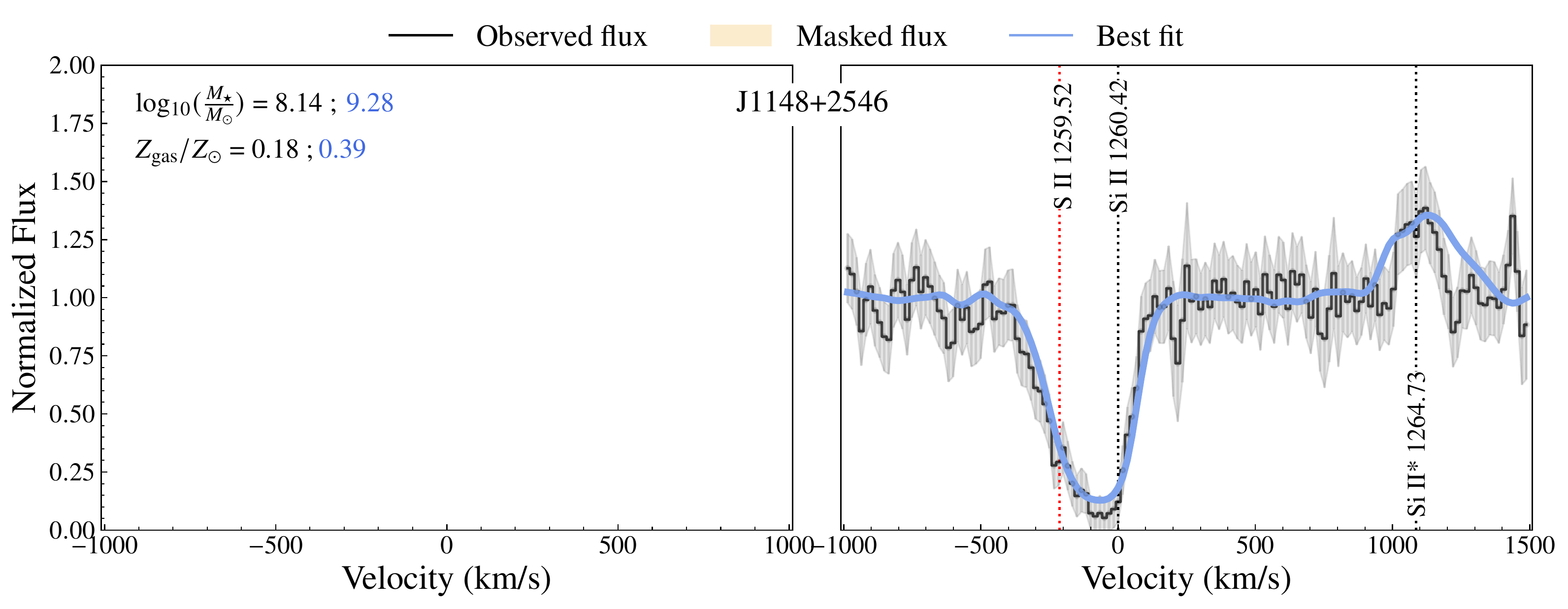}
    \includegraphics[width =0.45\textwidth,trim={0 0 0 1cm},clip]{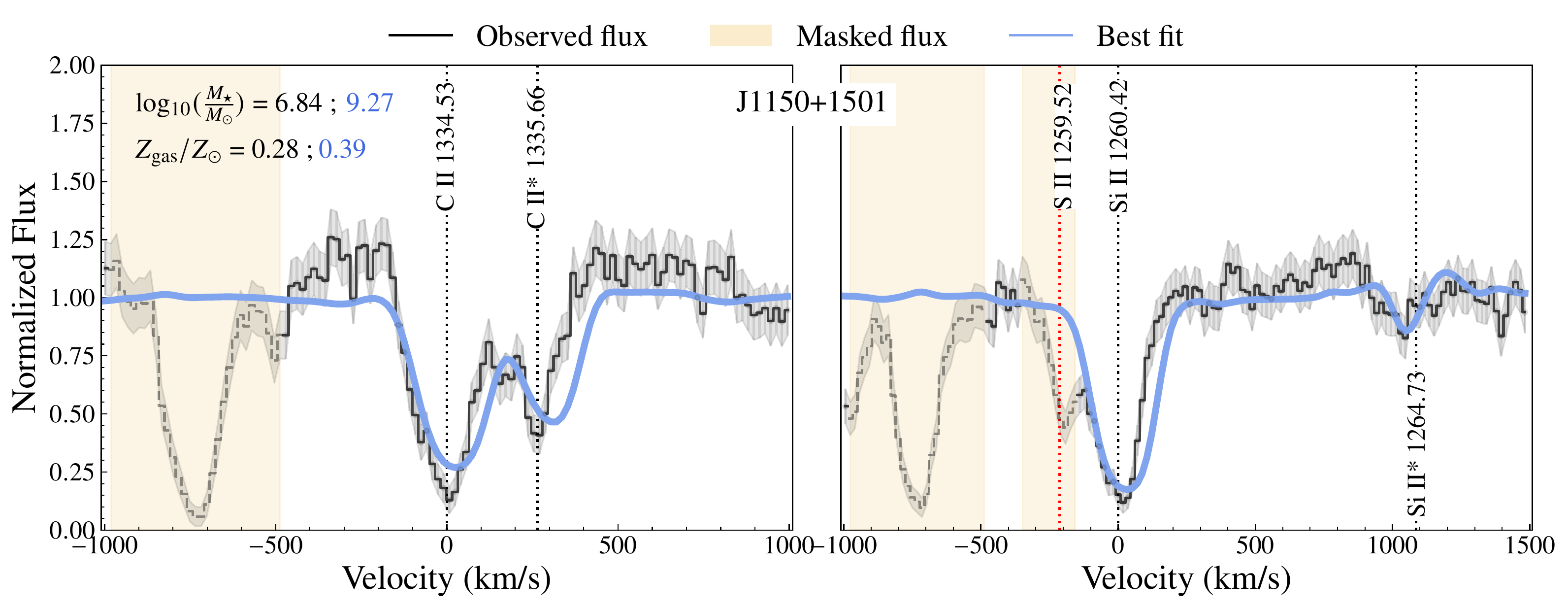}
    \includegraphics[width =0.45\textwidth,trim={0 0 0 1cm},clip]{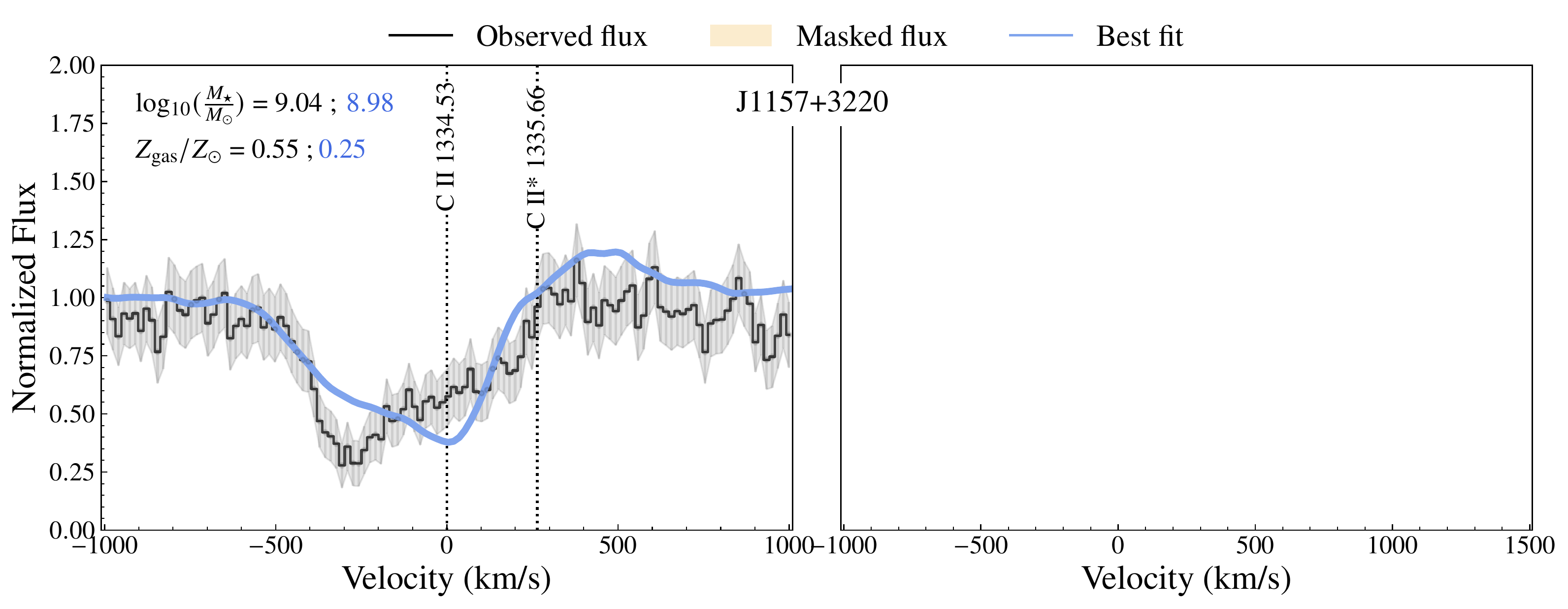}
    \includegraphics[width =0.45\textwidth,trim={0 0 0 1cm},clip]{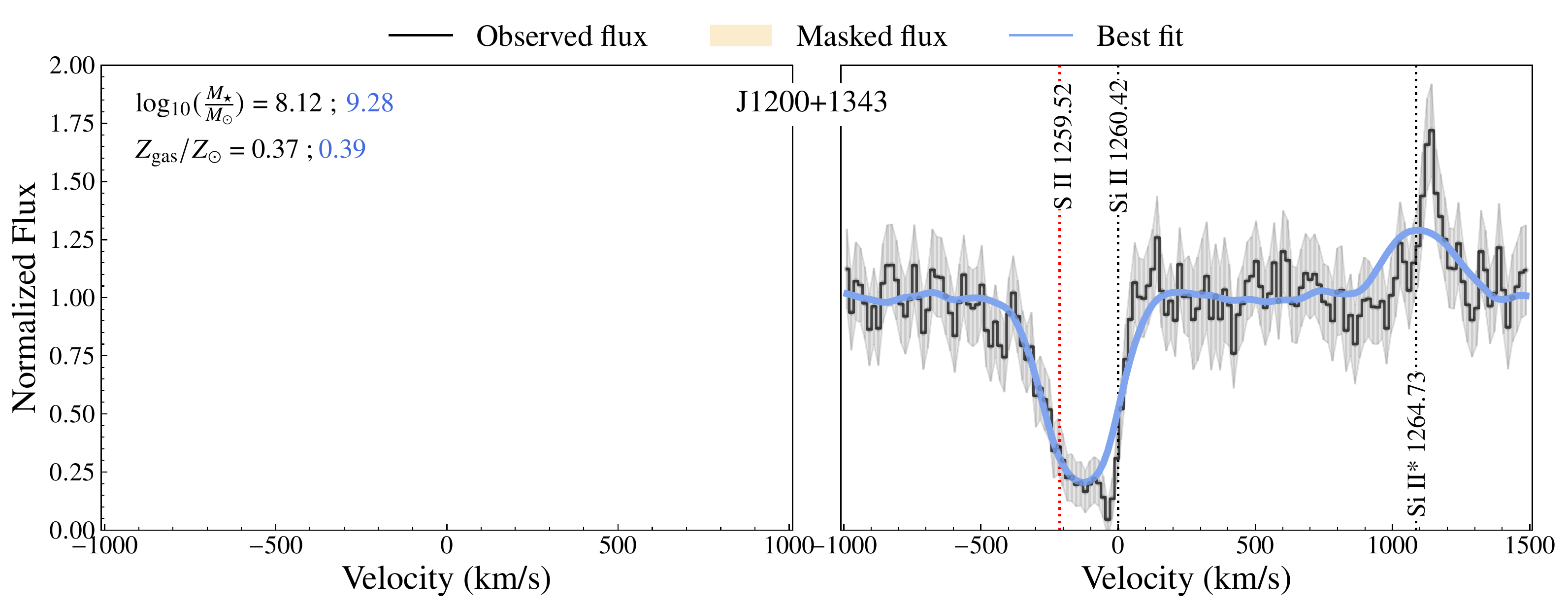}
    \ContinuedFloat
    \caption{\textit{(cont.)}}

\end{figure*}

\begin{figure*}[!htbp]
    \centering
    \includegraphics[width =0.45\textwidth]{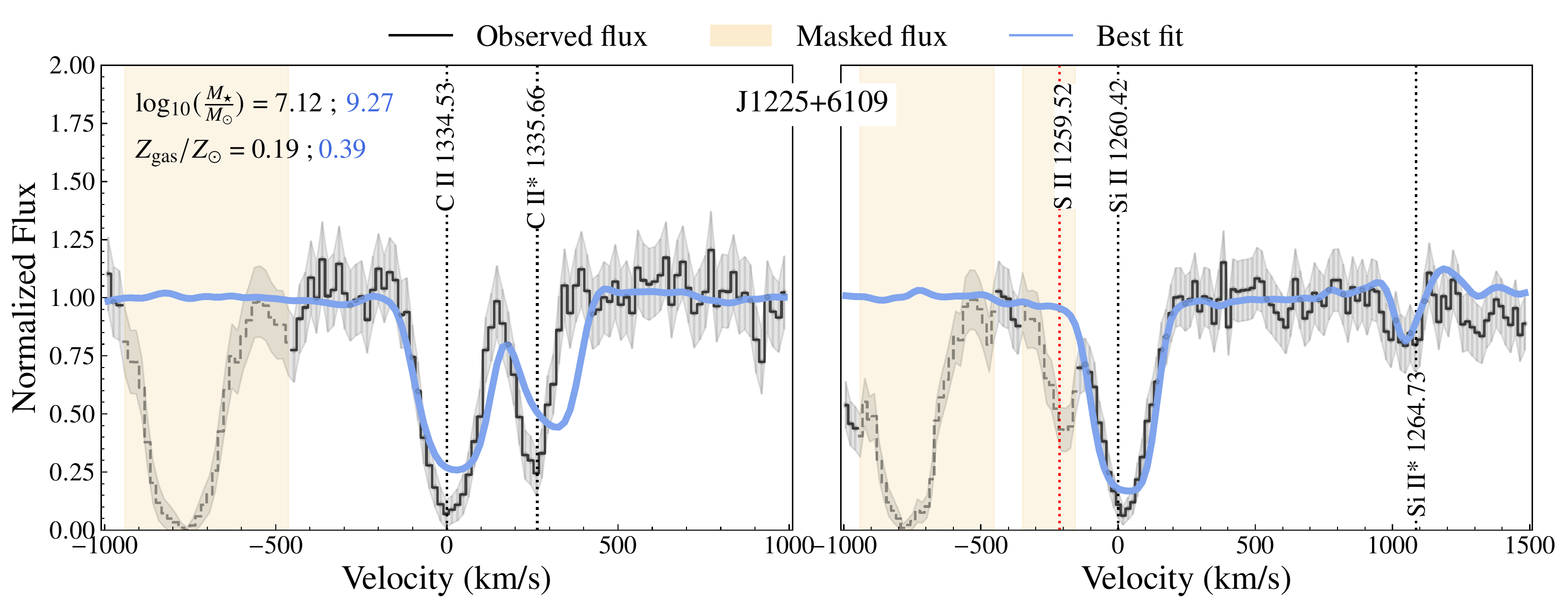}
    \includegraphics[width =0.45\textwidth]{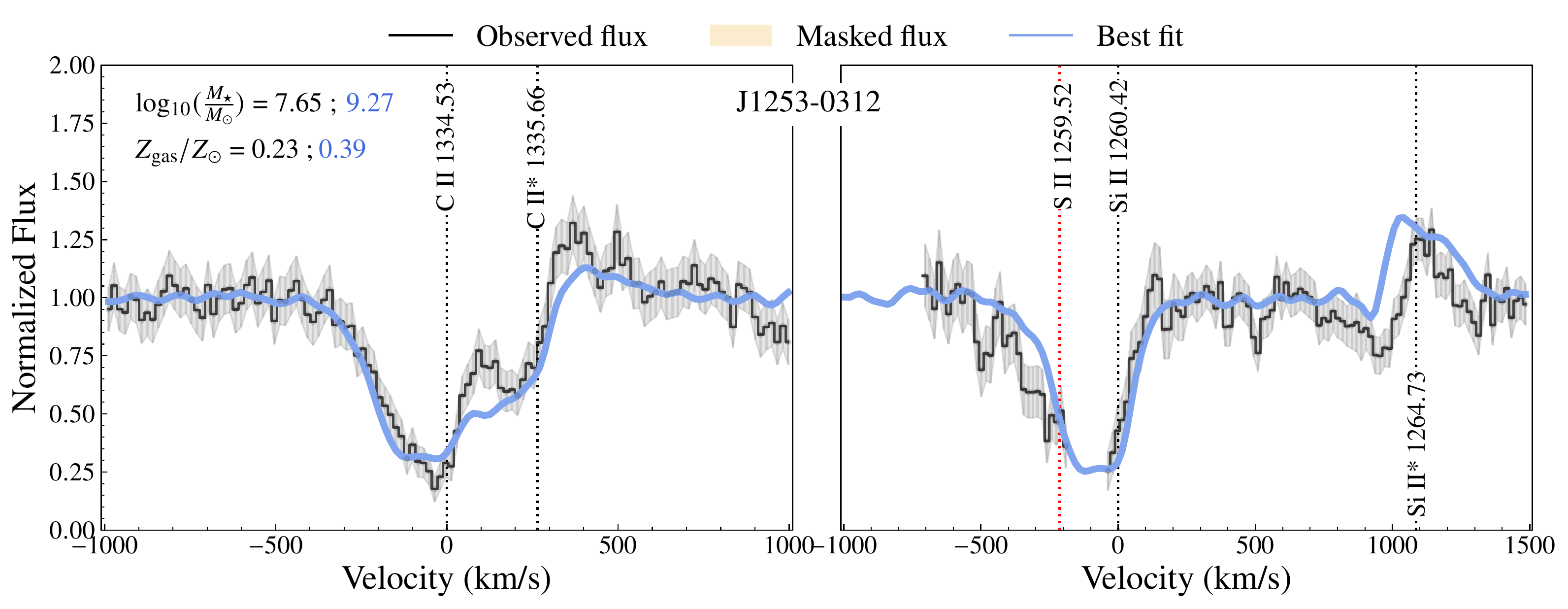}
    \includegraphics[width =0.45\textwidth,trim={0 0 0 1cm},clip]{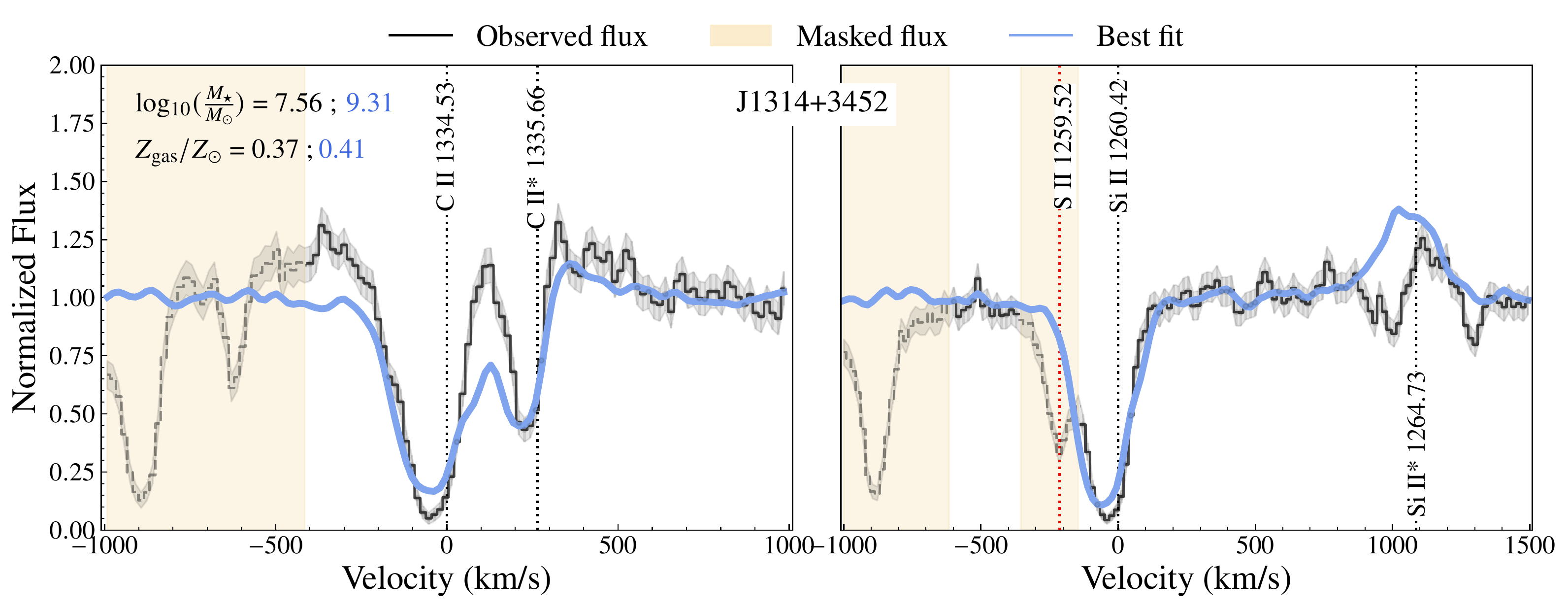}
    \includegraphics[width =0.45\textwidth,trim={0 0 0 1cm},clip]{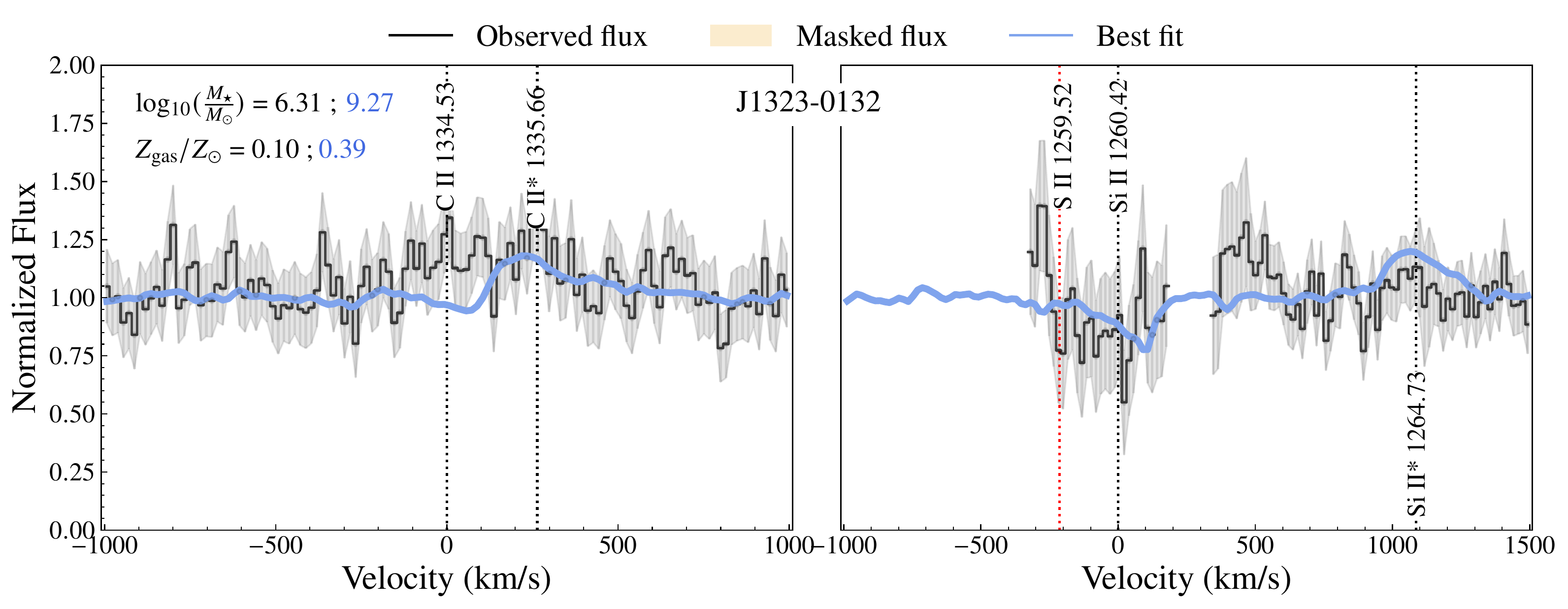}
    \includegraphics[width =0.45\textwidth,trim={0 0 0 1cm},clip]{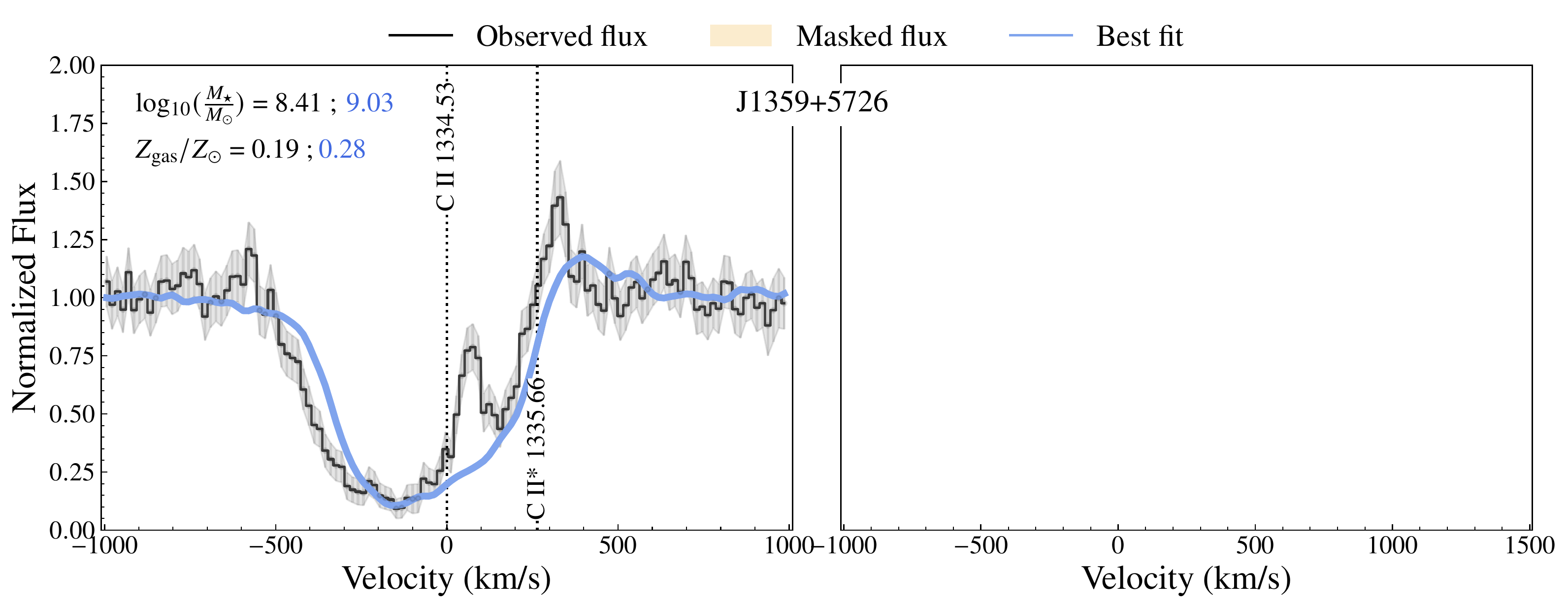}
    \includegraphics[width =0.45\textwidth,trim={0 0 0 1cm},clip]{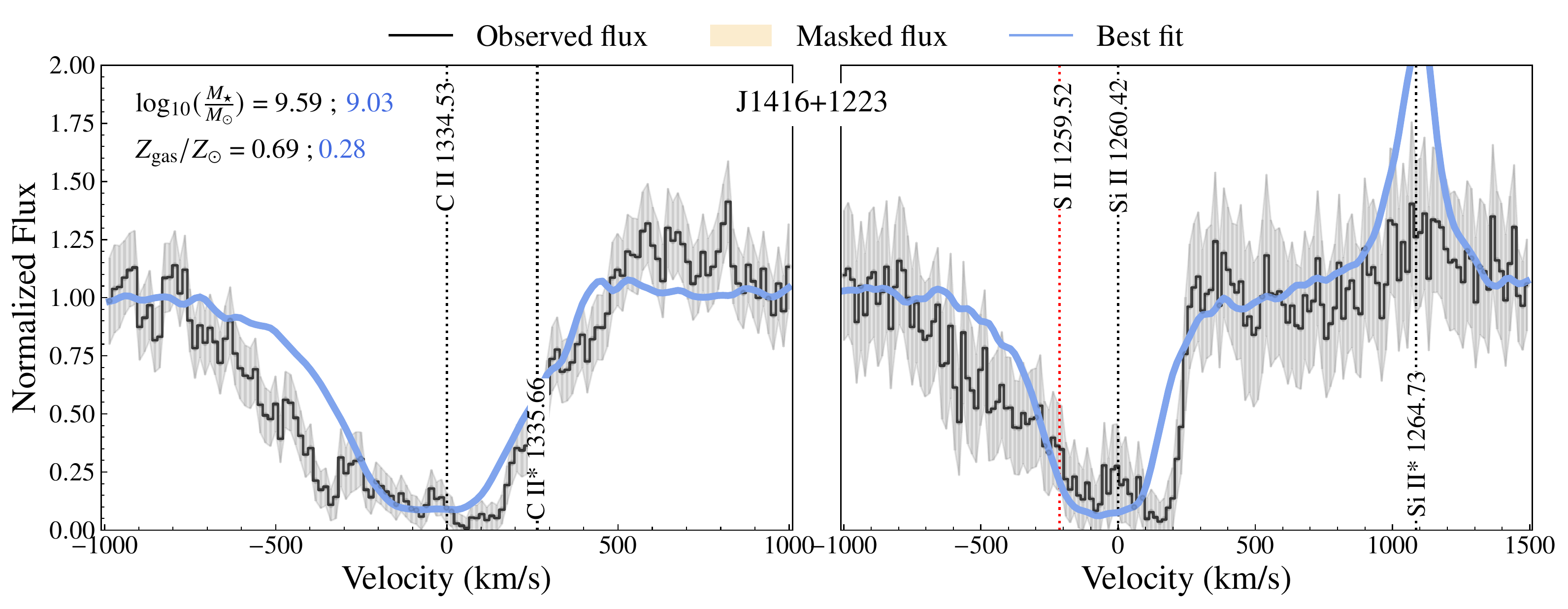}
    \includegraphics[width =0.45\textwidth,trim={0 0 0 1cm},clip]{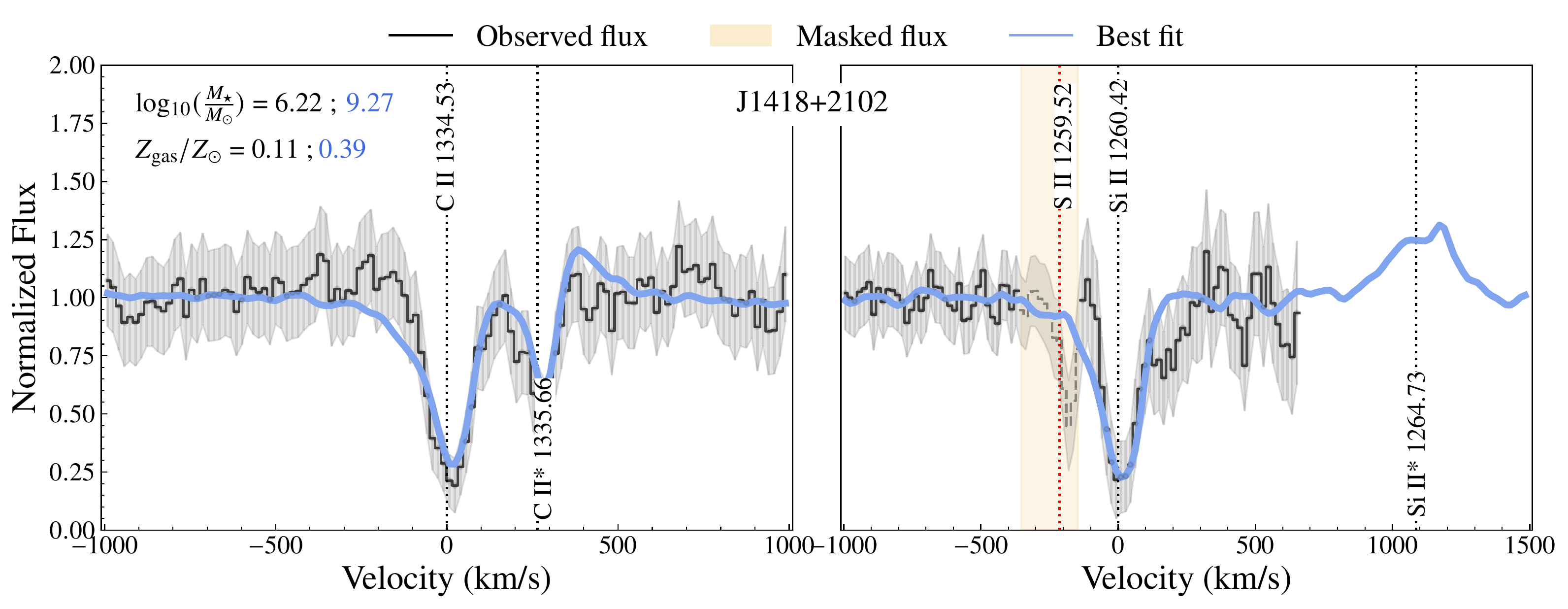}
    \includegraphics[width =0.45\textwidth,trim={0 0 0 1cm},clip]{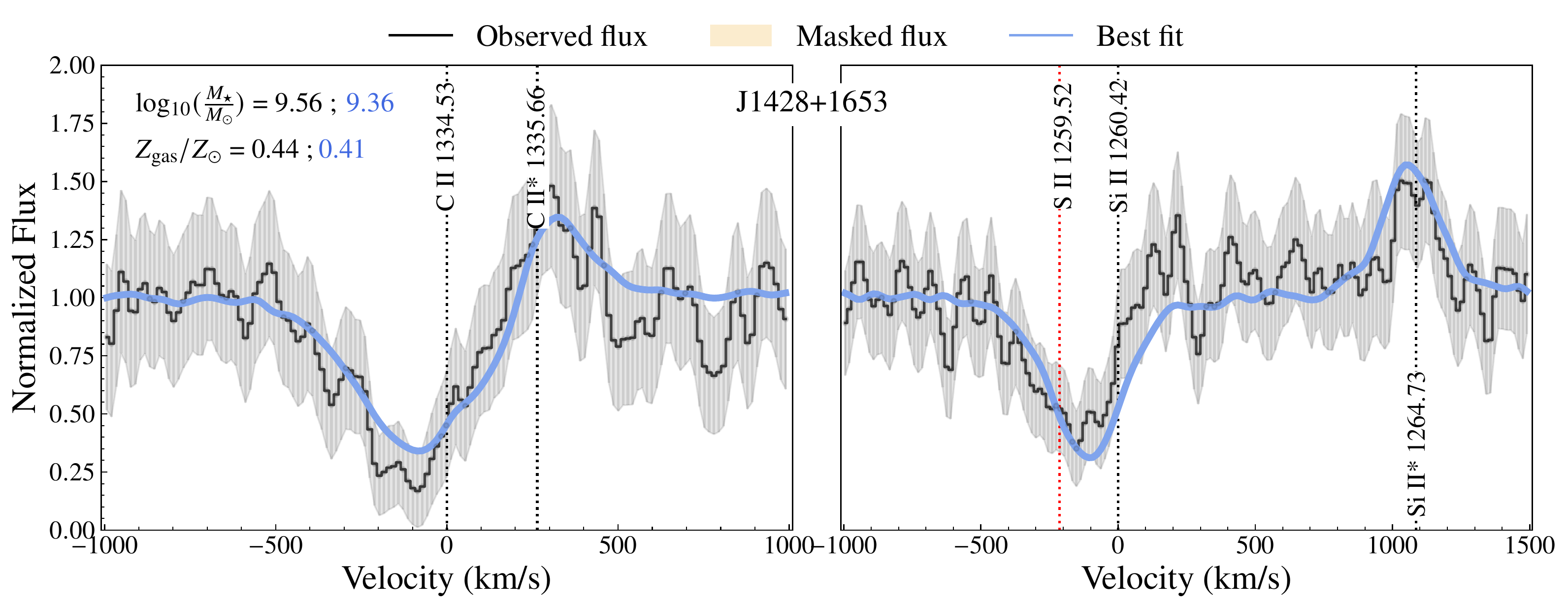}
    \includegraphics[width =0.45\textwidth,trim={0 0 0 1cm},clip]{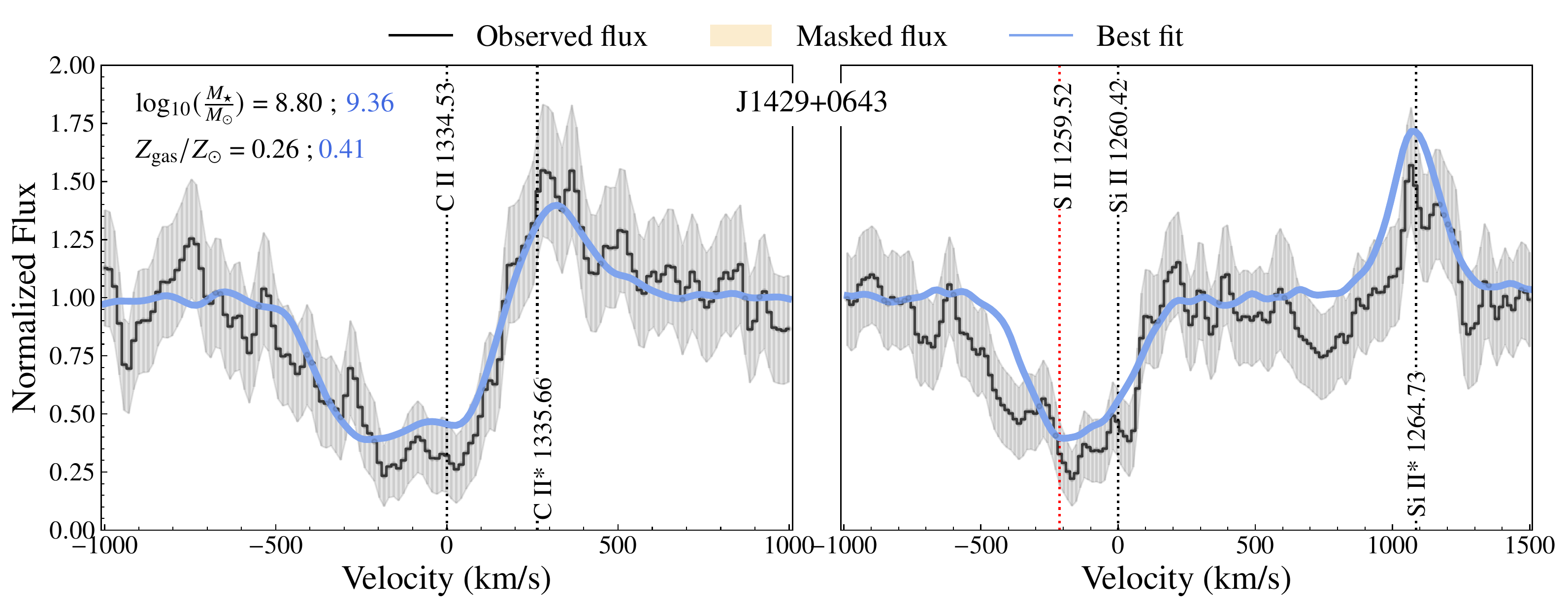}
    \includegraphics[width =0.45\textwidth,trim={0 0 0 1cm},clip]{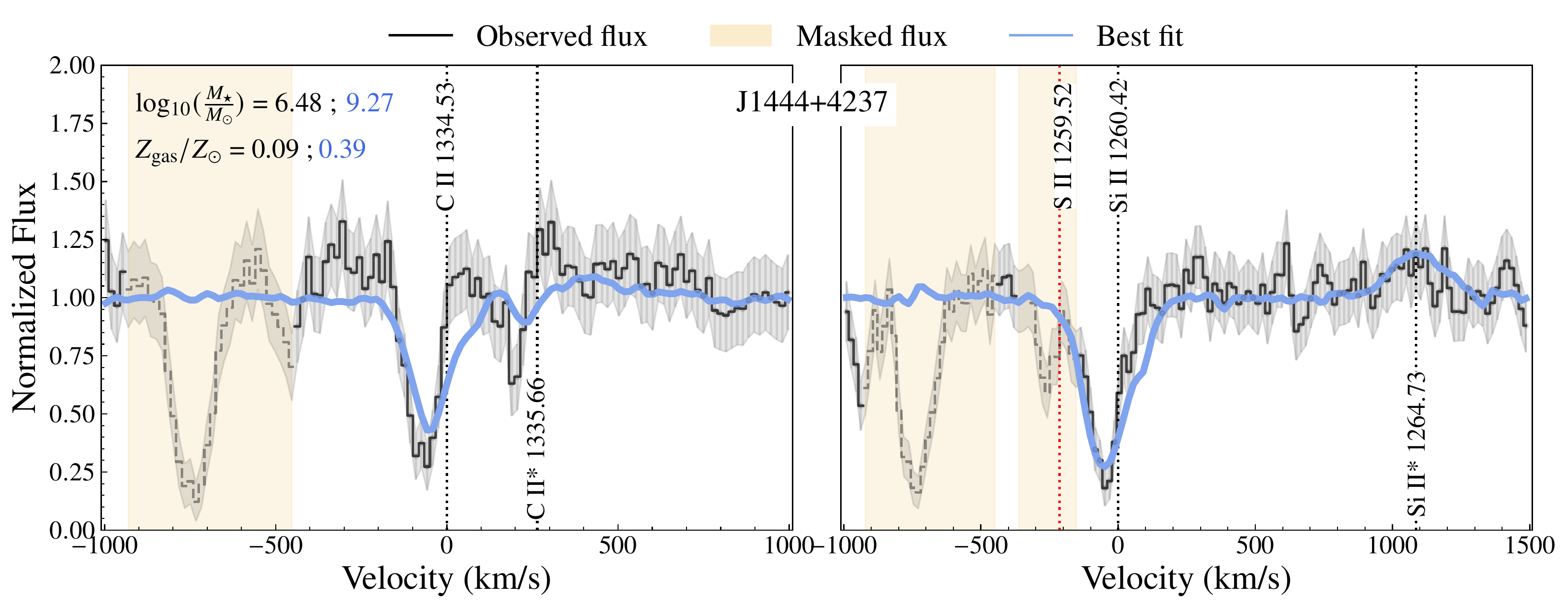}
    \includegraphics[width =0.45\textwidth,trim={0 0 0 1cm},clip]{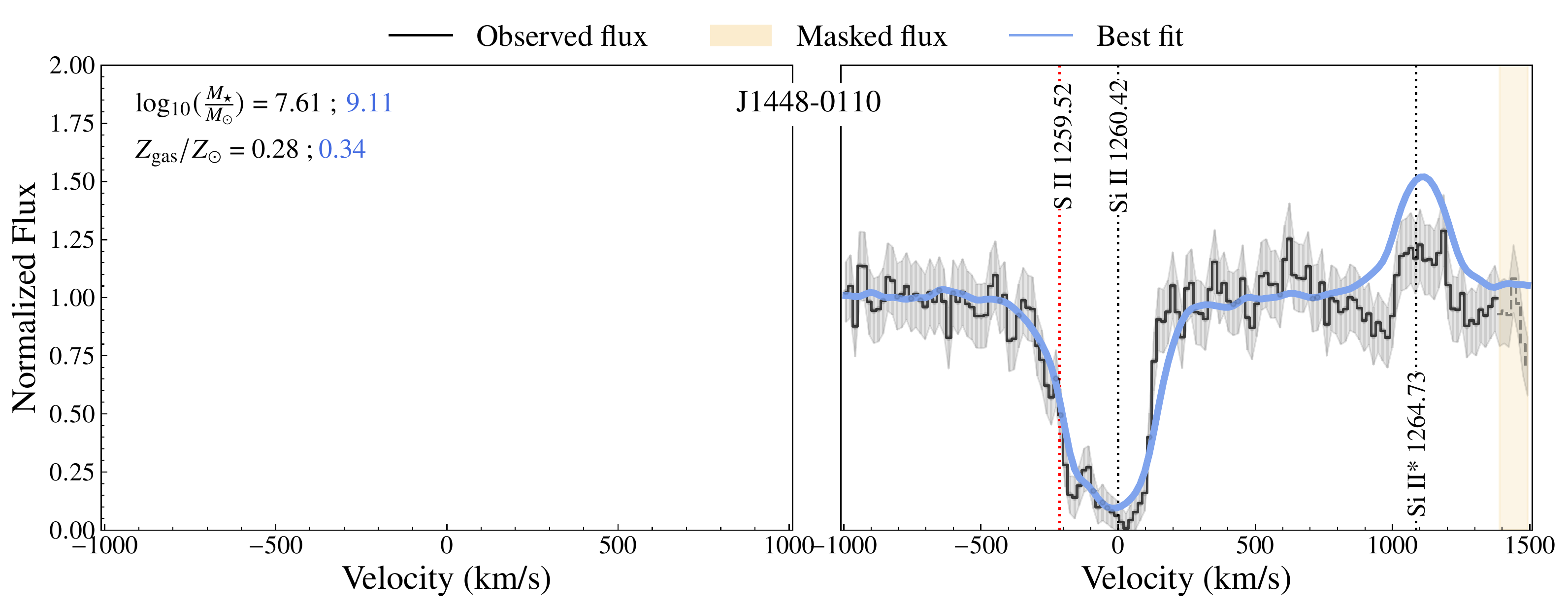}
    \includegraphics[width =0.45\textwidth,trim={0 0 0 1cm},clip]{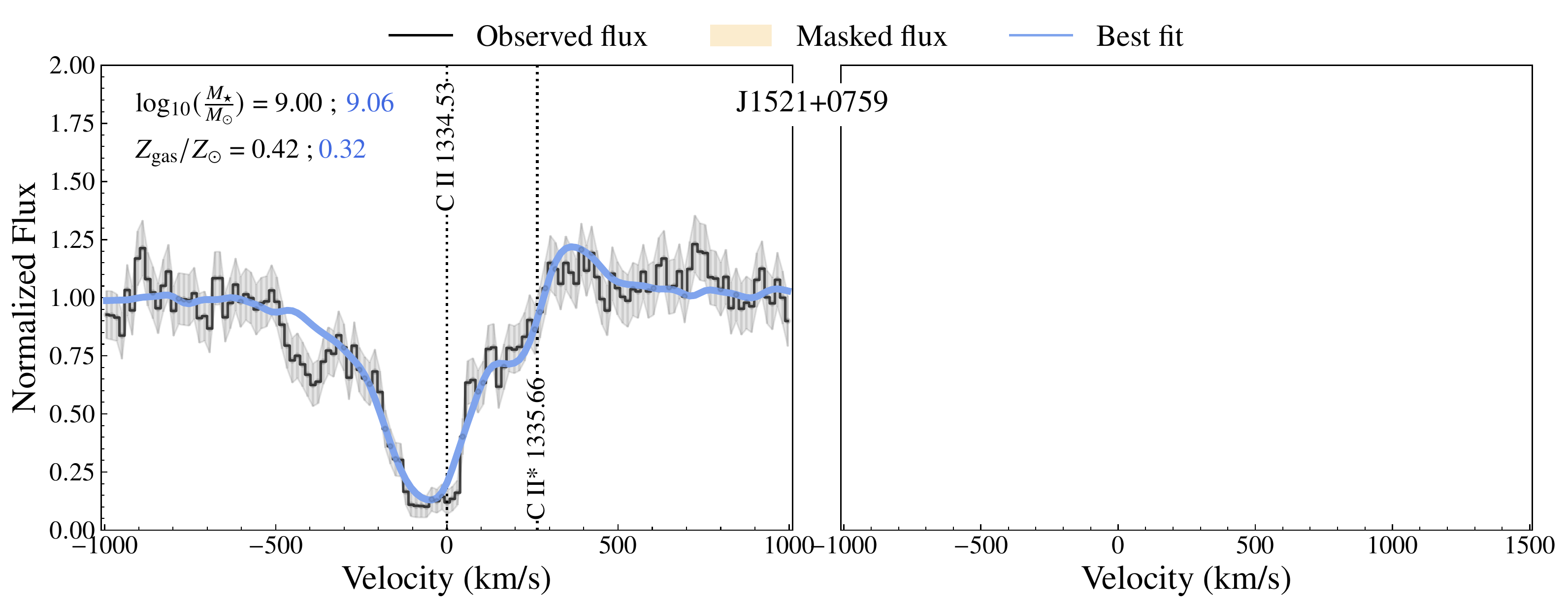}
    \includegraphics[width =0.45\textwidth,trim={0 0 0 1cm},clip]{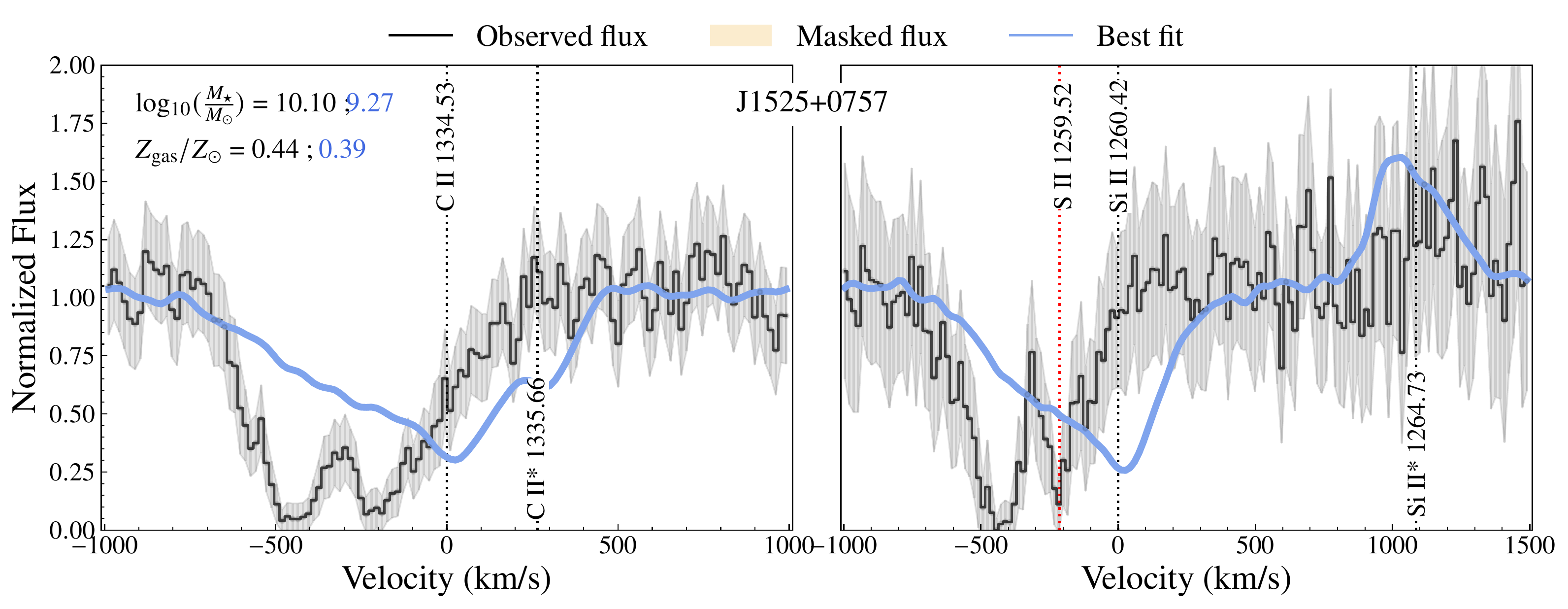}
    \includegraphics[width =0.45\textwidth,trim={0 0 0 1cm},clip]{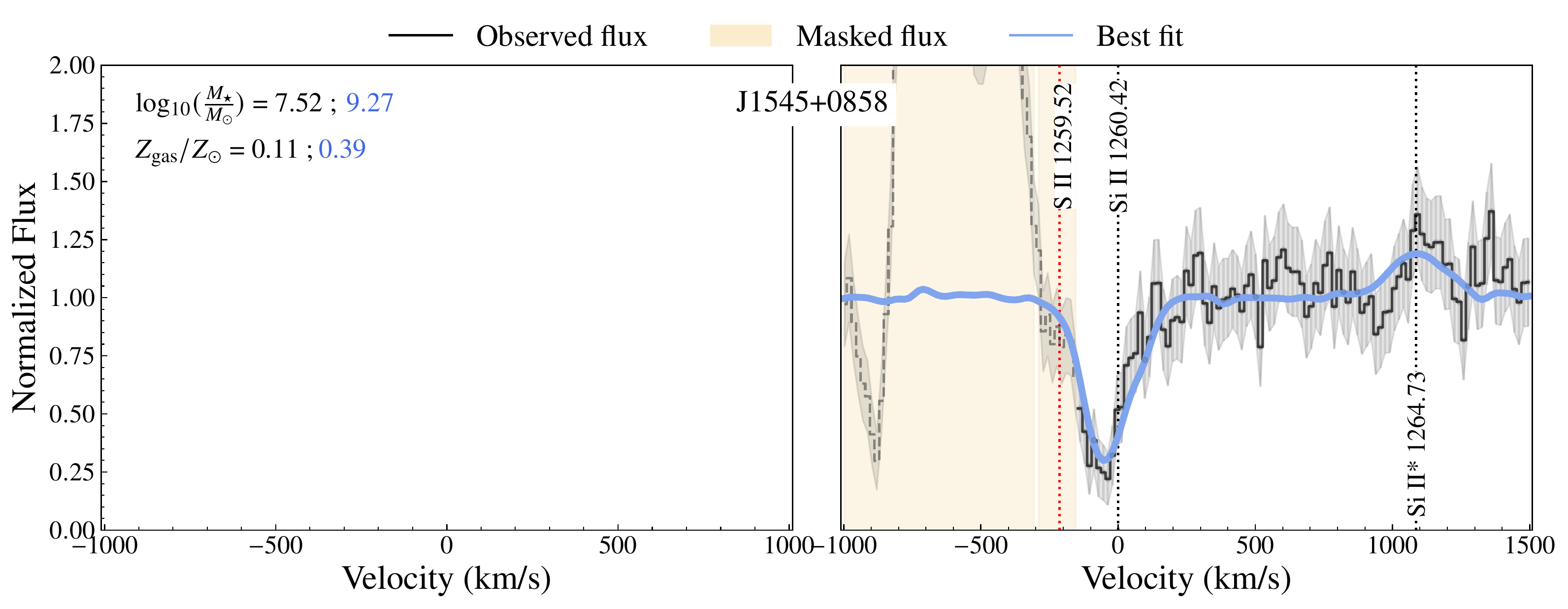}
    \includegraphics[width =0.45\textwidth,trim={0 0 0 1cm},clip]{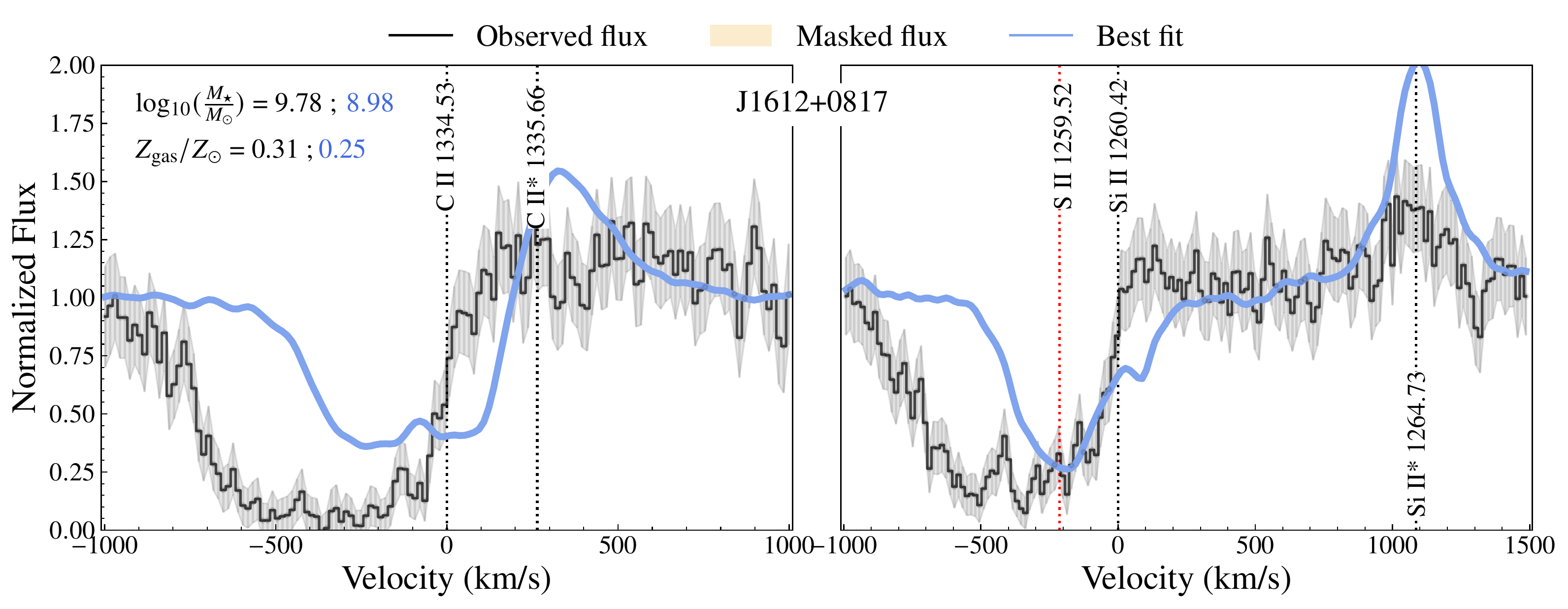}
    \ContinuedFloat
    \caption{\textit{(cont.)}}
\end{figure*}

\subsection{Using the fitting approach of Section~\ref{sec:aperfits}}
\label{app:fits2}

\begin{figure*}[!htbp]
    {\centering
    \includegraphics[width = 0.48\textwidth]{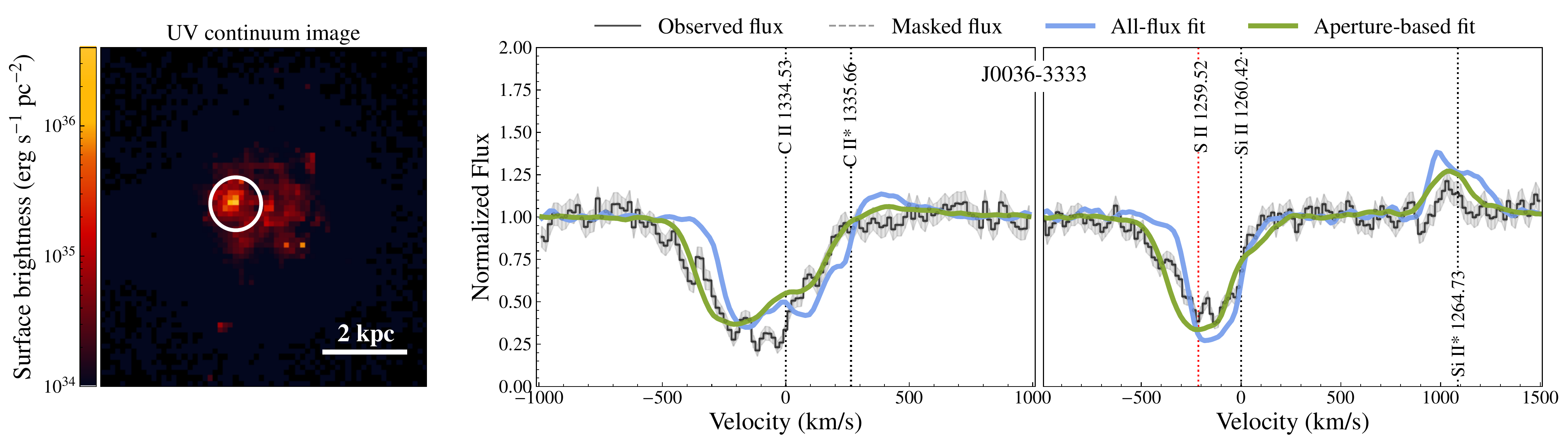}
    \includegraphics[width =0.48\textwidth]{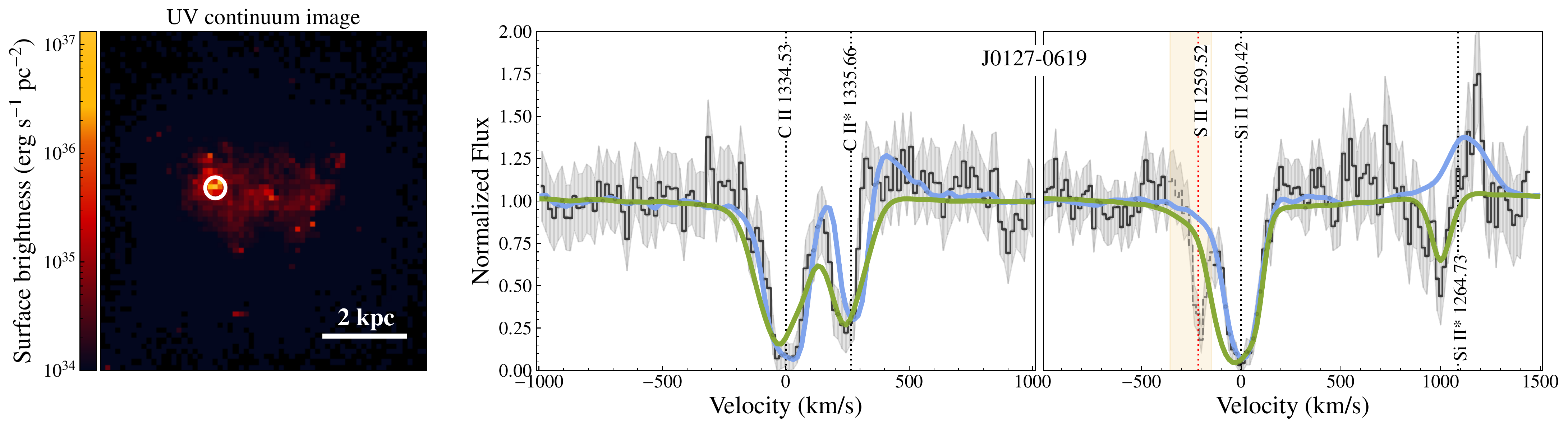}
    \includegraphics[width =0.48\textwidth]{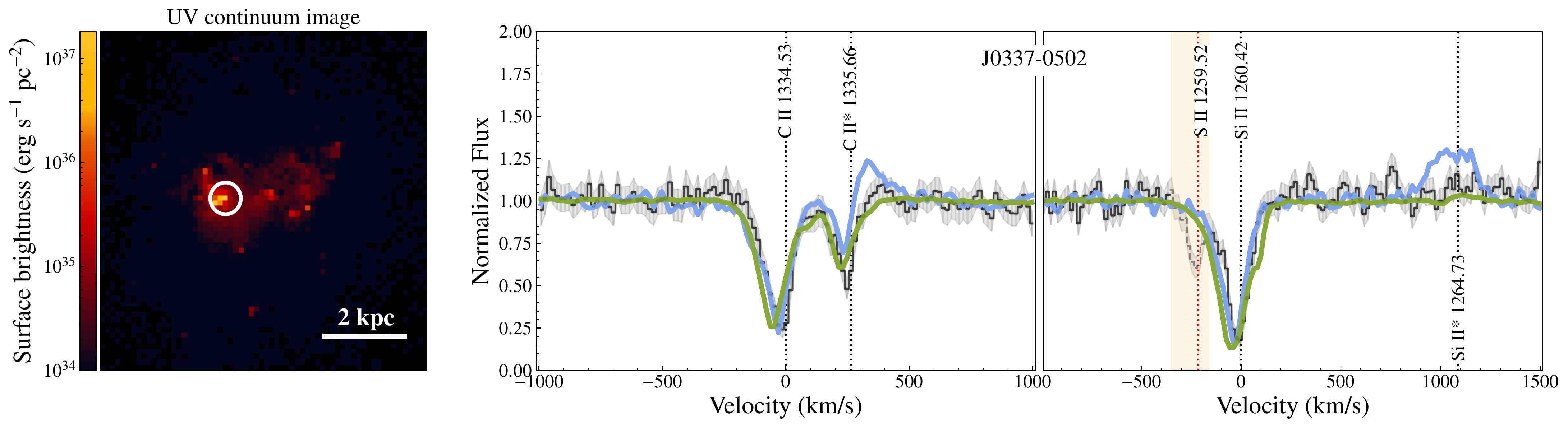}
    \includegraphics[width =0.48\textwidth]{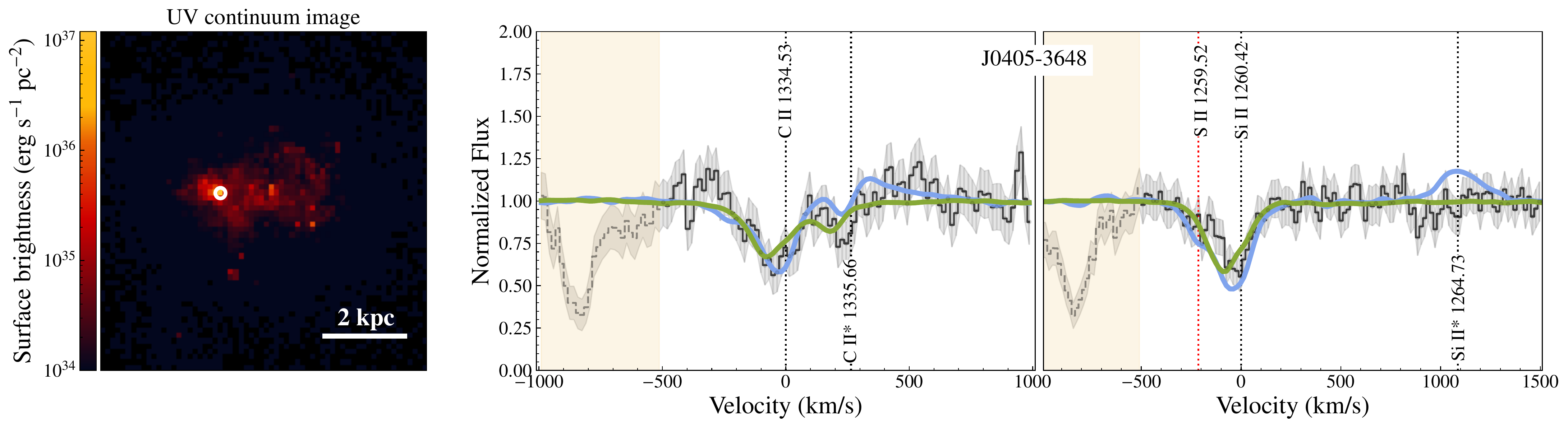}
    \includegraphics[width =0.48\textwidth]{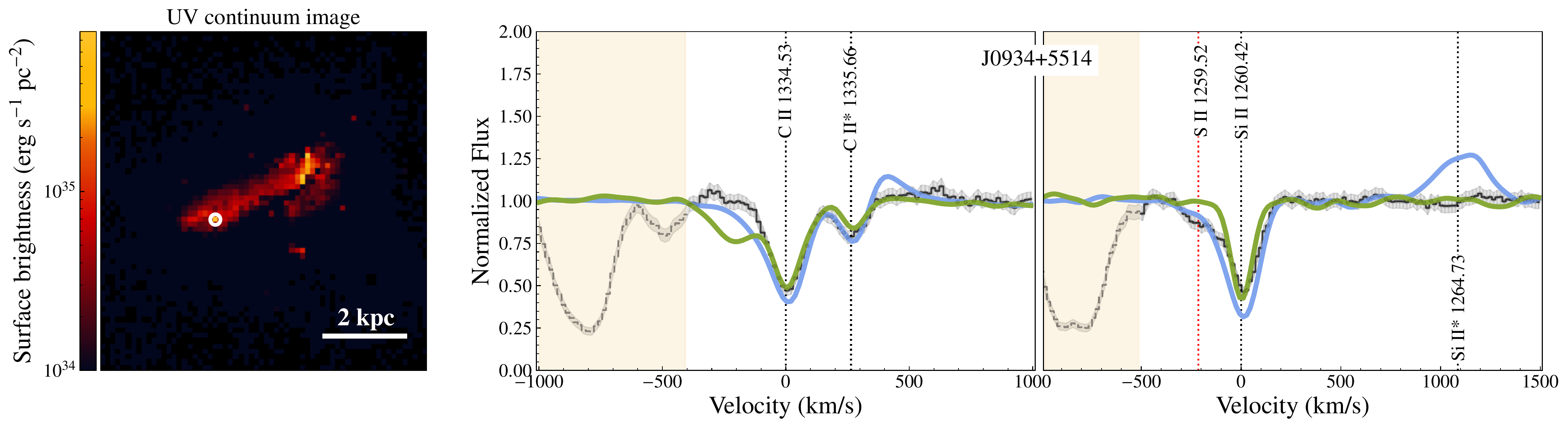}
    \includegraphics[width =0.48\textwidth]{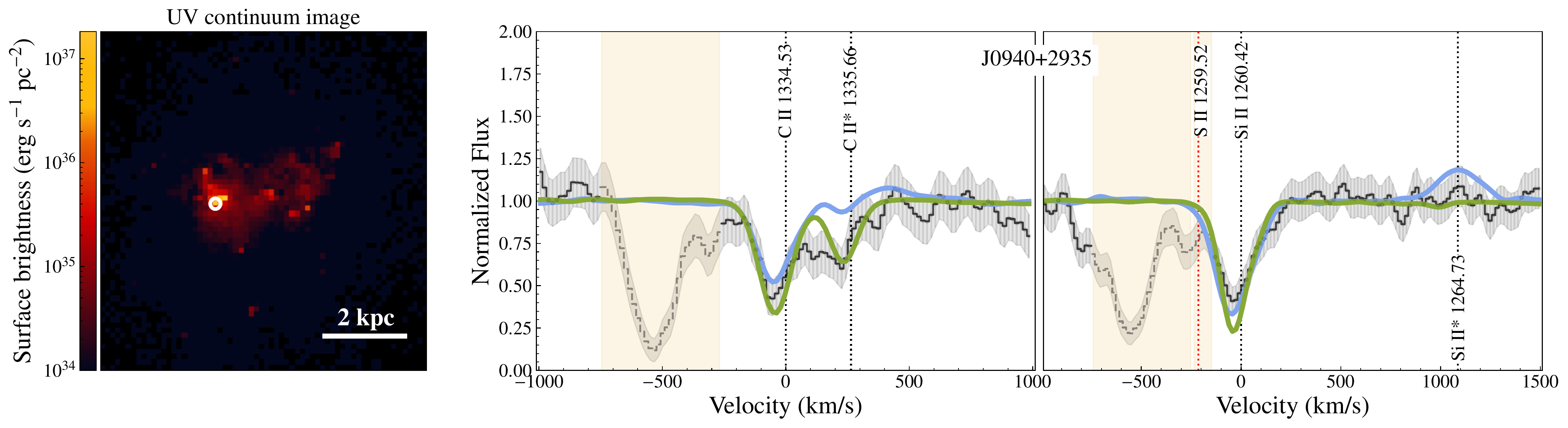}
    \includegraphics[width =0.48\textwidth]{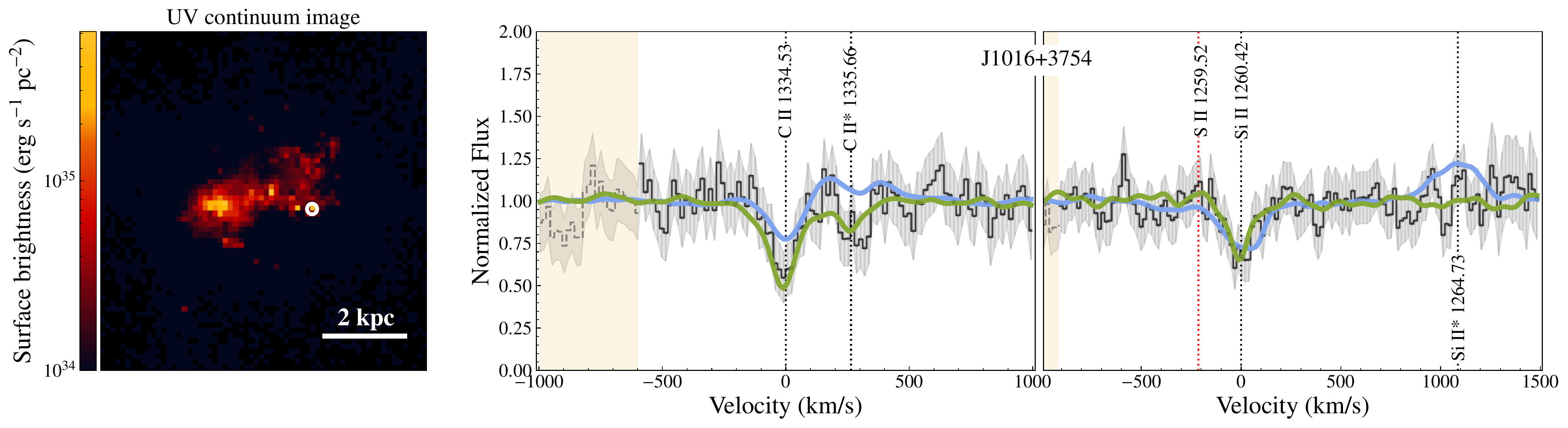}
    \includegraphics[width =0.48\textwidth]{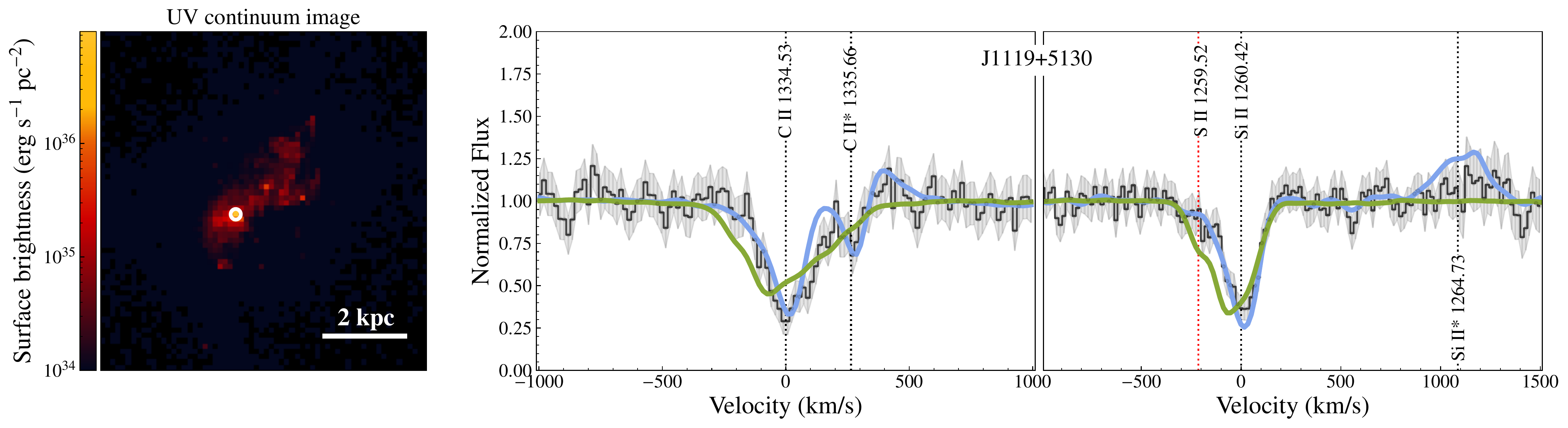}
    \includegraphics[width =0.48\textwidth]{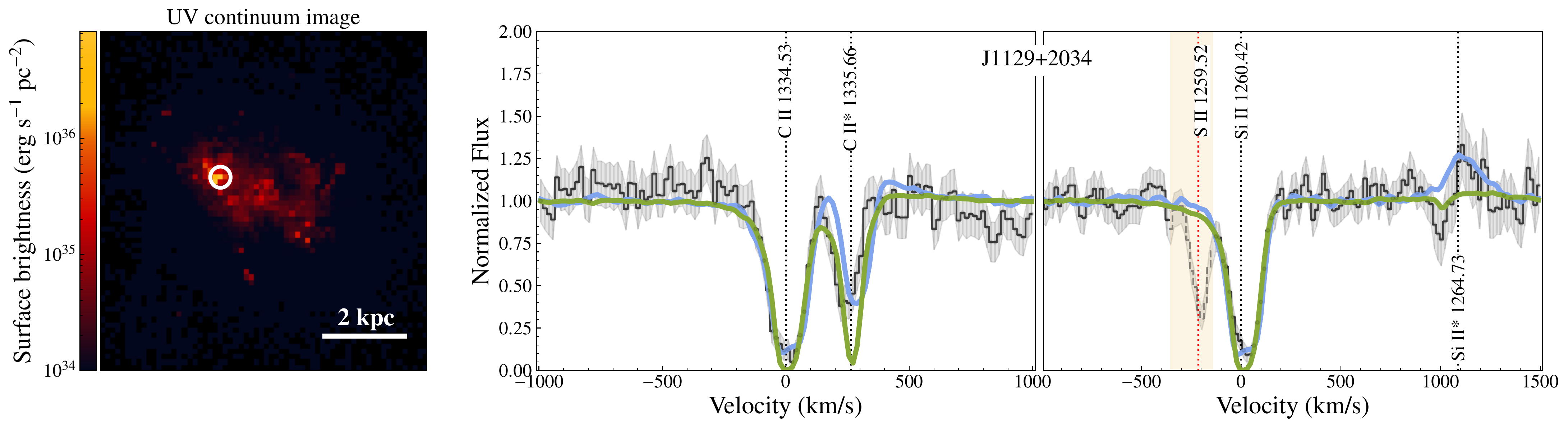}
    \includegraphics[width =0.48\textwidth]{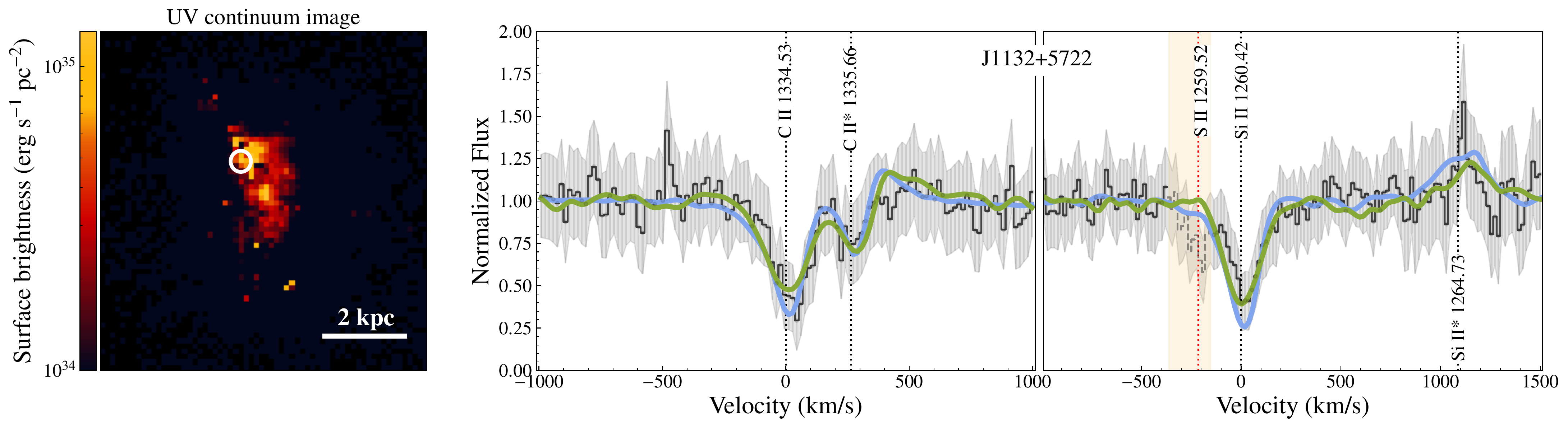}
    \includegraphics[width =0.48\textwidth]{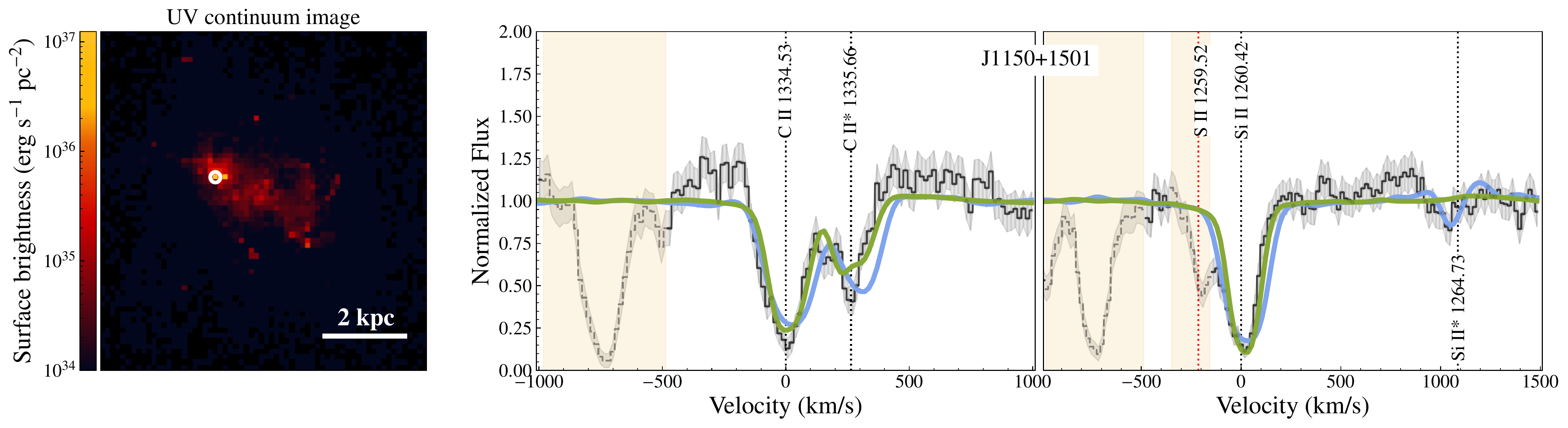}
    \includegraphics[width =0.48\textwidth]{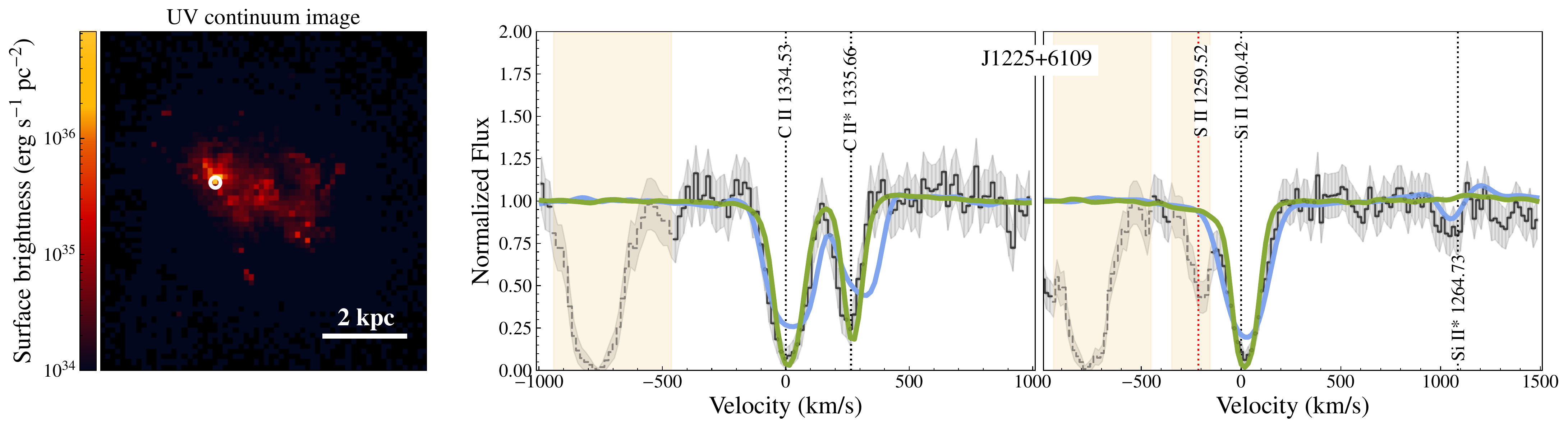}
    \includegraphics[width =0.48\textwidth]{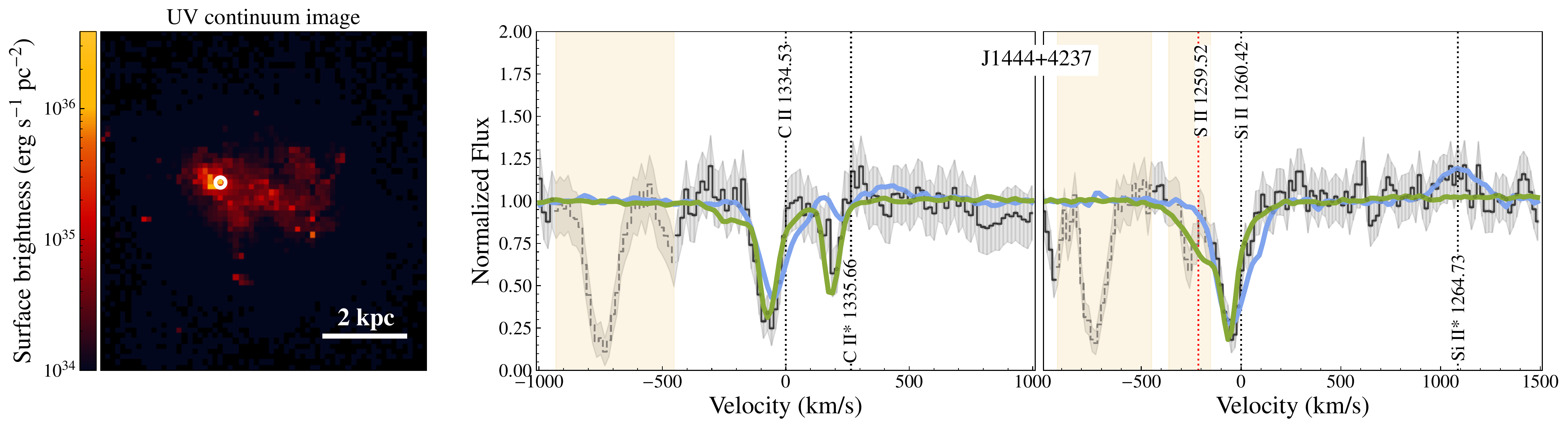}\hfill
    \includegraphics[width =0.48\textwidth]{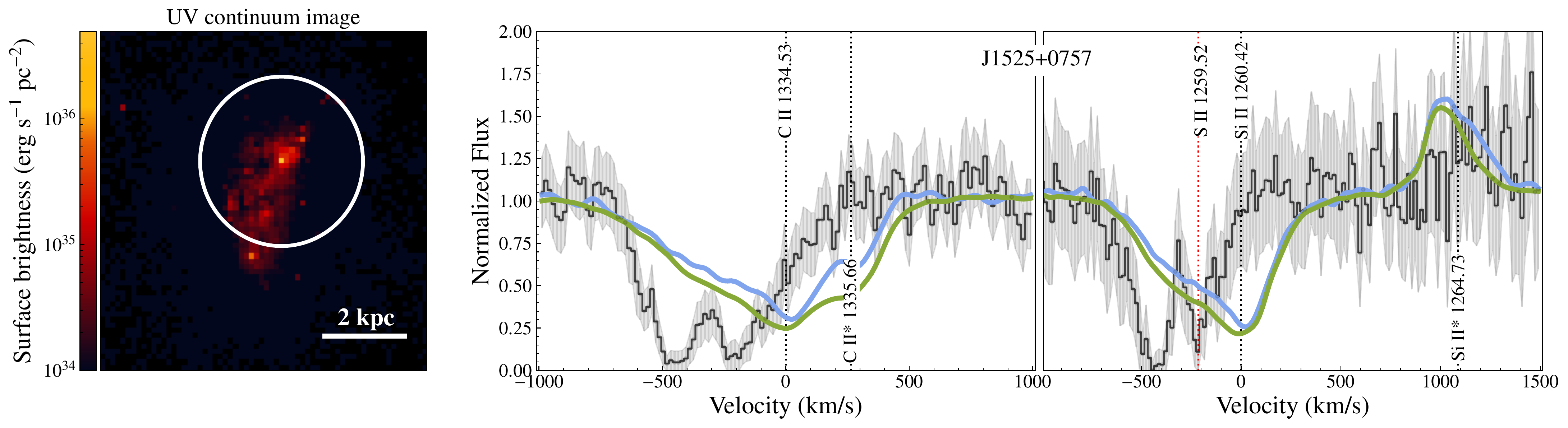}}
    \caption{Comparison of the \cIIl+\cIIlstar\ (middle panels) and \siIIl+\siIIlstar\ (right panels) fits for the 16 \classy\ galaxies which have $r_{50}/1.25 > 1$ and observations for both \siII\ and \cII. The blue line (``All-flux" fit) is the best-fit mock spectrum found in Section~\ref{sec:fitting}, the green line (``Aperture-based fit") is the best-fit mock spectrum obtained using a fiducial aperture size corresponding to the COS aperture size at the galaxy redshift. The location and size of the region from which the green spectrum is extracted are shown with a black circle on the left panel.  }

    \label{fig:fits2a}
\end{figure*}

Figure~\ref{fig:fits2a} presents the best fits of the \classy\ \siII\ and \cII\ spectra using the approach detailed in Section~\ref{sec:aperfits}. The fitting approach is done using a single simulation output where the mock spectra are extracted using circular apertures of which size corresponds to the size of the COS aperture at the galaxy redshift. As discussed in Section~\ref{sec:aperfits}, this refined approach enables improvement of the fits around the fluorescent wavelengths. For this experiment, we only consider galaxies that have observations for both \siII\ and \cII, and have $r_{50}/1.25$ values larger than 1 (i.e., potentially significantly impacted by aperture loss, see Section~\ref{sec:aperprop}).\\

\section{Comparison of the UV and Optical half-light radii}
\label{app:uvopt}

\begin{figure}[!htbp]
    \centering
    \includegraphics[width=\hsize]{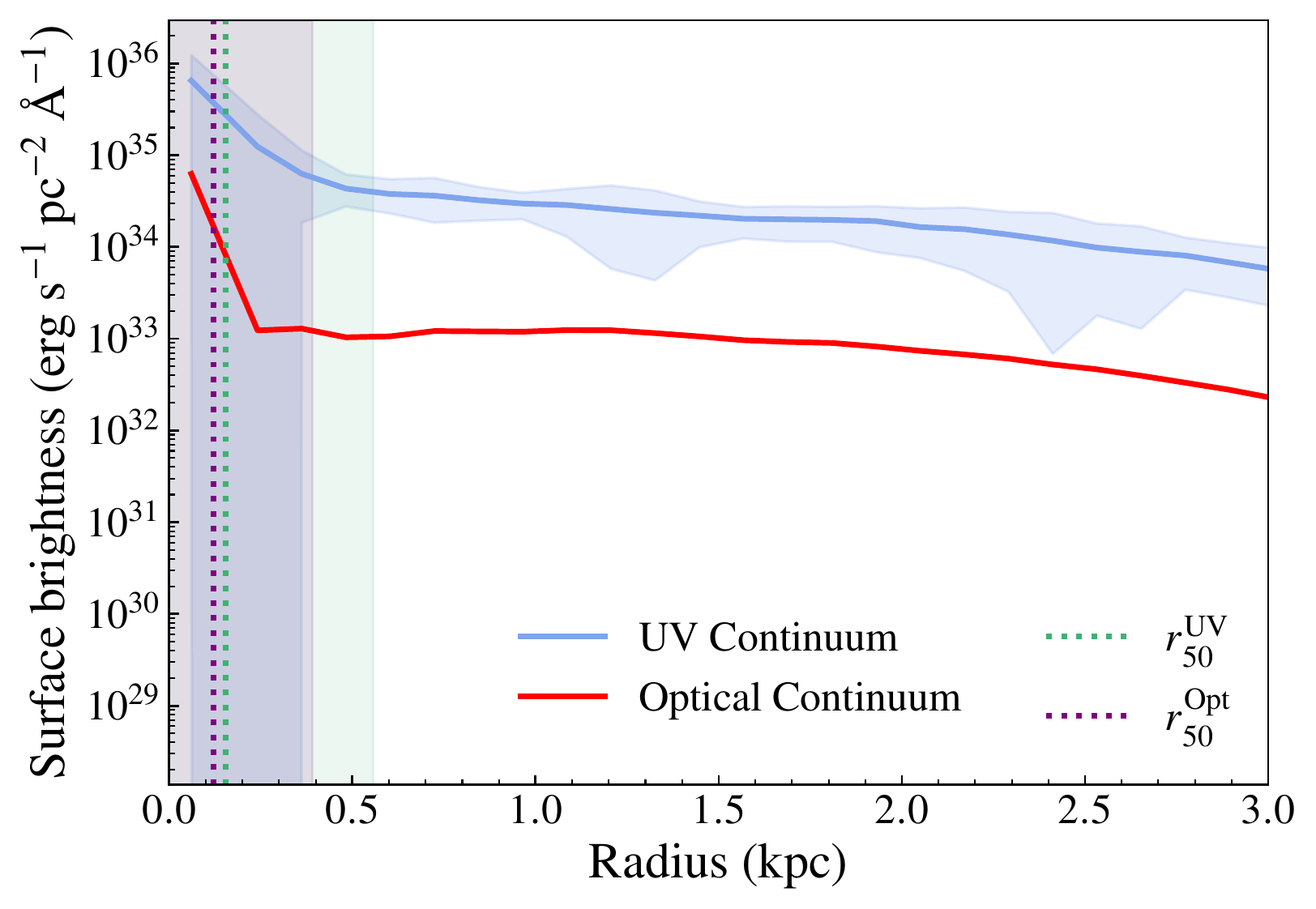}
    \caption{The light curves of the UV and optical continuum derived using the average and standard deviation of 300 line-of-sight images of the galaxy at 1270 and 3500 \AA, respectively. Despite differences regarding the total brightness seen at both wavelengths, the median UV and optical half-light radii determined from these curves only differ by a few parsecs.}
    \label{fig:compuvopt}
\end{figure}

In this section, we estimate and compare the half-light radii estimated from the UV continuum and from the optical continuum for the virtual galaxy. Using the simulation output at t $=$ 2 Gyrs, we extract the surface brightness (SB) profiles at $1270$ \AA\ and at 3500 \AA\ for the UV and optical, for each of the 300 lines of sight available. Figure~\ref{fig:compuvopt} shows the median and standard deviation of the 300 UV and optical surface brightness profiles. For each of the 300 SB profiles, we derive the corresponding $r_{50}^{\rm UV}$ and $r_{50}^{\rm opt}$ and derive the median and standard deviation of the 300 values. We find that $r_{50}^{\rm UV} \approx r_{50}^{\rm opt}$.

\bibliography{bibliographie}{}
\bibliographystyle{aasjournal}



\end{document}